\documentclass[usenatbib]{mn2e}       %
\usepackage{amsmath,amssymb,amsfonts} 
\usepackage{graphics}                 
\usepackage{color}                    
\usepackage{hyperref}                 
\usepackage{graphicx}
\usepackage{mathrsfs}
\usepackage{nicefrac}
\usepackage{topcapt}
\usepackage{supertabular}
\DeclareFontFamily{U}{euc}{}
\DeclareFontShape{U}{euc}{m}{n}{<-6>eurm5<6-8>eurm7<8->eurm10}{}%
\DeclareSymbolFont{AMSc}{U}{euc}{m}{n} 
\DeclareMathSymbol{\upmu}{\mathord}{AMSc}{"16} 

\newcommand{\epi}{E_{p,int}}
\newcommand{\epo}{E_{p,obs}}
\newcommand{\epobs}{{\it E}$_{p,obs}$~}
\newcommand{\eiso}{E_{iso}}

\newcommand{\liso}{L_{iso}}
\newcommand{\sbol}{S_{bol}}

\newcommand{\hr}{H\!R_{H}}
\newcommand{\hrh}{{\it HR}$_{H}$~}
\newcommand{\hrs}{\text{\it{HR}}_{S}}

\newcommand{\hrep}{{\it HR}$_{H}$--{\it E}$_{p,obs}$}

\title{Hardness as a Spectral Peak Estimator for Gamma-Ray Bursts}
\author[A. Shahmoradi and R. J. Nemiroff]{A. Shahmoradi$^{1}$\thanks{E-mail:
ashahmor@mtu.edu; nemiroff@mtu.edu} and R. J. Nemiroff$^{1}$\footnotemark[1]\\
$^{1}$Department of Physics, Michigan Technological University, Houghton, MI 49931}

\begin{document}

\date{\date{Accepted ... Received ... ; in original form ...}}

\pagerange{\pageref{firstpage}--\pageref{lastpage}} \pubyear{2009}

\maketitle

\label{firstpage}

\begin{abstract}
Simple hardness ratios are found to be a good estimator for the spectral peak energy in Gamma-Ray Bursts (GRBs).  Specifically, a high correlation strength is found between the $\nu F_{\nu}$ peak in the spectrum of BATSE GRBs, $\epo$, and the hardness of GRBs, $\hr$, as defined by the fluences in channels 3 and 4, divided by the combined fluences in channels 1 and 2 of the BATSE Large Area Detectors.  The correlation is independent of the type of the burst, whether Long-duration GRB (LGRB) or Short-duration (SGRB) and remains almost linear over the wide range of the BATSE energy window (20-2000 KeV).  Based on Bayes theorem and Markov Chain Monte Carlo techniques, we also present multivariate analyses of the observational data while accounting for data truncation and sample-incompleteness.  Prediction intervals for the proposed \hrep ~relation are derived.  Results and further simulations are used to compute $\epo$ estimates for nearly the entire BATSE catalog: 2130 GRBs.  These results may be useful for investigating the cosmological utility of the spectral peak in GRBs intrinsic luminosity estimates.  
\end{abstract}


\section{Introduction}

One of the most widely used spectral parameters in the studies of GRBs is the time-integrated $\nu F_{\nu}$ spectrum peak energy of these cosmic events.  Since the early 1990's there has been a growing trend (e.g. Liang 1989) to plot the GRBs spectra in the form of $E^{2}dE$ or $\nu F_{\nu}$ vs. energy, where $F_{\nu}$ is the spectral flux at the frequency $\nu$. This has the advantage of making it easy to discern the energy of the peak power from the burst. The $\nu F_{\nu}$ plot of many of the bursts' spectra shows a peak which is denoted by \epobs. 

Among all the gamma-ray observatories that have detected GRBs, the Burst and Transient Source Experiment (BATSE) onboard the now defunct Compton Gamma Ray Observatory (CGRO) has provided the largest GRB database, consisting of observational data for 2704 GRBs.  The \epobs of BATSE GRBs that have been spectrally analyzed indicates a narrow distribution extending from tens of KeV to a few MeV (Preece et al. 2000).  However, there has been great debate on whether the upper and lower cutoff in the tails of the \epobs distribution are intrinsic to GRBs or strongly affected by the sensitivity of detectors (e.g. Piran 2005; Lloyd \& Petrosian 1999; Higdon \& Lingenfelter 1998; Cohen et al. 1997; Piran \& Narayan 1996). 

The discovery of X-Ray Flashes (XRFs) with familiar temporal structure to GRBs but with typically lower peak energies (Kippen et al. 2004; Kippen et al. 2002; Heise et al., 2001) indicates that the lower cutoff observed in the distribution of \epobs is not real and is due to either the low sensitivity of BATSE LAD detectors in this energy range or sample incompleteness caused by the limitations of the spectral analysis. Similarly, the reality of the observed cutoff in the upper tail of \epobs distribution has also been questioned by the references above. However, a study of the Solar Maximum Mission (SMM) data (Harris \& Share, 1998) suggests that there is a deficiency - by at least a factor of 5 - of GRBs with \epobs above $3MeV$ relative to GRBs peaking at $\sim 0.5MeV$. But these data are consistent with a population of peak energy that extends up to $2MeV$.

There have also been reports on the existence of correlations between the rest frame spectral peak energies ($\epi$) of GRBs and their 1-second isotropic peak luminosity ($\liso$), as well as reports on correlations between $\epi$ and isotropic-equivalent radiated energy ($\eiso$) (Schaefer 2007; Ghirlanda et al. 2007; Amati 2006; Amati et al. 2002). However, one major problem with these relations is sample incompleteness. In particular, the Amati \& the proposed $\liso-\epi$ relations have been constructed from less than 100 bright GRBs.  Several studies have questioned the validity of these relations, arguing that some GRBs do not obey them (e.g. Shahmoradi \& Nemiroff 2009a; Butler et al. 2009a; Butler et al. 2007; Nakar \& Piran 2005; Band \& Preece 2005), and that it is plausible that some part of these relations are due to complex selection effects in the detection and measurement processes.

So far, the spectra of only 350 out of 2704 GRBs detected by BATSE have been analyzed in detail using a variety of spectral models (e.g. Kaneko et al. 2006, hereafter K06).   Although these 350 GRBs need to be especially bright to allow for accurate spectral analyses, inclusion rules, such as requiring a minimum flux or Signal-to-Noise Ratio ($S/N$), may carry significant limitations and biases (Sharmoradi and Nemiroff 2009a).

In this paper we propose a novel method that is designed to increase the number of usable GRBs in spectral correlations and population studies while reducing limitations and biases: using a hardness ratio to estimate $\epo$.  We define a new hardness ratio \hrh that is the sum of the fluences in channel 4 and 3 of BATSE LAD detectors, divided by the sum of the fluences in channel 2 and 1. Equivalently, this corresponds to dividing the energy fluence received by BATSE triggered LADs in the energy range (300 - 2000 KeV) by the energy fluence received in the energy range (20 - 300 KeV).   We will show here this and other hardness ratios are in fact good estimators of $\epo$.  
\\

The plan of the paper is as follows: \S2 describes the \hrep ~relation while briefly discussing the significant differences that exist between the reported values of \epobs of BATSE GRBs due to the use of different spectral models to fit the data. \S3 is spent on the derivation of the prediction intervals for the relation. The estimation and prediction power of the relation will also be discussed there. Some examples on the applications of the relation, in particular, the $\epo$ distribution of BATSE GRBs will be discussed and the results will be summarized in \S4.

\begin{figure}
\includegraphics[scale=0.31]{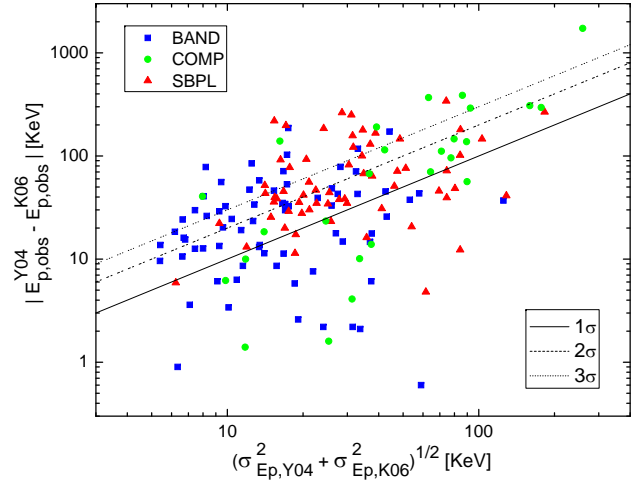}
\caption{Differences in \epobs vs. total 1$\sigma$ errors in \epobs for Y04 and K06 GRBs.  For any point that lies above the solid line$/$dashed line$/$dotted line, the two reported values of \epobs for that GRB are inconsistent at more than $1\sigma /2\sigma /3\sigma$ level.  Therefore, 69\% (26\%) of all bursts have an \epobs reported by K06 that is inconsistent at the $>1\sigma$ ($>3\sigma$) level with the \epobs reported by Y04. \label{Y04K06diff}}
\end{figure}

\section{\epobs ESTIMATION FOR BATSE BURSTS}

The fluences in different channels are taken from the current BATSE catalog available at the HEASARC archives \footnotemark \footnotetext{http://gammaray.msfc.nasa.gov/batse/grb/catalog/current/}.   Immediately, a strong correlation was found between \hrh and \epobs for 249 BATSE bright bursts that are common between the current BATSE catalog and 350 BATSE bursts analyzed by K06 (Kendall's $\tau_{K}=0.68, 16\sigma$).   

In fact, for many GRB detectors such as BATSE, \hrh might even be a preferred measure of GRB spectral peakedness in $\nu F_{\nu}$ than \epobs for several reasons.  First, \hrh is relatively immune to details of GRB spectral analyses, and therefore, quite possibly, hiding fewer unexpected thresholds and hidden selection effects.  Additionally, \hrh is easier to measure for faint GRBs, allowing \hrh to be computed with little statistical error for many more GRBs than 350 -- in fact for most BATSE GRBs. 

We use \hrh rather than any other definition of hardness ratio to estimate \epobs because of its relatively low statistical variance, the boundary energy between high and low energies (100 KeV), and its strong linear correlation with \epobs (Pearson's correlation coefficients: $0.89$,$0.88$ \& $0.88$ at $>13\sigma$, $>10\sigma$ \& $>14\sigma$ significances for the three Band, COMP(CPL), and SBPL spectral models respectively).    The existence of such a strong positive correlation might not be very surprising since both parameters \epobs \& \hrh are measures of the spectral hardness of a GRB.  An unexpected but useful perk is that the relation is nearly linear over a wide range of BATSE \epobs energies. 

\begin{figure}
\includegraphics[scale=0.31]{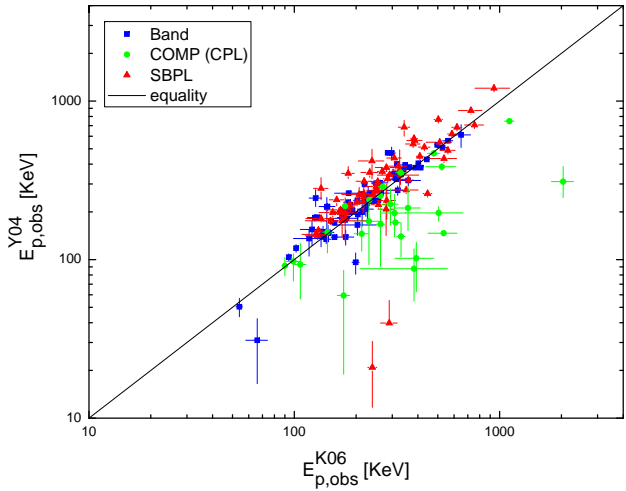}
\caption{Direct comparison of \epobs estimates given by K06 and Y04 for the same sample of BATSE LGRBs: The sample for which K06 find Band as the best fit model (blue squares), the sample for which K06 find COMP as the best fit model (green circles), and the sample for which K06 find SBPL as the best fit model (red triangles). Note that Y04 use Band model for all GRBs. The solid line in the graph delineates where the two $\epo$ estimates would be equal. Comparison of the two $\epo$ estimates reveals that the presumed spectral model can have a significant effect on the derived spectral parameters. \label{BandCOMPSBPL}}
\end{figure}

\subsection{Systematic Errors in \epobs Estimates}

Although a decade has passed since BATSE ended its mission, less than one quarter of the total number of GRBs detected by BATSE have been spectrally analyzed to date. Examples of such works include Ghirlanda et al. (2009), hereafter G09; K06; Yonetoku et al. (2004), hereafter Y04; Preece et al. (2000). In an impressive work, K06 have analyzed 350 brightest BATSE GRBs, and fit the spectra of the bursts with a variety of spectral models. BATSE GRBs spectra are most commonly fit with an empirically determined double power law connected by a smoothly fit transition region (Band et al. 1993, commonly referred to as the ``Band model").  When plotted as $\nu F_{\nu}$ versus $\nu$, a peak energy is evident, usually referred to as \epobs.  Other spectral models exist that exhibit a similar peak.  It has been noted, however, that spectral parameters including \epobs are highly dependent on which spectral model is being fit, leading to non-negligible differences in the various published values of the bursts properties by independent authors (e.g. Collazzi \& Schaefer 2008; K06; Preece et al. 2002; Ghirlanda et al. 2002). 

In order to show the biases produced by using different spectral models, we compared the peak energies of 161 BATSE GRBs obtained by Y04 using the Band model to the peak energies of the same bursts reported by K06 using three different models: Band, Comptonized Model (COMP, also known as CPL; Pendleton et al. 1997; Mazets et al. 1981), and Smoothly-Broken Power Law (SBPL) (Preece et al. 2000; Ryde 1998). The variations are seen graphically in Figure~\ref{Y04K06diff}, which shows the difference in \epobs values reported by Y04 and K06 vs. the square-root of the sum of the squares of their $1\sigma$ error bars for each individual burst. We divided the whole sample into three subsamples delineated by Band, CPL and SBPL in K06. For any point that lies above the solid line in Figure~\ref{Y04K06diff}, the two reported values of \epobs of the GRB are inconsistent at more than their given $1\sigma$ uncertainties in Y04 and K06.

Surprisingly, $63\%$ of the Y04 \epobs are inconsistent with K06 \epobs for the GRBs best described by the Band model in K06 at $>1\sigma$ level. This ratio is even higher for the case of GRBs best described by COMP model ($65\%$) and by SBPL model ($78\%$). Overall, we find that 69\% (26\%) of all bursts have reported \epobs by K06 and Y04 that are inconsistent with each other at $>1\sigma$ ($>3\sigma$) level.  Moreover, the \epobs differences do not show a Gaussian distribution about zero, particularly those described by CPL, and to a lesser extent those described by SBPL. This implies the existence of a systematic bias in \epobs of Y04 due to the use of the Band model as an `a priori' best fit to all LGRBs.

Inspection of Figure~\ref{BandCOMPSBPL} indicates that Y04 underestimates \epobs, relative to K06, for those LGRBs that are best fit by CPL, while Y04 slightly overestimates \epobs for GRBs best fit by SBPL. For one burst in the CPL subsample (BATSE trigger 6539) there is an extremely large difference between the two peak energies ($E_{p,K06}=2039\pm251$, $E_{p,Y04}=310.4 _{76.9}^{64.2}$). Excluding this burst, the average difference of the peak energies for COMP subsample is 109 KeV. Overall, the similarity of Y04 and K06 \epobs estimates can be rejected at $>46\sigma$, $>35\sigma$ \& $>74\sigma$ for the three Band, COMP \& SBPL spectral models respectively, according to $\chi^2$ test. These discrepancies indicate that systematic errors and uncertainties due to model biases are significant.  In addition, K06 find that very frequently, some time-resolved spectra cannot be adequately fit by the Band model.

\subsection{\hrep ~Correlation}

As shown in the previous section, the use of a specific spectral model as an `a priori' best fit can result in a significant underestimation or overestimation of \epobs of GRBs. In order to nullify the possible effects of the spectral model, several authors attempted to simultaneously fit the spectra of GRBs with different models and choose the one with the least $\chi^2$ value as the best fit. The most comprehensive analysis of this type for BATSE GRBs has been done by K06 who consider five different spectral models in their analyses and spectral fits. Although other authors (e.g. G09; Nava et al. 2008; Y04; Band \& Preece 2005) have extended the number of GRBs with measured \epobs, for the reasons discussed above, we rely only on the sample of GRBs with measured \epobs reported by K06 to find the best linear fit for the \hrep ~relation. K06 provide the most precise \epobs measurement for the bright BATSE GRBs to date.

\begin{figure}
\includegraphics[scale=0.31]{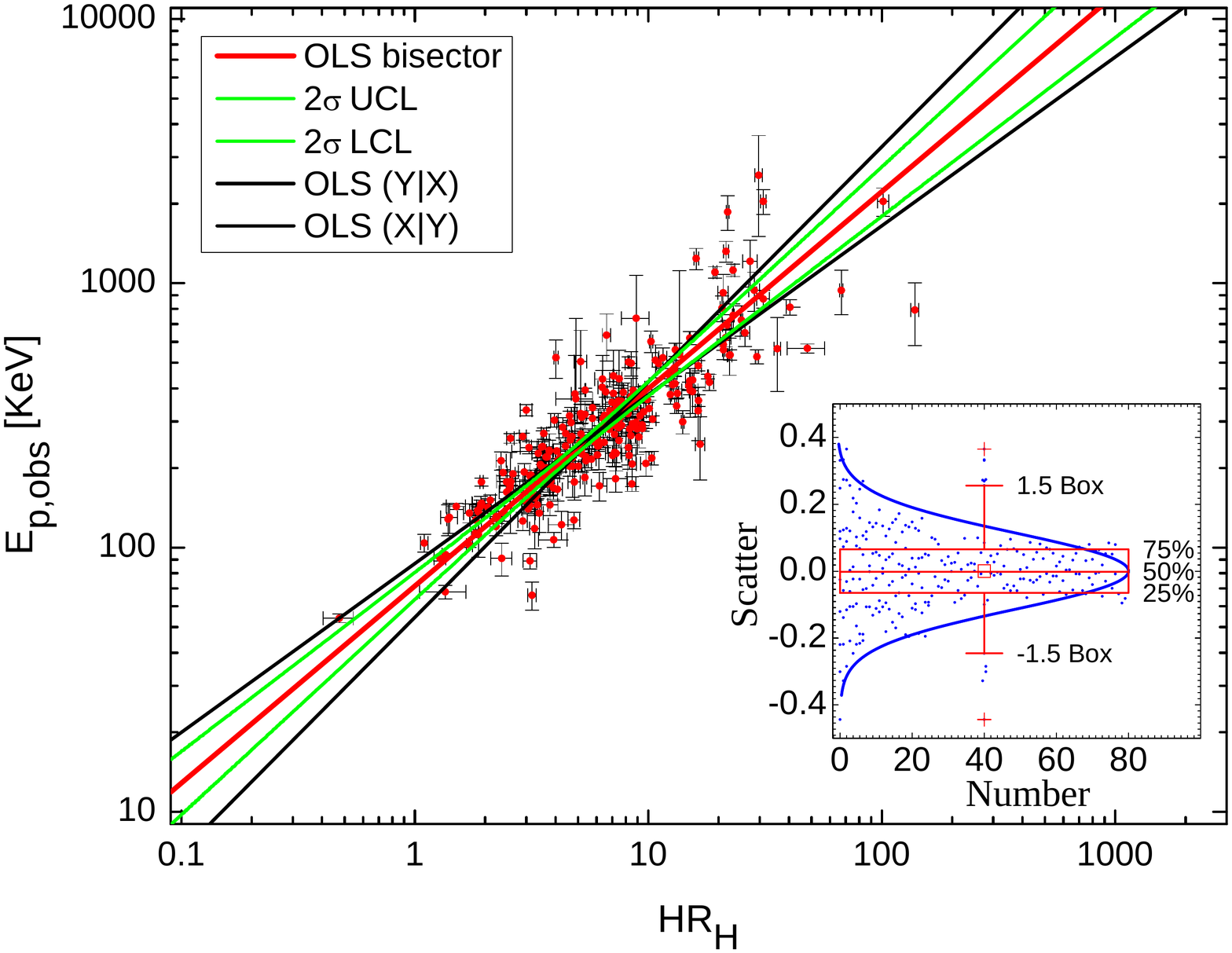}
\includegraphics[scale=0.31]{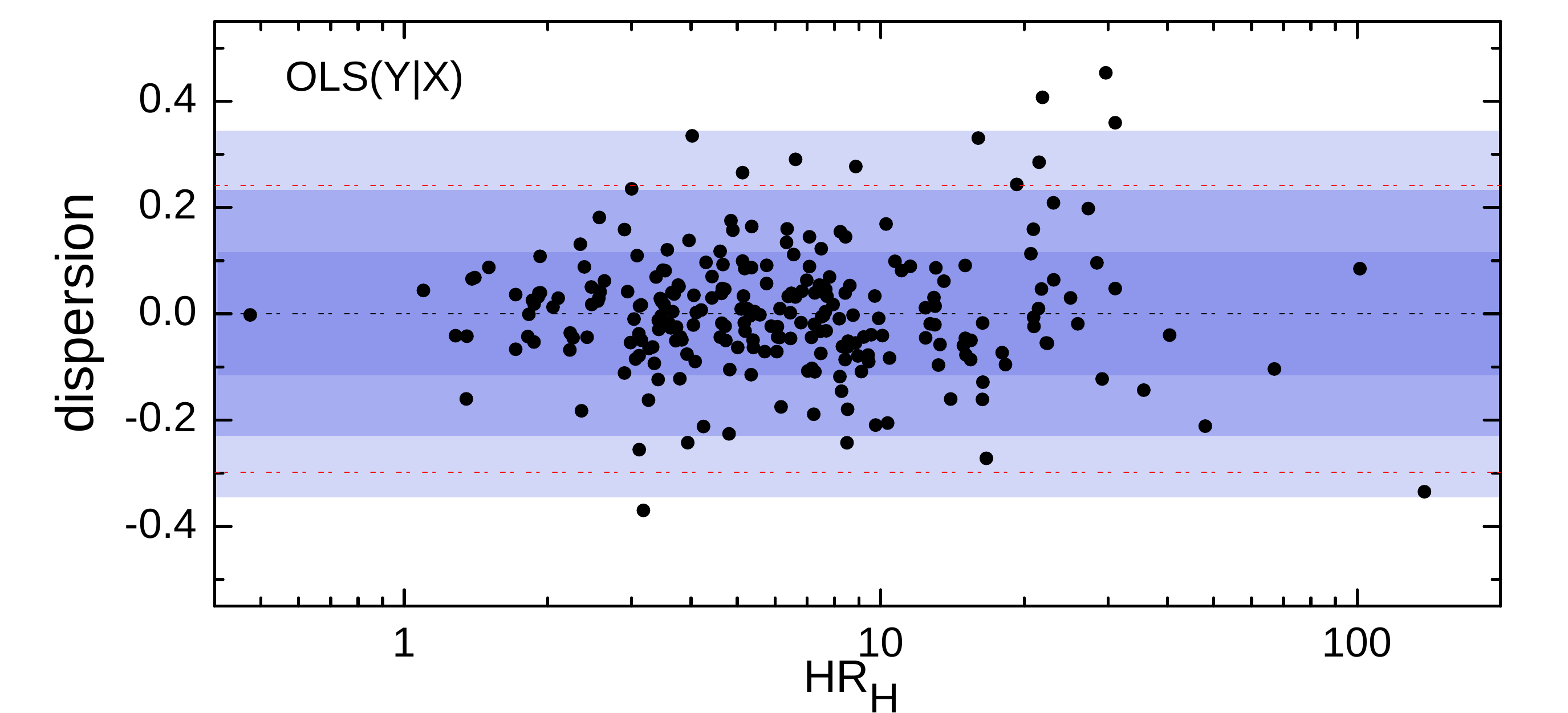}
\includegraphics[scale=0.31]{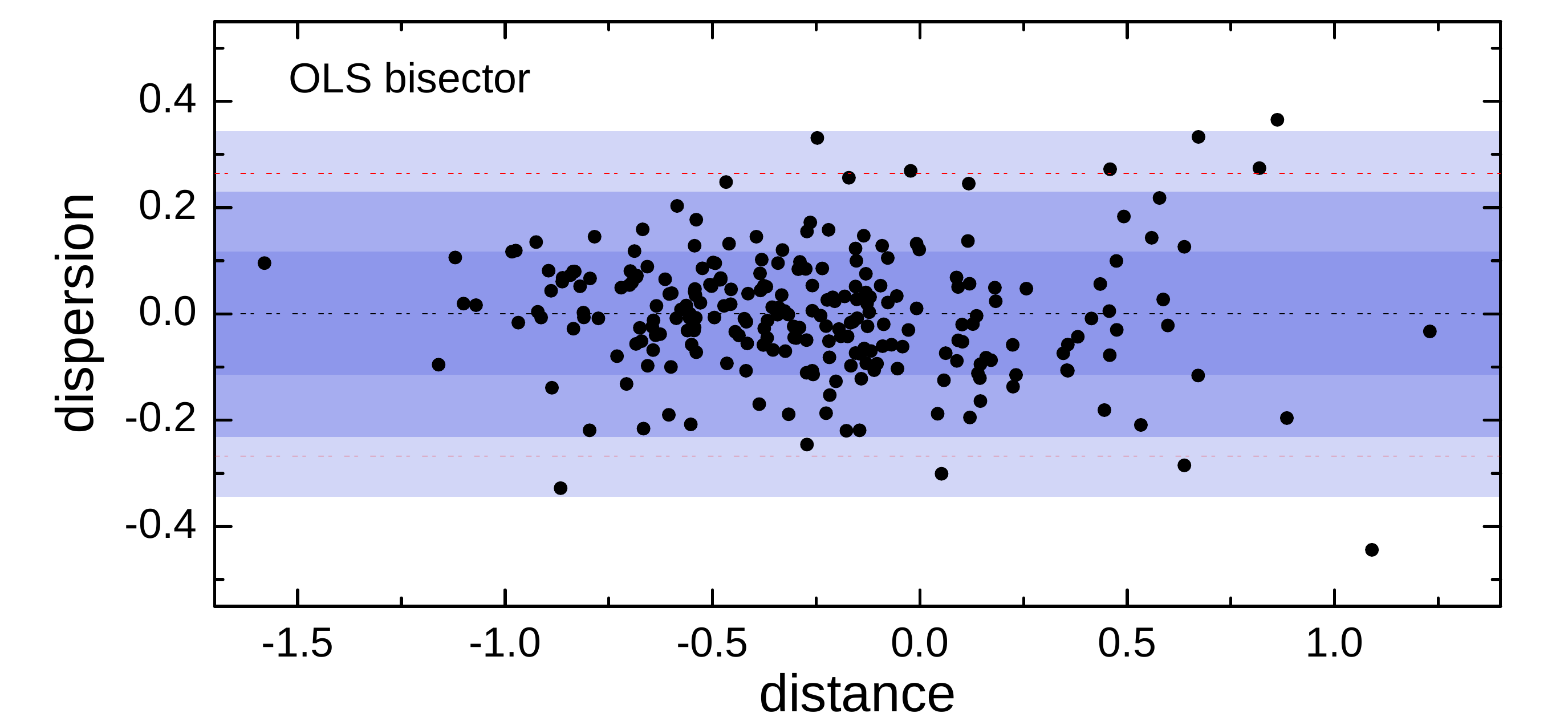}
\caption{{\bf Top:} Plot of \epobs versus \hrh as defined in \S 1. \epobs of 249 bright BATSE bursts are taken from K06. The two green solid lines represent $2\sigma$ upper and lower confidence intervals for the mean response of OLS-bisector. The inset graph is the box-\&-whisker plot (Tukey 1977; Tukey 1970) and histogram of the residuals of OLS-bisector. The distribution of the residuals (blue dots), while being symmetric about zero, is slightly heavy-tailed compared to Gaussian distribution shown as the solid blue curve. {\bf Middle:} Plot of OLS(Y\textbar X) residuals versus the explanatory variable $\hr$. {\bf Bottom:} Plot of OLS-bisector residuals versus their distances from the barycenter of the sample projected on the bisector line. The three different background colors in the {\it middle} and the {\it bottom} plots are $1\sigma$, $2\sigma$ \& $3\sigma$ regions of the residuals from dark to light blue respectively. The red-dashed lines represent the interquartile ranges, beyond which points are generally considered as outliers. The only $>4\sigma$ outlier of OLS-bisector fit is GRB trigger 2679 (Figures~\ref{leverage},~\ref{robust},~\ref{hrshrh}). \label{hrepkplot}}
\end{figure}

Figure~\ref{hrepkplot}, shows the \hrep ~relation for 249 bursts from K06 with measured \epobs that also have reported fluences in the current BATSE catalog.  In a standard ``Ordinary-Least-Squares" fit, here abbreviated as OLS(Y\textbar X), it is assumed that there is one {\it dependent} and one {\it independent} variable. In the present case, however, there is no priority in assigning either of \epobs or \hrh as the dependent or independent variables.  For such occasions, alternative regression methods have been discussed by several authors (e.g. Babu \& Feigelson 1996; Akritas \& Bershadi 1996; Feigelson \& Babu 1992; Isobe et al. 1990).  Following these references, we rely only on the ordinary least squares, OLS(Y\textbar X) \& OLS(X\textbar Y), and the bisector of these two lines (Isobe et al. 1990). Nevertheless, we also provide fits to the sample using other regression methods, including robust regressions that are powerful tools for outlier diagnostics. As will be seen in the later sections, the simulation of the \hrep ~relation indicates that {\it structural} regression models should be used for the calibration of the relation, since the systematic scatter in the data dominates random measurement noise. Therefore with no weighting for the error bars we find,
\begin{equation}
\label{eq:olsyx}
Log\left( \frac{\epo}{300 [KeV]} \right) = 0.10 + 0.63 Log\left( \frac{\hr}{10}\right) , \\
\end{equation}
for OLS(Y\textbar X), with intercept and slope uncertainties of $\sigma_{a} = 0.02$ and $\sigma_{b} = 0.05$ respectively, and,
\begin{equation}
\label{eq:olsxy}
Log\left( \frac{\epo}{300 [KeV]} \right) = 0.15 + 0.87 Log\left( \frac{\hr}{10}\right) , \\
\end{equation}
for OLS(X\textbar Y), with intercept and slope uncertainties of $\sigma_{a} = 0.02$ and $\sigma_{b} = 0.002$ respectively, and,
\begin{equation}
\label{eq:olsbisector}
Log\left( \frac{\epo}{300 [KeV]} \right) = 0.12 + 0.75 Log\left( \frac{\hr}{10}\right) , \\
\end{equation}
for the bisector line, with intercept and slope uncertainties of $\sigma_{a} = 0.01$ and $\sigma_{b} = 0.04$ respectively.

The uncertainties in the slopes and the intercepts are calculated using the formulae given in Table 1 of Isobe et al. (1990). The formulae are obtained via the {\it delta method} which is based on the central limit theorem and therefore applies only to large samples ($N\gtrsim30$). Here, there are $N=249$ GRBs in the calibration sample. The uncertainties derived above are also in excellent agreement with the uncertainties we obtained via bootstrapping (Davison \& Hinkley 1997) which is a superior method when Gauss-Markov assumptions in linear modeling do not hold (e.g. Placket 1950). The sample shows approximately the same dispersion around the three best linear fits ($0.11$ dex).  

The common lack-of-fit F-test that is used together with linear regression in order to test for possible nonlinearities in the data (e.g. Neter et al. 1996), requires the existence of replicates in the observations (i.e. observations with the same explanatory variable).  Since the \hrep ~relation has no clear independent variable, we also fit the data with a second-order polynomial, and test the coefficient of the second-order term against the null hypothesis of being zero using t-statistic. There is only a very weak $0.7\sigma$ evidence ($p=0.48$) to reject the null hypothesis and therefore, the linearity of the relation, cannot be rejected {\it within the range of} the calibration sample. Following Feigelson \& Babu (1992), the $1\sigma$ confidence intervals for the mean responses of the three OLS lines given a hardness ratio $x_0$ can also be derived as,
\begin{eqnarray}
\label{eq:responseCI}
\hat{\sigma}_{1,2,3}^{2} &=& \frac{1}{N^2}\sum^{N}_{i=1}\{y_{i}-\bar{y}-\hat{\beta}_{1,2,3}(x_{i}-\bar{x}) \\ \nonumber
&+& \hat{a}_{j}(x_{i}-\bar{x})[y_{i}-\bar{y}-\hat{\beta}_{1}(x_{i}-\bar{x})] \\ \nonumber
&+& \hat{b}_{j}(y_{i}-\bar{y})[y_{i}-\bar{y}-\hat{\beta}_{2}(x_{i}-\bar{x})]\}^{2}.
\end{eqnarray}
The indices 1,2,3 of the slope of the regression line, $\hat{\beta}$, correspond to OLS(Y\textbar X), OLS(X\textbar Y) \& OLS-bisector, respectively, and $x_{i}$ \& $y_{i}$ are $\log(\hr)$ \& $\log(\epo)$ of $N=249$ GRBs in the calibration sample given in Table~\ref{Calsample}. Also $\bar{x}$ \& $\bar{y}$ represent respectively the sample means of the regressor and the regressand in the calibration sample. The rest of the parameters are as follows,
\begin{equation}
\label{eq:parameters}
\hat{a}_{1} = \hat{\psi}, ~~~;~~~  \hat{b}_{1} = 0. \\
\end{equation}
\begin{equation}
\hat{a}_{2} = 0, ~~~;~~~  \hat{b}_{2} = \hat{\omega}. \\
\end{equation}
\begin{equation}
\hat{a}_{3} = \hat{\psi}\hat{\beta}_{3}[(1+\hat{\beta}_{2}^{2})/(1+\hat{\beta}_{1}^{2})]^{1/2}(\hat{\beta}_{1}+\hat{\beta}_{2})^{-1},
\end{equation}
\begin{equation}
\hat{b}_{3} = \hat{\omega}\hat{\beta}_{3}[(1+\hat{\beta}_{1}^{2})/(1+\hat{\beta}_{2}^{2})]^{1/2}(\hat{\beta}_{1}+\hat{\beta}_{2})^{-1}. \\ \nonumber
\end{equation}
where,
\begin{equation}
\hat{\psi} = \frac{N(x_{0}-\bar{x})}{\sum_{i=1}^{N}(x_{i}-\bar{x})^{2}}, ~~;~~ \hat{\omega} = \frac{N(x_{0}-\bar{x})\hat{\beta}_{2}}{\sum_{i=1}^{N}(y_{i}-\bar{y})^{2}}.
\end{equation}
The $2\sigma$ upper \& lower confidence limits on the mean response for OLS-bisector line are depicted as green solid lines in Figure~\ref{hrepkplot}.

To obtain the slope variances and the confidence intervals we did not rely on the standard formulae (e.g. Bevington \& Robinson 2003) that are derived based on the restrictive Gauss-Markov assumptions, including the presumption of the homoscedasticity of residuals (i.e. uniform dispersion of data along the regression line) which should hold in order for OLS to be the Best Linear Unbiased Estimator (BLUE).  In the presence of heteroscedasticity (i.e. nonuniform dispersion of data along the regression line), although OLS still remains unbiased, it may not be an efficient estimator. This means that although the slope and intercept of OLS line remain unbiased, the errors in the slope and intercept could be significantly underestimated, which in turn could result in the overestimation of the $t$-scores of the estimated coefficients and increase the chance of making type I error in hypothesis testings of the following sections. In addition, the confidence intervals on the mean response might not be reliable. 

The classic assumption of homoscedasticity appears not to hold in \hrep ~relation. In order to identify heteroscedasticity in the residuals of OLS-bisector fit, we employ three classic heteroscedasticity tests: White test (White 1980), Glejser test (Glejser 1969) and the modified Levene's test (Brown \& Forsythe 1974; Levene 1960) which is more robust against departures of the residuals from normality as compared to Glejser test. According to these tests, the null hypothesis of homoscedasticity is rejected at $6\sigma$ ($p=0.0024$), $3\sigma$ ($p=0.0034$) \& $2.4\sigma$ ($p=0.019$) respectively. It should be noted that the residuals of OLS(Y\textbar X) also show a significant evidence of heteroscedasticity at about the same significance levels as for the OLS-bisector fit.

\begin{figure}
\includegraphics[scale=0.31]{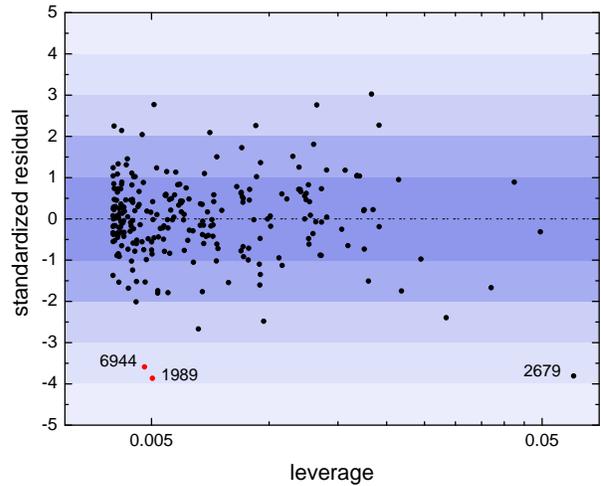}
\caption{Plot of the standard residuals of OLS-bisector fit to the calibration sample versus the their corresponding hat values (leverages) depicting the effects of individual data points on OLS-bisector line. The five differently colored regions represent the $1\sigma$, $2\sigma$, $3\sigma$, $4\sigma$ \& $5\sigma$ of the standardized residuals from dark to light blue respectively. Points with high leverages and standard residuals are considered to be {\it influential} observations that have significant effects on the slope and intercept of the bisector line. The two red points represent GRB triggers 1989 \& 6944 that were excluded from the calibration sample due to the reasons described in \S2 (also Figure~\ref{hrshrh}). Although far outliers to the bisector line, their exclusion has a very weak effect on the slope \& intercept due to their closeness to the barycenter of the data. In contrast, GRB trigger 2679, although a high-leverage outlier to the bisector fit and influential point, was not excluded since it cannot be labeled as an sporadic outlier (\S2 \& Figures~\ref{robust} \& \ref{hrshrh}). \label{leverage}}
\end{figure}

\begin{figure}
\includegraphics[scale=0.31]{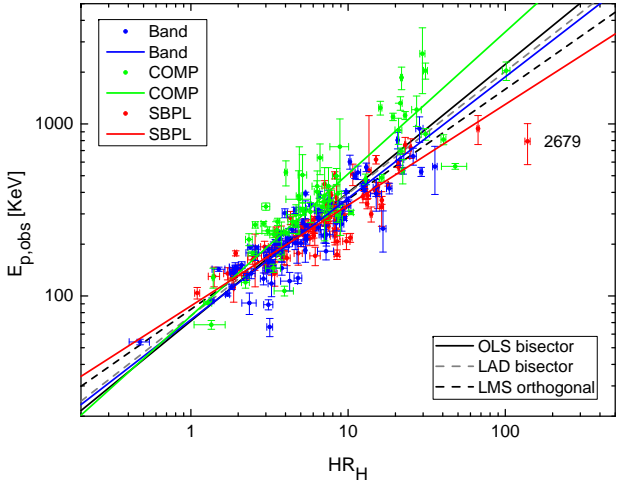}
\caption{Plot of \epobs versus \hrh for the same sample of GRBs as in Figure~\ref{hrepkplot}. The three colors correspond to the three different spectral models that K06 have used to analyze the spectra of the bursts. The colorful lines (blue, green \& red) are OLS-bisector fits to corresponding samples. The slopes of the three lines and their corresponding $1\sigma$ uncertainties are $\hat{\beta}_{1,Band} = 0.71$, $\hat{\sigma} = 0.04$, $\hat{\beta}_{1,COMP} = 0.83$, $\hat{\sigma} = 0.06$, $\hat{\beta}_{1,SBPL} = 0.59$, $\hat{\sigma} = 0.04$. As implied by the variances, the slopes of the three lines are statistically, significantly different from each other. This causes the residuals of OLS-bisector fit to the entire sample (solid-black line) to be heteroscedastic as mentioned in \S2.  The two gray \& black dashed lines, represent respectively $L_{1}$ regression line, also known as Least Absolute Deviations, and the high-breakdown LMS  regression which uses the median of the squared residuals as the breakdown value. $L_{1}$ regression, although less sensitive to outliers compared to OLS, does not show any significant difference with OLS-bisector fit. However, since more than half of the sample are fit by either the Band or SBPL model, LMS line follows the trend in the sample of these two models and treats the COMP model bursts as outliers. This causes the slope of LMS regression (black dashed line) to be significantly different from OLS-bisector's, but similar to the bisector fits for the GRB samples of the Band \& SBPL models. The labeled data point is GRB trigger 2679. Although a high-leverage $>4\sigma$ outlier (Figure~\ref{leverage}) to OLS-bisector (black solid) line and an {\it influential} observation in the regression, it was not excluded from the calibration sample due to the reasons discussed in \S2, also Figure~\ref{hrshrh}. \label{robust}}
\end{figure}

Although the hypothesis of homoscedasticity is rejected, the general pattern of the heteroscedasticity is not clear in the residuals of the OLS fits to the calibration sample (Figure~\ref{hrepkplot}). It was therefore useful to simulate the relation beyond the ranges of \epobs \& \hrh of the calibration sample GRBs to find the pattern. As shown in Figure~\ref{hrepkplot} (Middle \& Bottom), there is also evidence for the existence of $>3\sigma$ outliers, though they constitute a very small fraction ($<1\%$) of the calibration sample. 

Although OLS-bisector's residuals resemble a Gaussian distribution (Figure~\ref{hrepkplot}), careful analysis indicates that it is heavy-tailed compared to a Normal distribution.  First, the Kolmogorov-Smirnov (K-S) (Kolmogorov 1941; Smironov 1948) test of normality for OLS-bisector residuals does not indicate a significant deviation from normality ($p=0.28$). However, according to Anderson-Darling test (Anderson \& Darling 1952), a test that is more sensitive to the tails of the distribution than K-S test, the null hypothesis of normality is rejected at 1\% significance level. 

Symmetry \& normality of the residual distribution is of particular importance for works that study the population distribution of \epobs (e.g. Shahmoradi \& Nemiroff 2009a). If the symmetry and normality criteria do not hold, \hrh might be a biased indicator of the value of \epobs, particularly in the large data sample outside of the small calibration data set.  It is therefore useful to examine the validity of the symmetry and normality criteria beyond the range of the calibration sample by simulation.

Figure~\ref{robust} shows the calibration sample of Figure~\ref{hrepkplot}, but fit to three different spectral models used by K06.  Each model is shown by a different color.  As seen, GRBs that are fit by COMP (CPL) model (green dots) are on average above the OLS-bisector fit for the entire sample (black solid line).  Conversely, the two other groups of GRBs fit by the Band and SBPL models show trends that are different from COMP model (green solid) line, but similar to each other. This means that for a given hardness ratio \hrh reported in the BATSE catalog, there can be different \epobs associated with the burst, according to what model has been used for the spectral analysis. 

In addition, there is an intrinsic scatter in each of the three samples which is not due to measurement error. The `unexplained variances' of \epobs might be attributed to the fact that \hrh is not uniquely determined by \epobs and other free parameters of the 3 spectral models, such as the high and low energy photon indices, create the observed dispersions in data. Another reason for the scatter in the data could be the fact that the two GRB parameters, \hrh \& \epobs come from different sources, i.e. BATSE catalog \& K06, who consider different time \& energy interval and background fits in their spectral analyses of the bursts (e.g. K06: Table 1). 

In order to show this effect more clearly, we also calculate the hardness ratios for the calibration sample using the spectral parameters given by K06.  The new hardness ratio $\hrs$ is defined in the same way as \hrh: The fluence in 100-2000 KeV energy range divided by the fluence in 20-100 KeV energy range, corresponding to the ratio of the sum of the fluences in channels 4 \& 3 of BATSE LADs to the sum of the fluences in channels 2 \& 1.

\begin{figure}
\includegraphics[scale=0.31]{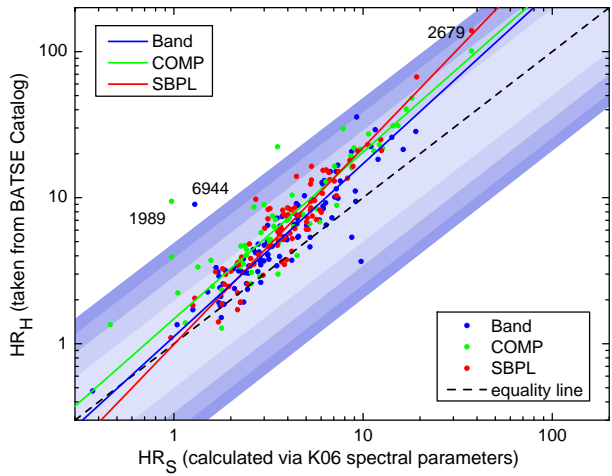}
\caption{Plot of \hrh taken from BATSE catalog, versus $\hrs$ calculated via the spectral parameters given by K06 for the calibration sample, in the same energy range as for \hrh. As seen the hardness ratios calculated via BATSE catalog data are on average always greater than their counterparts calculated using the spectral parameters. In addition, the deviation from the equality line increases as the two hardness ratios increase. The reason for such discrepancy should be probably sought in the way the DRMs have been included in the spectral analysis (\S2). The colored regions from dark to light represent the region of $\hrs/5<\hr<5\hrs$, $\hrs/4<\hr<4\hrs$, $\hrs/3<\hr<3\hrs$ \& $\hrs/2<\hr<2\hrs$ respectively. The three blue, green \& red colored points and their OLS-bisector fits (the blue, green \& red solid lines) correspond to the three Band, COMP \& SBPL spectral models used by K06 to analyze the bursts. GRB trigger 1989 \& 6944 are the two low-leverage outliers of \hrep ~relation (Figure~\ref{leverage}) that were excluded from the calibration as their reported spectral parameters in K06 \& BATSE catalog are  in stark contrast to each other. The high-leverage $>4\sigma$ outlier to OLS-bisector line (Figure~\ref{hrepkplot}), GRB trigger 2679, however, was not excluded from the calibration sample, since it lies along the general trend (the red OLS-bisector line) visible in the plot, though it is an {\it influential} point in the regression and its \hrh is 5 times larger than its $\hrs$. \label{hrshrh}}
\end{figure}

Contrary to our initial guess, there is a systematic difference between these two hardness ratio estimates which tends to correlate positively with both hardness ratios \hrh \& $\hrs$. \hrh from the BATSE catalog appears to be, on average, greater than $\hrs$. The reason might be sought in the way the Detector Response Matrices (DRMs) have been considered by K06 and BATSE team in their analyses (Preece, private communication). Nevertheless, the primary reasons for the biases in \hrh \& $\hrs$ will be investigated in a separate work (Shahmoradi \& Nemiroff 2009c). Figure~\ref{hrshrh} shows graphically the systematic differences between these two hardness ratio estimates.

For the three spectral models we find the below relations for \hrh as a function of $\hrs$:
\begin{equation}
\label{eq:olsband}
Log\left( \hr \right) = 0.09 + 1.12 Log\left( \hrs \right) , \\
\end{equation}
for the Band model GRBs, and,
\begin{equation}
\label{eq:olscomp}
Log\left( \hr \right) = 0.22 + 1.06 Log\left( \hrs \right) , \\
\end{equation}
for the COMP model GRBs, and,
\begin{equation}
\label{eq:olssbpl}
Log\left( \hr \right) = 0.06 + 1.26 Log\left( \hrs \right) , \\
\end{equation}
for the SBPL model GRBs.

Heteroscedastic residuals and the presence of outliers are strong motivations to consider also robust regression methods such as $L_1$ (Karst 1958) -- also known as the Least Absolute Deviation (LAD) regression -- \& high-breakdown regressions (e.g. Rousseeuw 1984) that are resistant to high-leverage outliers. These regression techniques effectively check that the OLS methods are not significantly affected by the presence of hidden outliers. Applying these regression methods to the entire 249 GRBs in the calibration sample, we find a weak evidence for the significance of the difference between slopes of $L_1$ and OLS regression lines, according to the uncertainties in the slopes. However, the Least Median of Squares (LMS) line (Rousseeuw 1984) which uses the highest breakdown value ($50\%$) shows a significantly different slope compared to OLS-bisector fit to the entire calibration sample (Figures~\ref{hrepkplot} \& \ref{robust}). Recall that the entire sample of 249 GRBs was divided into three groups by K06 -- those GRBs best fit by the Band model, those GRBs best fit by the COMP model, and those GRBs best fit by the SBPL model.  Now the slope of the LMS line for all 249 GRBs in the K06 calibration sample appears more similar to the slopes of OLS-bisector lines for the sample of the Band \& SBPL spectral models (Figure~\ref{robust}). This is due partly to the fact that there is an intrinsic difference between the slopes of the bisector and orthogonal regression lines. But the main reason for the difference is the high breakdown value of LMS estimator. Since more than half of the calibration sample GRBs are fit by the Band \& SBPL models, the LMS line would treat the COMP model GRBs as outliers and therefore, the LMS line follows approximately OLS-bisector lines for the two Band \& SBPL GRB samples (blue \& red solid lines in Figure~\ref{robust}).

However, we expect that as more BATSE GRBs are added, the slopes of the two OLS and LMS regression lines would converge to each other. The reason is that the rest of BATSE catalog GRBs are generally much fainter than 249 bright GRBs in the calibration sample and therefore of much lower signal-to-noise ratios compared to the K06 sample. In these cases, COMP model appears to be the preferred spectral model due to its simplicity and low number of free parameters compared to the Band \& SBPL models, since having an additional free parameter usually results in highly cross-correlated, unconstrained parameter determinations (e.g. K06). This can in turn, balance the number of COMP model GRBs compared to the number of GRBs well fit by the two other spectral models as more spectrally analyzed bursts are added to the sample.

\section{prediction intervals for \hrep ~relation}

Unlike the analytical method used to derive confidence intervals for the mean responses of OLS regression lines in \S2.2, the derivation of prediction intervals cannot be achieved without recourse to a detailed and accurate knowledge of the pattern of the heteroscedasticity in \hrep~ relation, in particular, beyond the range of the calibration sample where the relation might not have a linear behavior. The prediction interval is, by definition, the interval within which the individual values of $\epo$ of BATSE GRBs would most likely fall given their hardness ratios. This is in contrast to the definition of the confidence interval {\it on the mean response} of OLS regression lines derived in \S 2.2. 

To achieve this goal, we run extensive simulations of the relation over a wide range \hrh \& \epobs for the Band, COMP \& SBPL spectral models. The simulation algorithm is fairly simple and includes the following steps:
\begin{enumerate}
	\item Simulate a GRB by drawing randomly its spectral parameters from their parent distributions for each of the 3 GRB models. These include \epobs for the three models and their corresponding high and low-energy photon indices.\\
	
	\item Calculate the hardness ratio $\hrs$ as defined in \S2.2. The spectral forms of the 3 GRB models that are required to calculate $\hrs$, have been given and discussed in detail by K06. \\
	
	\item Map $\hrs$ to \hrh via Eqns.~\eqref{eq:olsband},\eqref{eq:olscomp} \& \eqref{eq:olssbpl}. Here, we have essentially assumed that the bias seen between \hrh and $\hrs$ in \S2.2 (Figure~\ref{hrshrh}) remains linear over the entire range of BATSE GRBs hardness ratios. The assumption appears reasonable as we find that \hrh \& $\hrs$ of BATSE GRBs other than 249 GRBs considered here also follow the same patterns. These include GRBs analyzed by Nava et al. (2008) -- hereafter N08 -- \& G09.\\
	
	\item Repeat steps (i) to (iii) until the desired accuracy in the prediction intervals is achieved.
\end{enumerate}
	Steps (ii)-(iv) in the above algorithm are straightforward. The first step, however, requires the knowledge of the distributions of GRB spectral parameters for the three spectral models. To derive the parent distributions from which the random spectral parameters of the simulated GRB should be drawn, we present in the following subsection, a comprehensive Bayesian multivariate analysis of the observational GRB data, mainly taken from K06, subject to truncations and sample-incompleteness effects that exists in the data sets.
\subsection{Multivariate Analysis of GRB Spectral Parameters}
\subsubsection{Observational Data and Sample Selection}
So far, the spectra of 342 BATSE GRBs have been simultaneously fit by a variety of spectral models (K06) -- the most important models being Band, COMP, SBPL -- to find the best fit model of GRBs. The sample includes the observational data for 118 GRBs best fit by the Band, 70 GRBs best fit by COMP \& 137 GRBs best fit by SBPL models. This data set might therefore be used as an accurate proxy to derive the underlying probability density functions (pdf) of GRB spectral parameters. However, for the reason of better statistics, we increase the number of GRBs in the Band sample of K06, by including 32 BATSE GRBs that are well fit by the Band model in G09, also 16 LGRBs fit by the Band model in N08.

Due to low Signal-to-Noise ratio (S/N) of GRBs analyzed by G09 \& N08, they consider only two spectral models of the Band and COMP. However, although low S/N GRBs spectra might be well fit by the Band or CPL (COMP), these two might have not been necessarily the best-fit spectral models, had GRBs had high enough S/N to perform a 5-parameter model (e.g. SBPL) fit (K06; Band et al. 1993). Therefore, to ensure accuracy in the selected sample of GRBs from G09 and N08, we compared their derived fluences and hardness ratios to those given in the BATSE catalog and excluded inconsistent bursts. In addition, since we ignore the uncertainties of the spectral data in the following analyses, we also require that the spectral parameters of GRBs taken from G09 \& N08 have $1\sigma$ uncertainties that are at least as small as the largest $1\sigma$ uncertainties in K06 data. 

In the same way, because of the small number of COMP model GRBs in K06 sample, we include 39 BATSE LGRBs from G09 that are well fit by COMP model. Also, 27 SWIFT (Gehrels et al. 2004) GRBs from Cabrera et al. (2007), 36 HETE-II GRBs from Pelangeon et al. (2008) and 22 HETE-II GRBs from Sakamoto et al. (2005) are included. By contrast, due to the significant inconsistencies that exist between the fluences and hardness ratios of G09 SGRB sample and their corresponding values reported in BATSE catalog, none of G09 SGRBs were included in the sample of COMP model GRBs considered here.

Except K06, no one has attempted to fit the spectra of a significant number of GRBs by SBPL model. We therefore rely only on 137 GRBs of K06 to constrain the parameters of this model.

\subsubsection{Bayesian Multivariate Analysis of Data}

So far, only univariate analyses of the time-integrated spectral parameters of BATSE GRBs have been discussed in the literature (e.g. G09; Sakamoto et al. 2009) ignoring the possible underlying covariances that might exist between the parameters of a GRB model. This might not in general be true. Therefore, to explore the possible interrelationships between the GRB parameters we present below a multivariate analysis of the observational data given above.

According to the Bayes' theorem, the joint posterior probability density function of the set of parameters {\boldmath ${\Theta}$} of each spectral model, given the observed data {\boldmath ${D}$}, the assumed statistical model {\boldmath ${M}$} and the prior information {\boldmath ${I}$} is,
\begin{equation}
\label{eq:bayesrule}
P(\mathbf{\Theta}| {\mathbf D},{\mathbf M},{\mathbf I}) \propto P(\mathbf{\Theta} | {\mathbf M},{\mathbf I})\times P({\mathbf D} | \mathbf{\Theta},{\mathbf M},{\mathbf I})
\end{equation}

The parameters to be estimated, are the moments of the assumed statistical model. For example, if the assumed statistical model is multivariate normal distribution, the parameters to be constrained in each of the spectral models would be the means and variances of the low \& high-energy indices and the mean and variance of $\ln(\epo)$ for the Band, COMP \& SBPL models, as well as he possible correlations among the spectral parameters. The break scale $\Lambda$ of SBPL model appears to be a discrete variable in K06 sample. Therefore, the distribution of the break scale is considered separately. Also the normalization factor of all 3 models need not to be considered in the analysis, since the hardness ratio is independent of this parameter. Because of the large number of free parameters involved and the relatively small size of the observational data, comparison of a variety of statistical models to find the best seems impractical. To simplify, we assume that the spectral parameters of different GRB models -- denoted by $\mathbf{X}$ -- are drawn from a $p$-dimensional truncated multivariate normal distribution $\mathbf{D}\sim NT_{p}(\mathbf{\upmu},\mathbf{\Sigma})$ where $\mathbf{X}$ is constrained to the region of the $p$-dimensional parameter space $\mathbf{R}=\{a_i<\mathbf{X}_i<b_i,i=1,p\}$. Here, the truncation on the observational data is set by physical constraints on GRB models. For example, the high-energy photon index of the Band and SBPL models should always be $<0$, to avoid energy catastrophe in the GRB prompt emission. Also in COMP model, the photon index $\alpha$ must be $>-2$ by definition. In addition, SBPL model has \epobs defined only when its low-energy index $\alpha>-2$ and its high-energy index $\beta<-2$. There is, however, no physical constraint on the possible values that $\ln(\epo)$ could take and it can range from $-\infty$ to $+\infty$.

The likelihood of data for a GRB model with $p$ free parameters (ignoring the normalization factor) can then be written as,
\begin{eqnarray}
\label{eq:likelihood}
P({\mathbf D}|\mathbf{\Theta}) &=& \big(A(\mathbf{\upmu},\mathbf{\Sigma})\big)^{-1} (2\pi)^{-Np/2}|\mathbf{\Sigma}|^{N/2} \\ \nonumber
&\times& \exp\bigg(-\displaystyle\sum_{i=1}^{N}(\mathbf{D}_i-\mathbf{\upmu})^{T}\mathbf{\Sigma^{-1}}(\mathbf{D}_i-\mathbf{\upmu})/2\bigg) \\ \nonumber
&\times& \mathbf{I}_{R}\big(\mathbf{D}\big)
\end{eqnarray}

The $p\times 1$ vector $\mathbf{\upmu}$ and the $p\times p$ matrix $\mathbf{\Sigma}=\{\sigma_{ij}\}$ represent the mean vector and variance-covariance matrix of the truncated multivariate normal distribution, to be estimated from the data ${\mathbf D}=\left[{\mathbf D_{1}},...,{\mathbf D_{N}}\right]$, where $\mathbf{D}_i$ is the $i$th $p\times 1$ observation vector which depending on the spectral model can be written as,
\begin{eqnarray}
\label{eq:SP1}
&&\mathbf{D}_{\text{Band},i} = \left[\ln(E_{p,obs,i}),\alpha_i,\beta_i\right]^{T} \\
\label{eq:SP2}
&&\mathbf{D}_{\text{COMP},i} = \left[\ln(E_{p,obs,i}),\alpha_i\right]^{T} \\
\label{eq:SP3}
&&\mathbf{D}_{\text{SBPL},i} = \left[\ln(E_{p,obs,i}),\alpha_i,\beta_i\right]^{T}
\end{eqnarray}

   The subscript $T$ stands for `Transpose'. Also, the term $\mathbf{I}_{R}\big(\mathbf{D}\big)$ in the likelihood function (Eqn.~\eqref{eq:likelihood}) is an indicator function that sets the likelihood to zero when $\mathbf{D}$ is outside the truncated region $\mathbf{R}$ and it is unity otherwise. Also, the term $\big(A(\mathbf{\upmu},\mathbf{\Sigma})\big)^{-1}$ is a normalization factor which is due to the presence of truncation on data.

As the prior, we adopt the standard choice of the non-informative joint prior density for $\mathbf{\upmu}$ \& $\mathbf{\Sigma}$ (Dickey, Lindley \& Press 1985; Box \& Tiao 1973; Geisser \& Cornfield 1963),
\begin{eqnarray}
\label{eq:priors}
p(\mathbf{\upmu},\mathbf{\Sigma}) &=& p(\mathbf{\upmu})\times p(\mathbf{\Sigma}) \\ \nonumber
&\propto& 1\times |\mathbf{\Sigma}|^{-(p+1)/2} \\ \nonumber
&\propto& |\mathbf{\Sigma}|^{-(p+1)/2}
\end{eqnarray}
In the absence of truncation, the analytical expressions for the resulting marginal posterior densities have been derived by Geisser \& Cornfield (1963). In the presence of truncation, however, there are no analytical expressions for the marginals, due to the term $\big(A(\mathbf{\upmu},\mathbf{\Sigma})\big)^{-1}$ in the likelihood (Eqn.~\eqref{eq:likelihood}). Therefore, we set up a Markov-Chain Monte Carlo algorithm, widely known as Gibbs sampler (Geman \& Geman 1984) to obtian the marginal posterior densities. This is done by sampling iteratively from the conditional distributions of $\mathbf{\upmu}$ ($p$-dimensional multivariate normal distribution) \& $\mathbf{\Sigma}$ ($p$-dimensional inverse-Wishart distribution), while updating the conditional variables at each iteration by the previous values (e.g. Rodriguez-Yam et al. 2004; Griffiths 2004; Geweke 1991). Iteration is then continued until convergence to the target density is assured (Raftery \& Lewis 1992). The entire simulation algorithms are written in FORTRAN. The resulting marginal pdfs for the means and variances of the spectral parameters of the three GRB models in addition to the marginal posteriors of the correlation coefficients among the parameters are given in Figure~\ref{pdfs}.

\begin{figure}
\includegraphics[scale=0.31]{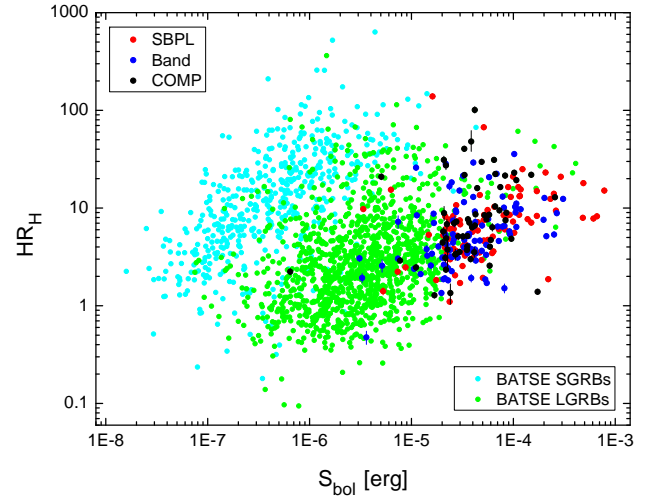}
\caption{Plot of \hrh vs. the bolometric fluence $\sbol$ of the entire BATSE catalog GRBs illustrating the effects of the flux-limit in the K06's analyzed sample of 350 bright BATSE GRBs. The flux-limit set by K06 was to ensure an accurate derivation of the spectral parameters. Although the derived parameters of GRBs in K06 have little uncertainties, the distribution of the spectral parameters of K06 sample might not represent the underlying distribution of the spectral parameters of the entire BATSE sample of GRBs, due to the effects of data truncation and sample incompleteness imposed by requiring a minimum fluence ($\gtrsim10^{-5}$ ergs) for the spectral analysis of GRBs. In particular, the variance of $\ln(\epo)$ distribution of BATSE GRBs, is likely significantly underestimated by the variance of $\ln(\epo)$ in the sample of K06 GRBs.\label{flux-limit}}
\end{figure}

The above analysis, though mathematically accurate, is based on an erroneous presumption that the spectral parameters of the samples being studied can be regarded as purely {\it random} variables. Such an assumption is evidently not true. Since different sources (e.g. K06, G09, N08) have been used to collect information on the spectral parameters, the sample is very heterogeneous and suffers from various selection and truncation effects that vary from burst to burst. The simplest of these truncation effects is due to the flux or fluence limits set by these authors (e.g. K06, G09, N08) to ensure a minimum signal-to-noise ratio in their spectral analyses. An example of such data truncation is well illustrated in Figure~\ref{flux-limit}. Therefore, it is likely that the posteriors derived based on these data sets would be biased, an important point that is generally overlooked in the literature (e.g. Sakamoto et al. 2009). 

In particular, the distribution of $\ln(\epo)$ is likely greatly affected in these flux-limited samples, which in turn could affect the distributions of other spectral parameters having nonzero covariances with $\ln(\epo)$. To identify these potential biases, we also run an extensive set of Monte Carlo simulations for each of the three spectral models to obtain the likelihood functions numerically, subject to data truncation due to sample-incompleteness, contrary to the above Bayesian approach where the likelihood could be written in analytical form (Eqn.~\eqref{eq:likelihood})

\subsubsection{Minimum $\chi ^2$ / Minimum KS-Distance Estimates of Parameters Subject to Sample-Incompleteness}

\begin{figure*}
\includegraphics[scale=0.20]{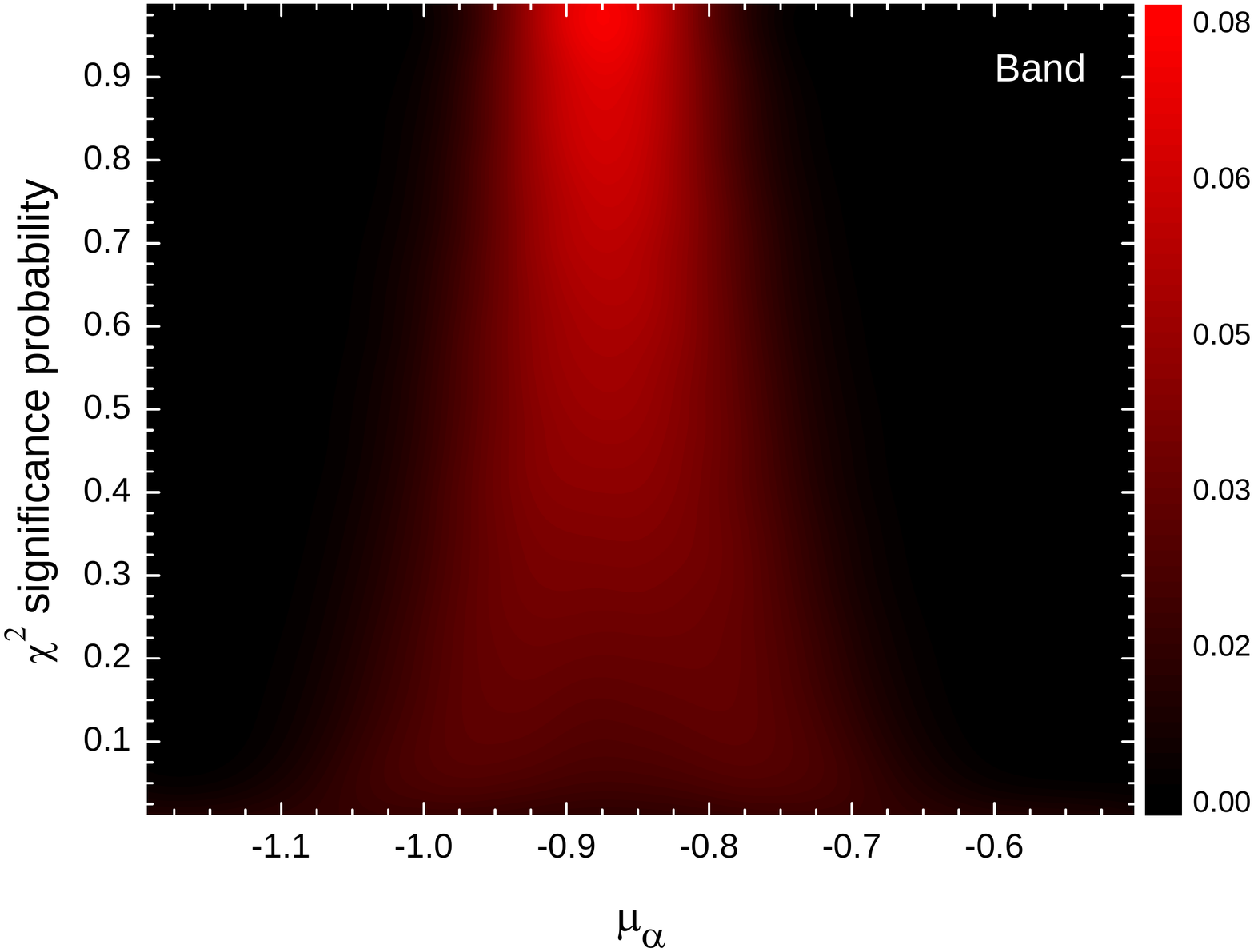}
\includegraphics[scale=0.20]{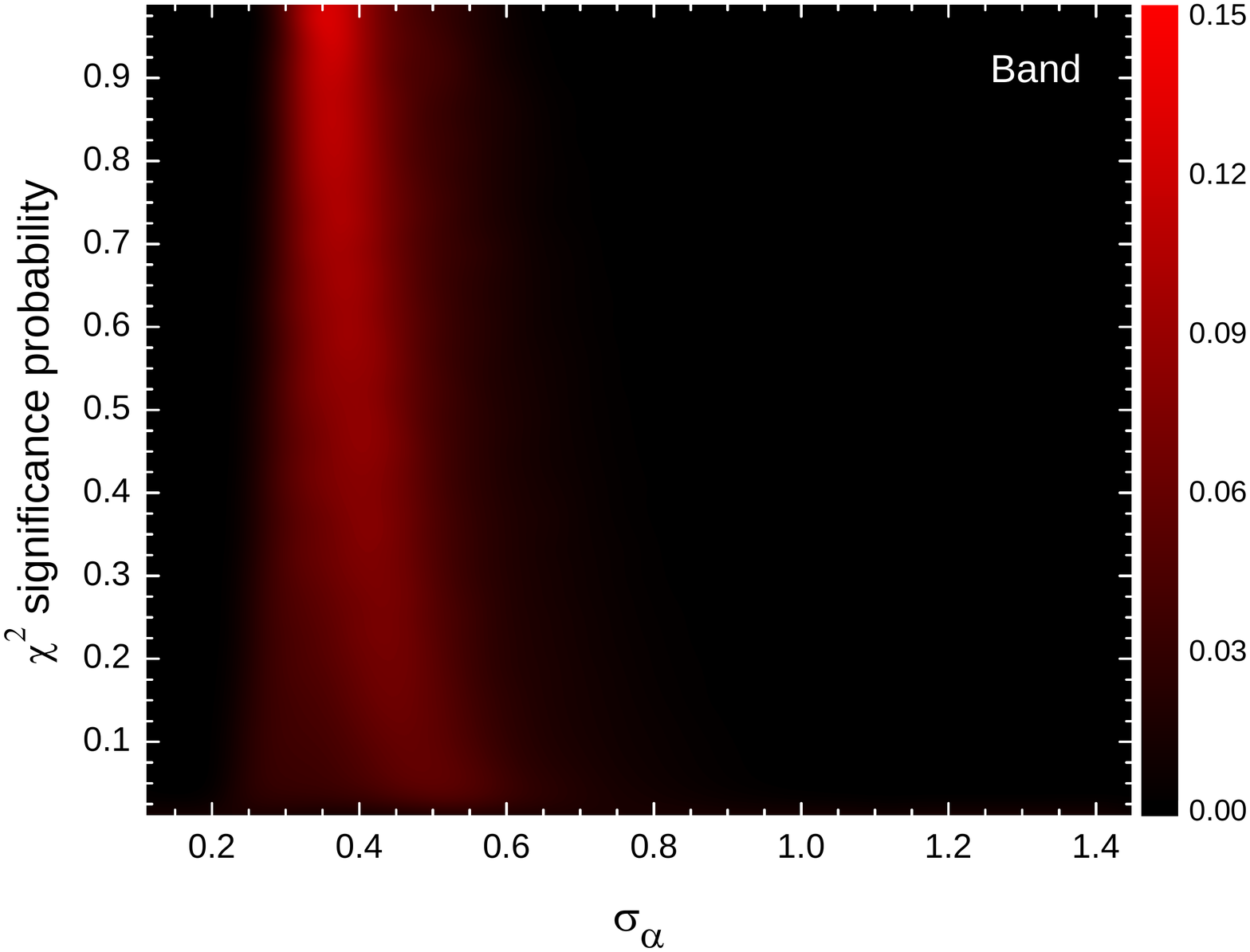}
\includegraphics[scale=0.20]{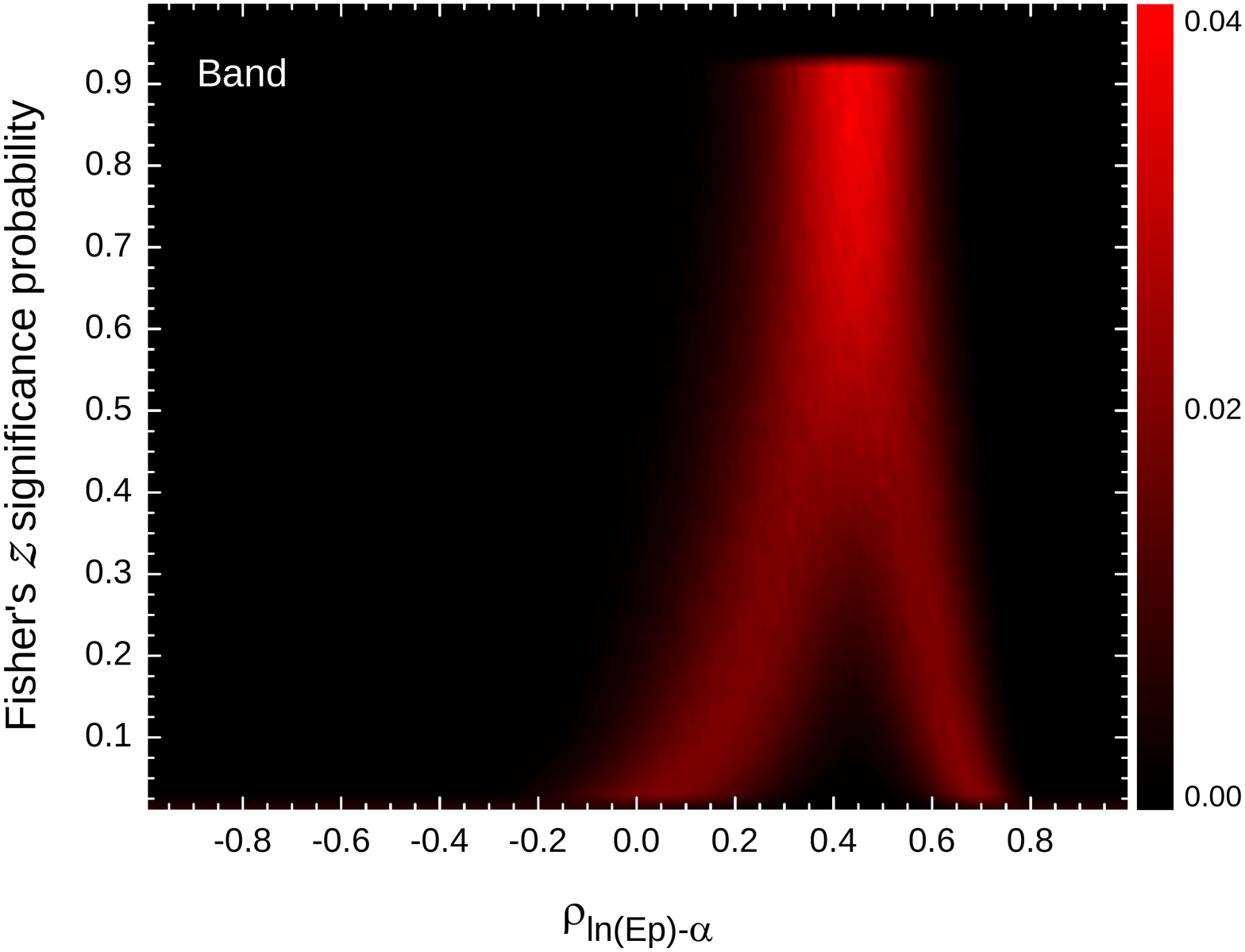}
\includegraphics[scale=0.20]{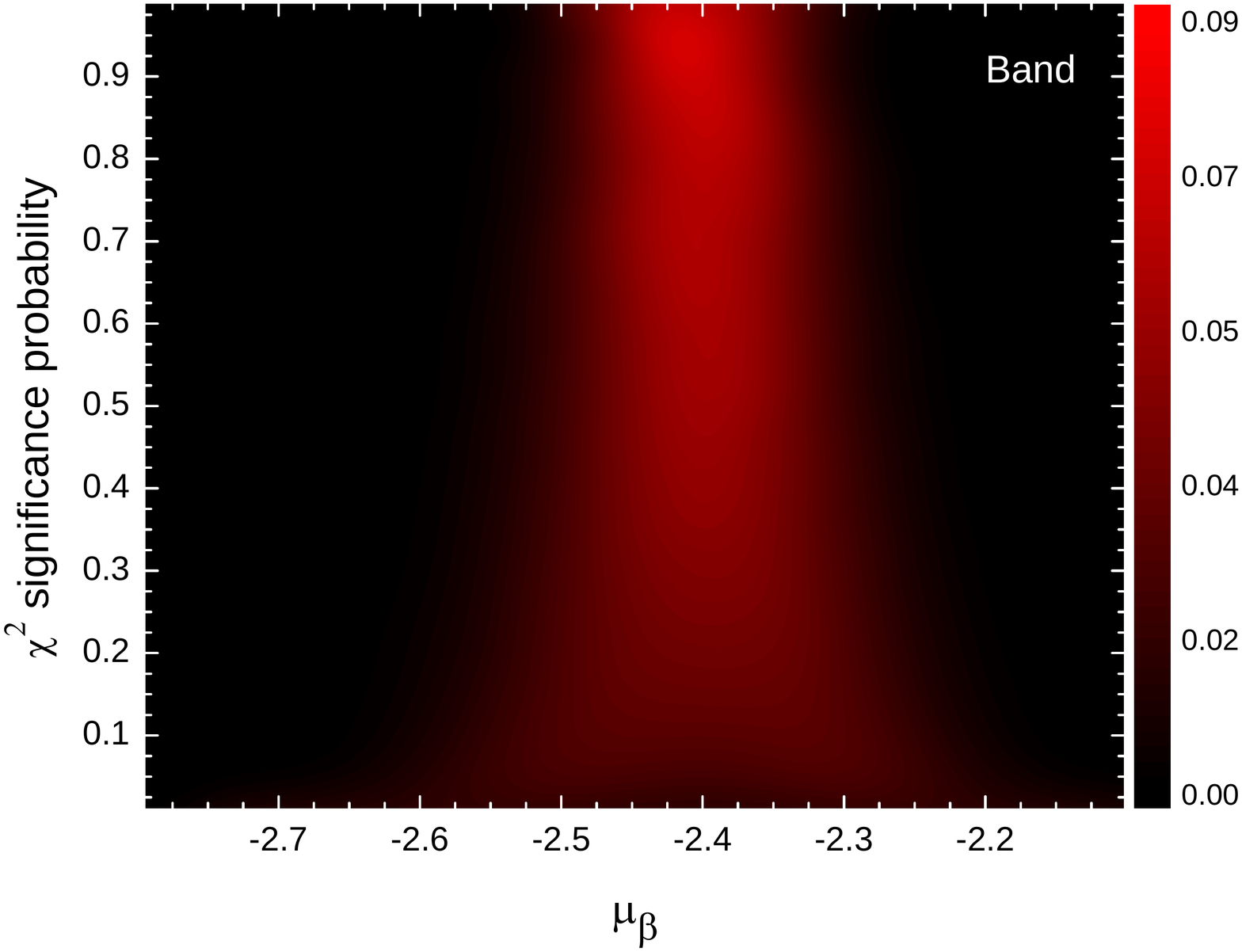}
\includegraphics[scale=0.20]{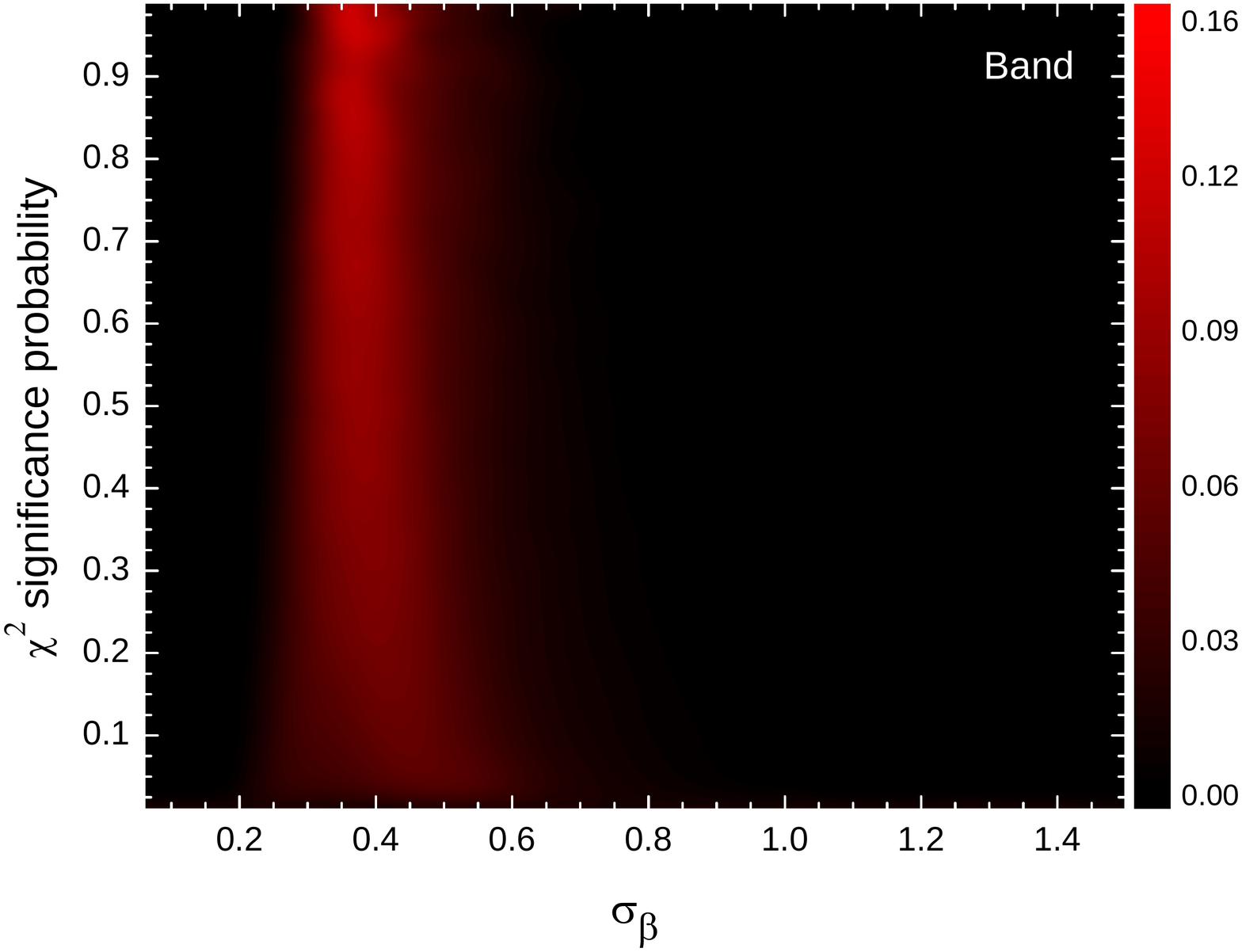}
\includegraphics[scale=0.20]{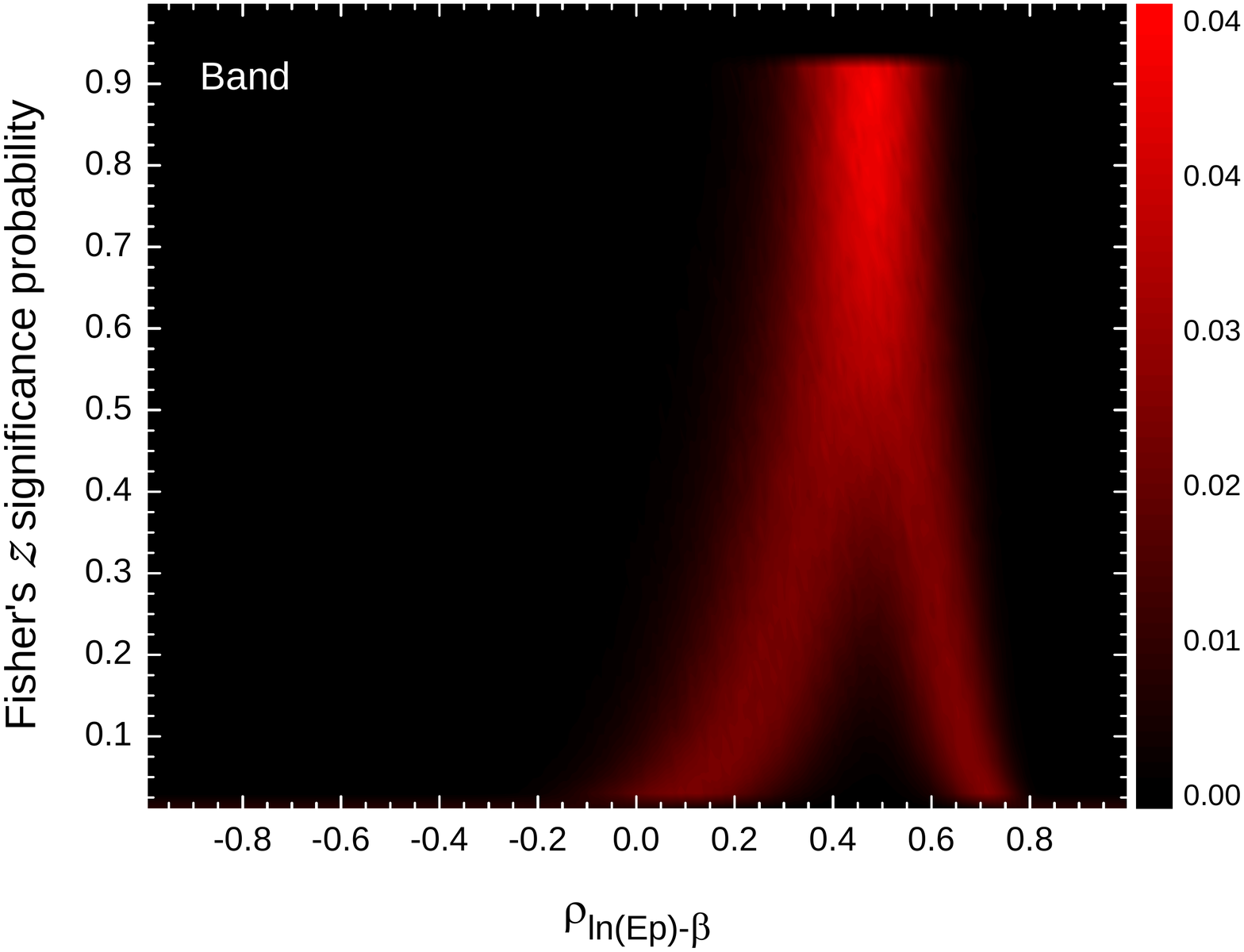}
\includegraphics[scale=0.20]{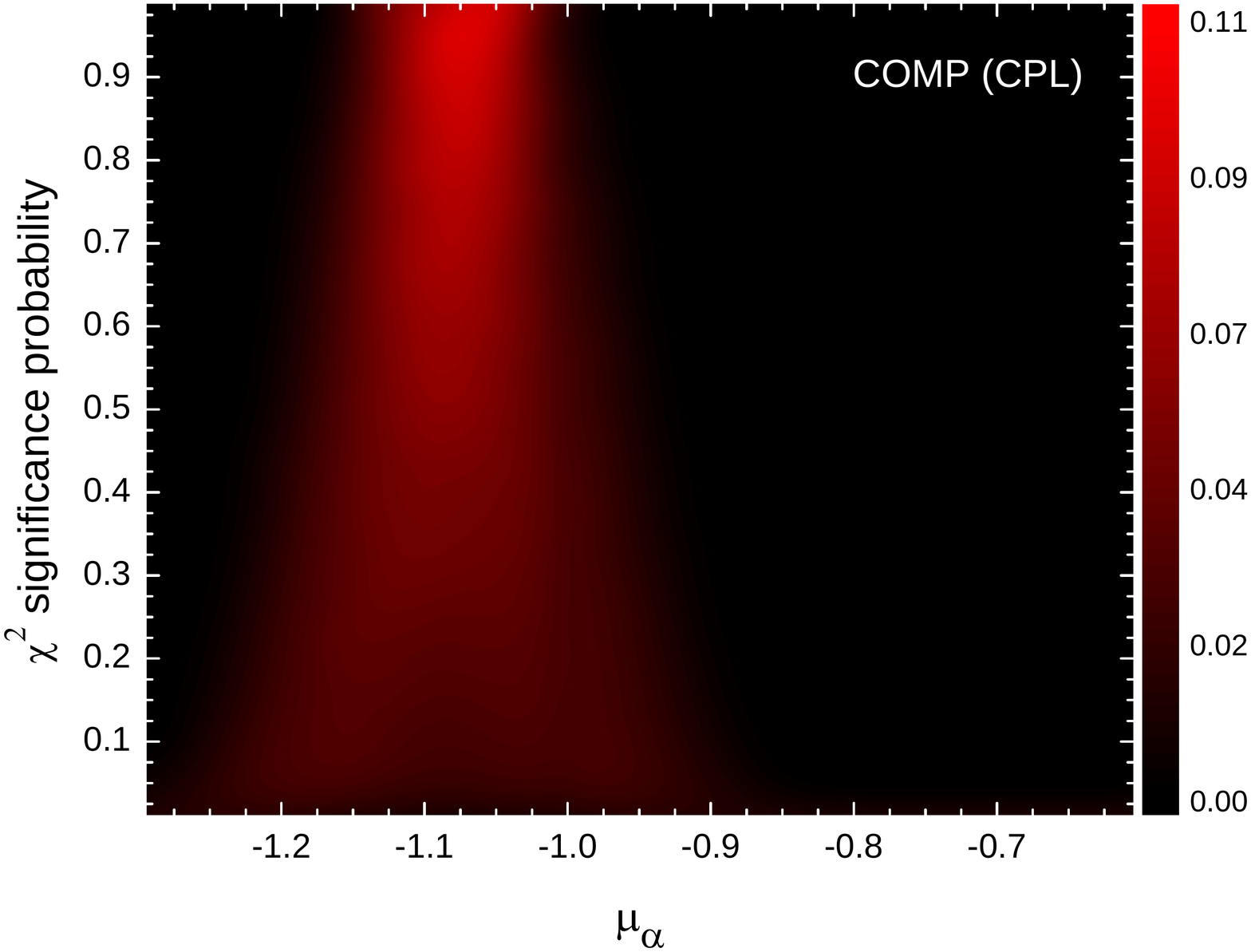}
\includegraphics[scale=0.20]{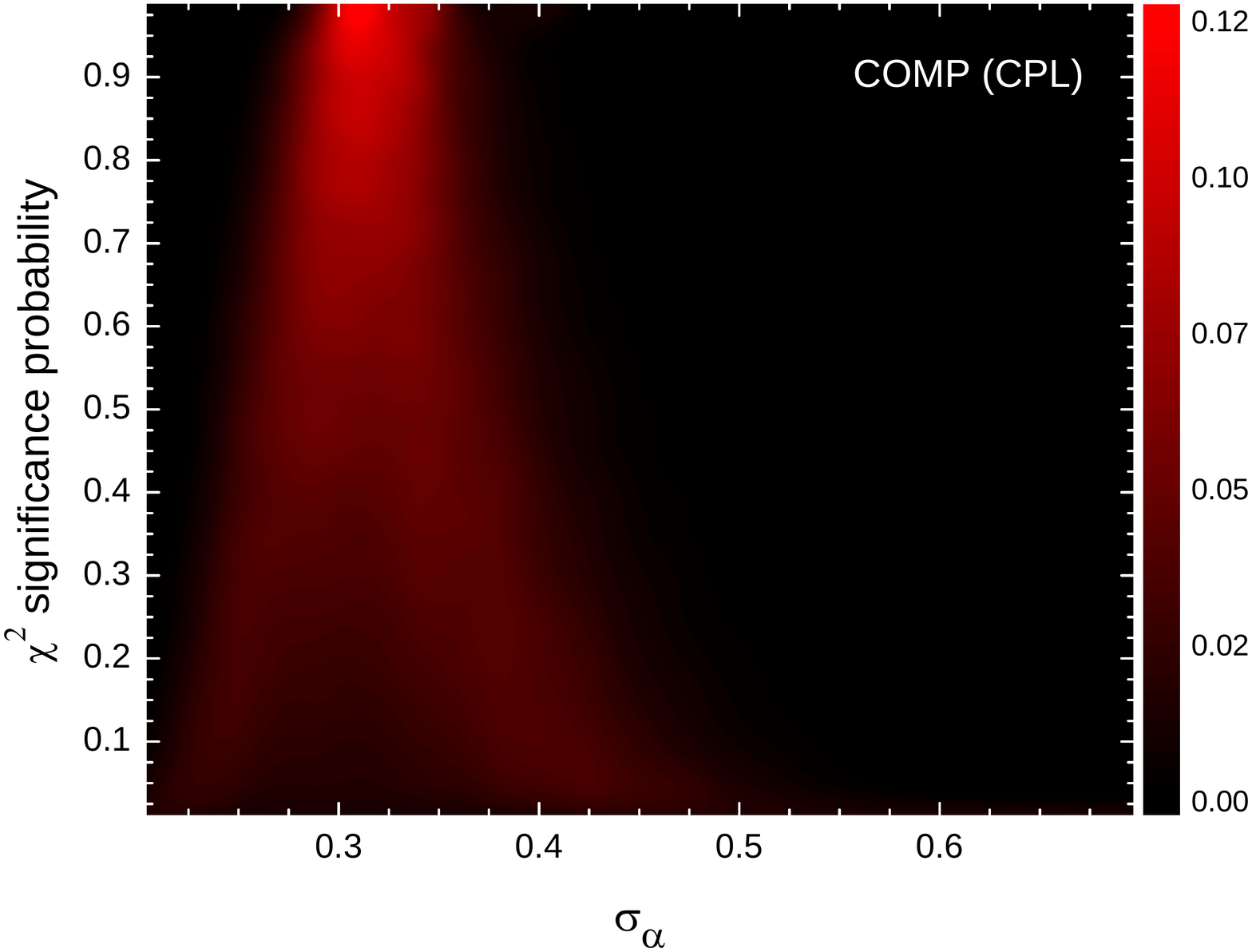}
\includegraphics[scale=0.20]{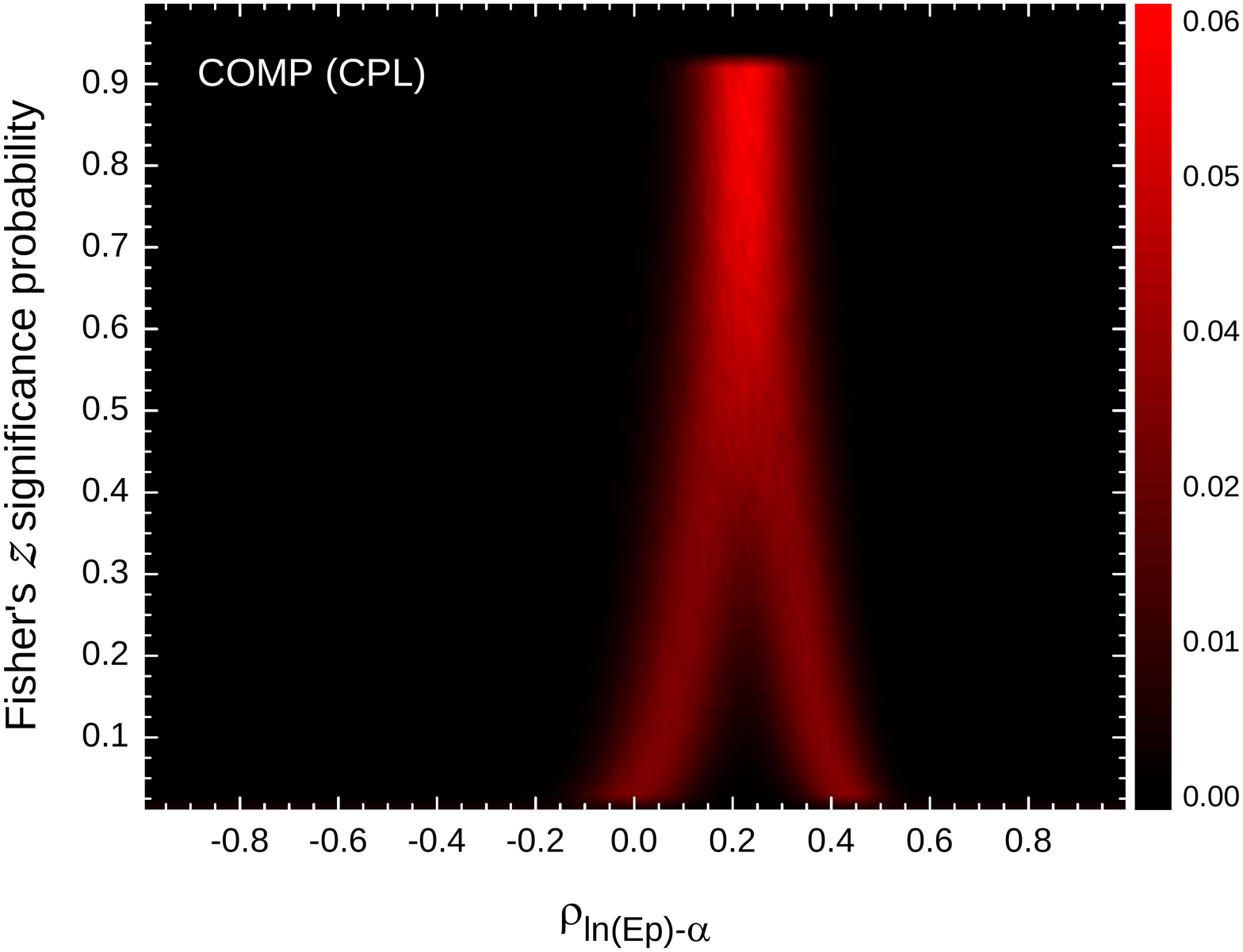}
\includegraphics[scale=0.20]{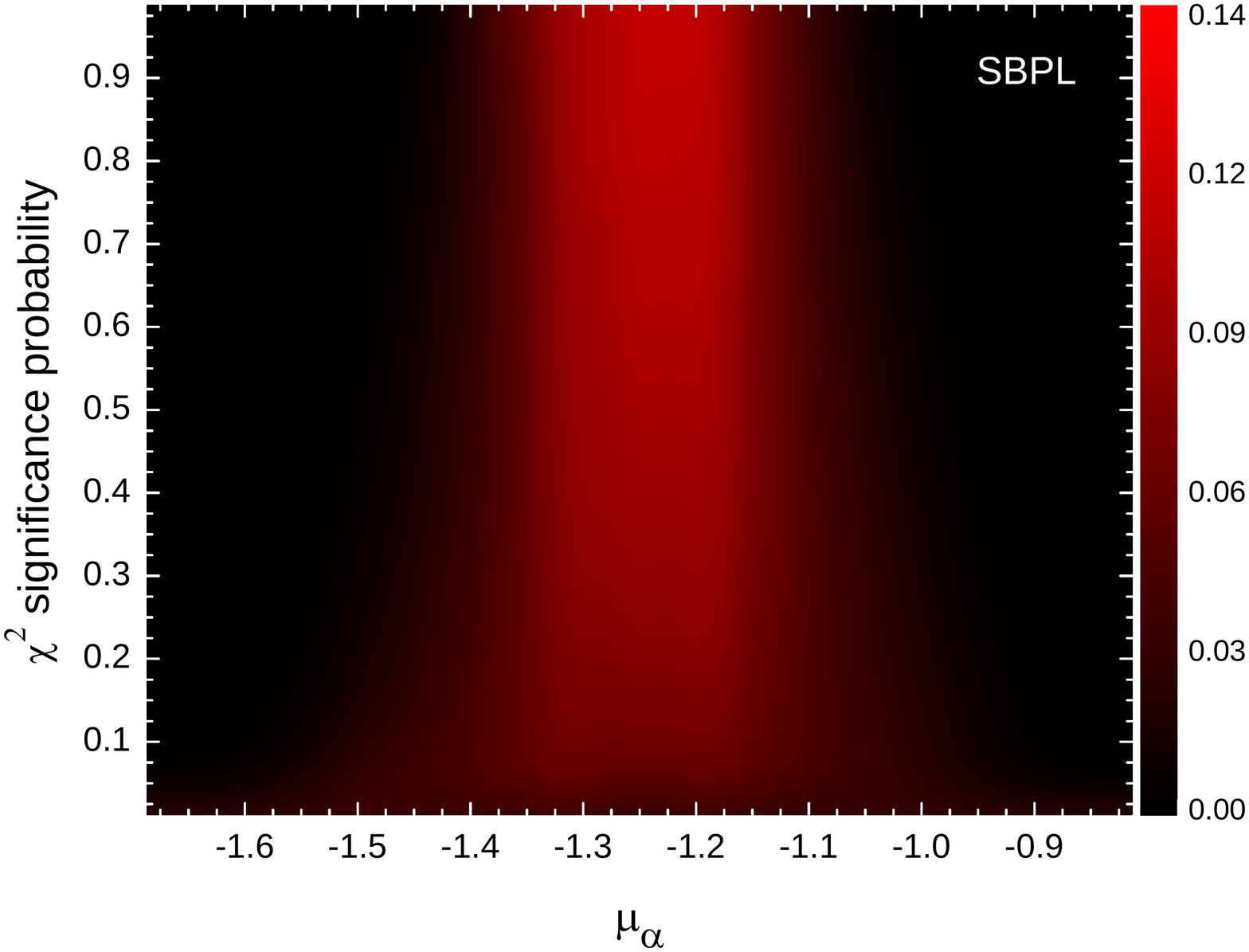}
\includegraphics[scale=0.20]{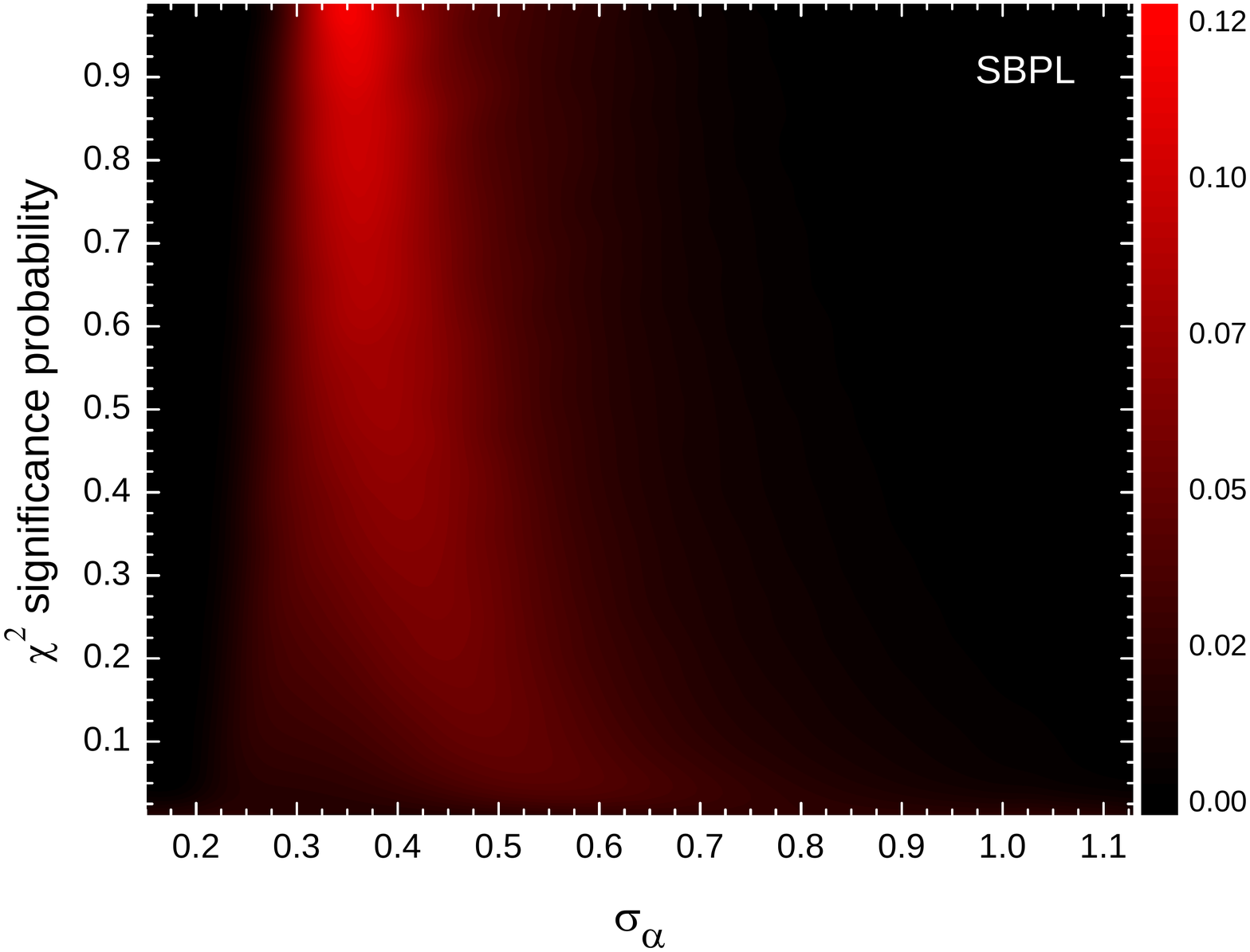}
\includegraphics[scale=0.20]{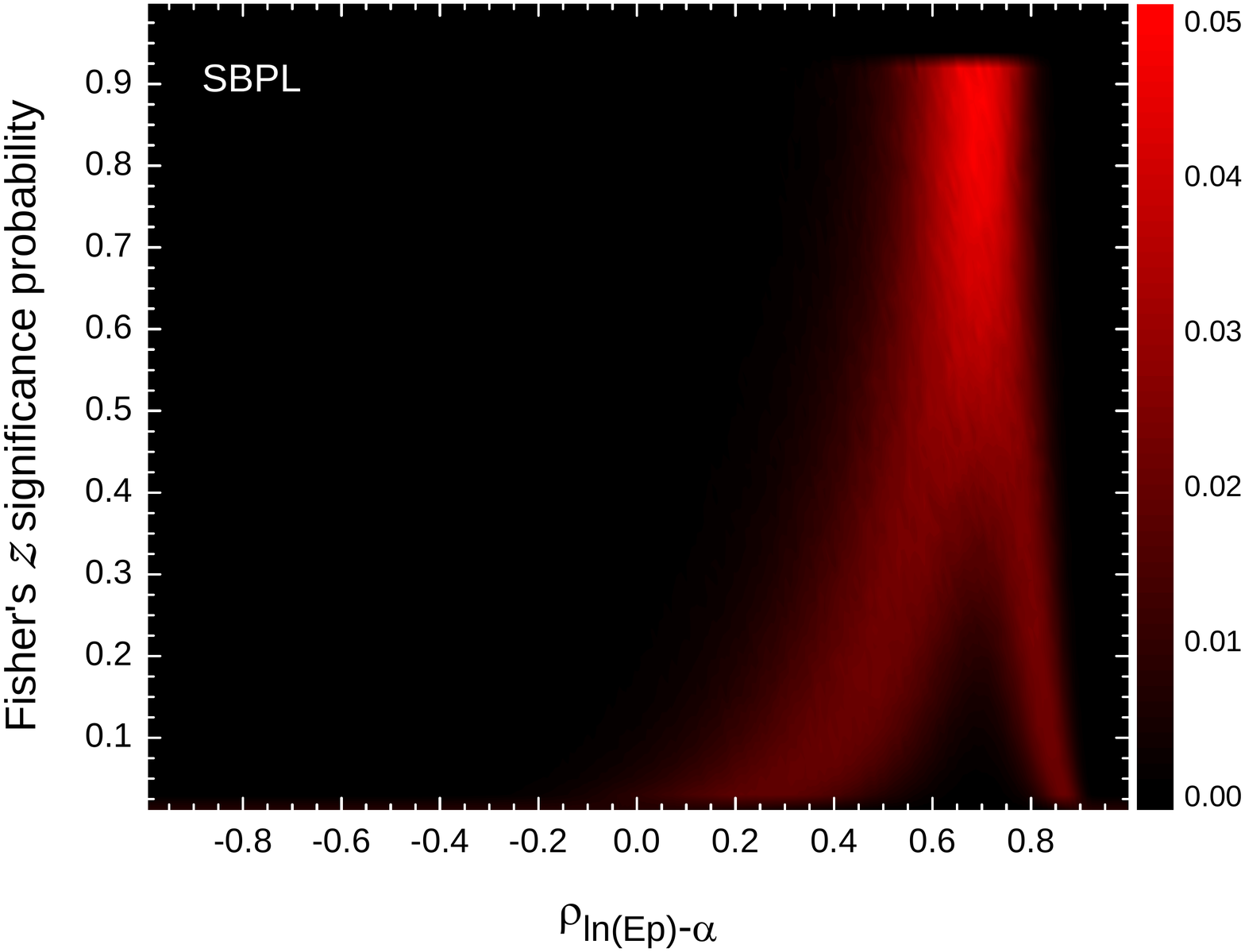}
\includegraphics[scale=0.20]{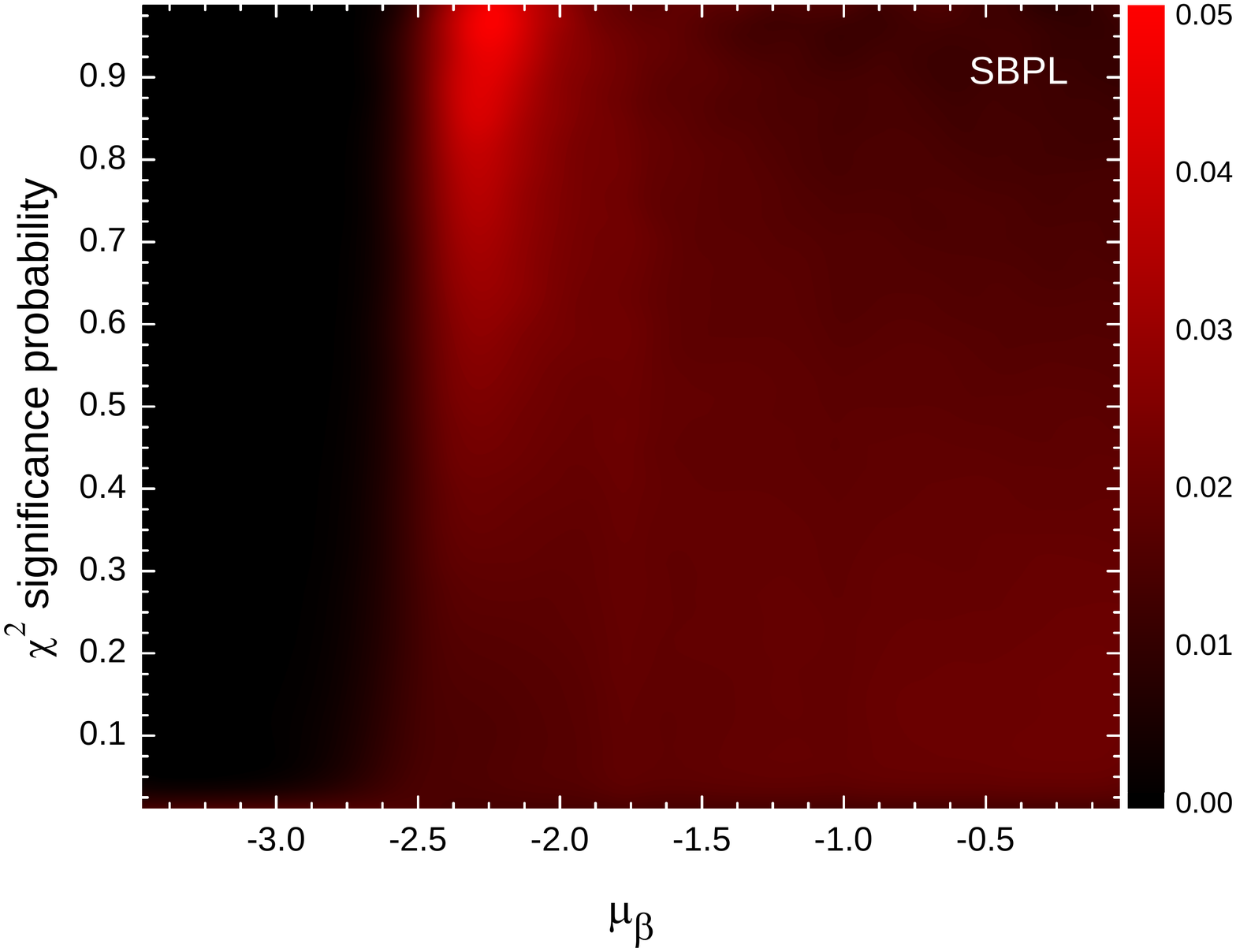}
\includegraphics[scale=0.20]{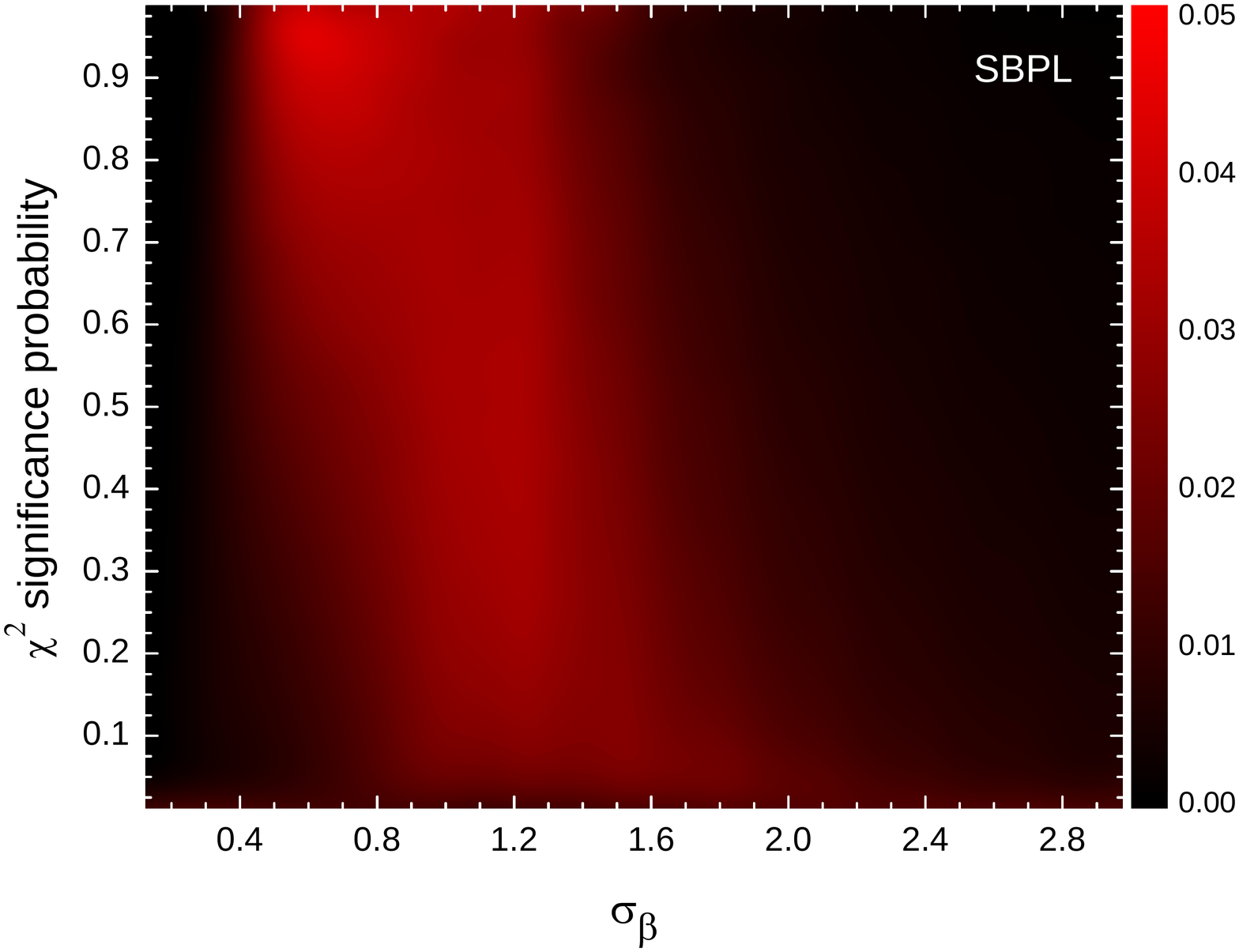}
\includegraphics[scale=0.20]{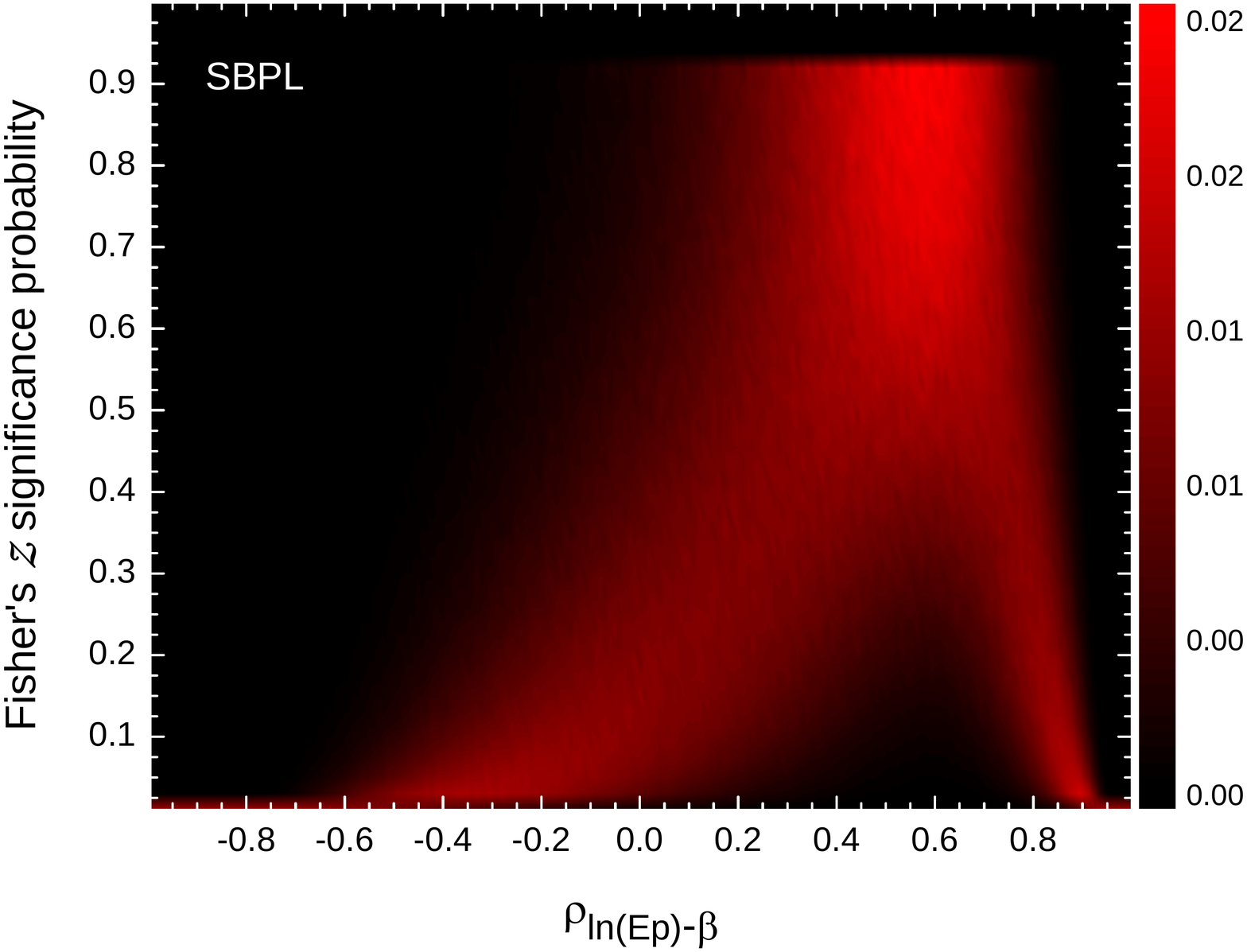}
\caption{Marginalized likelihood contour plots of the observed data given different parameter values of the truncated multivariate normal distribution assumed for the spectral parameters of the three GRB models. The likelihoods are obtained via simulation including the effects of sample-incompleteness as described in \S3.1.3. \label{Chisqsigs}}
\end{figure*}
\begin{figure*}
\includegraphics[scale=0.20]{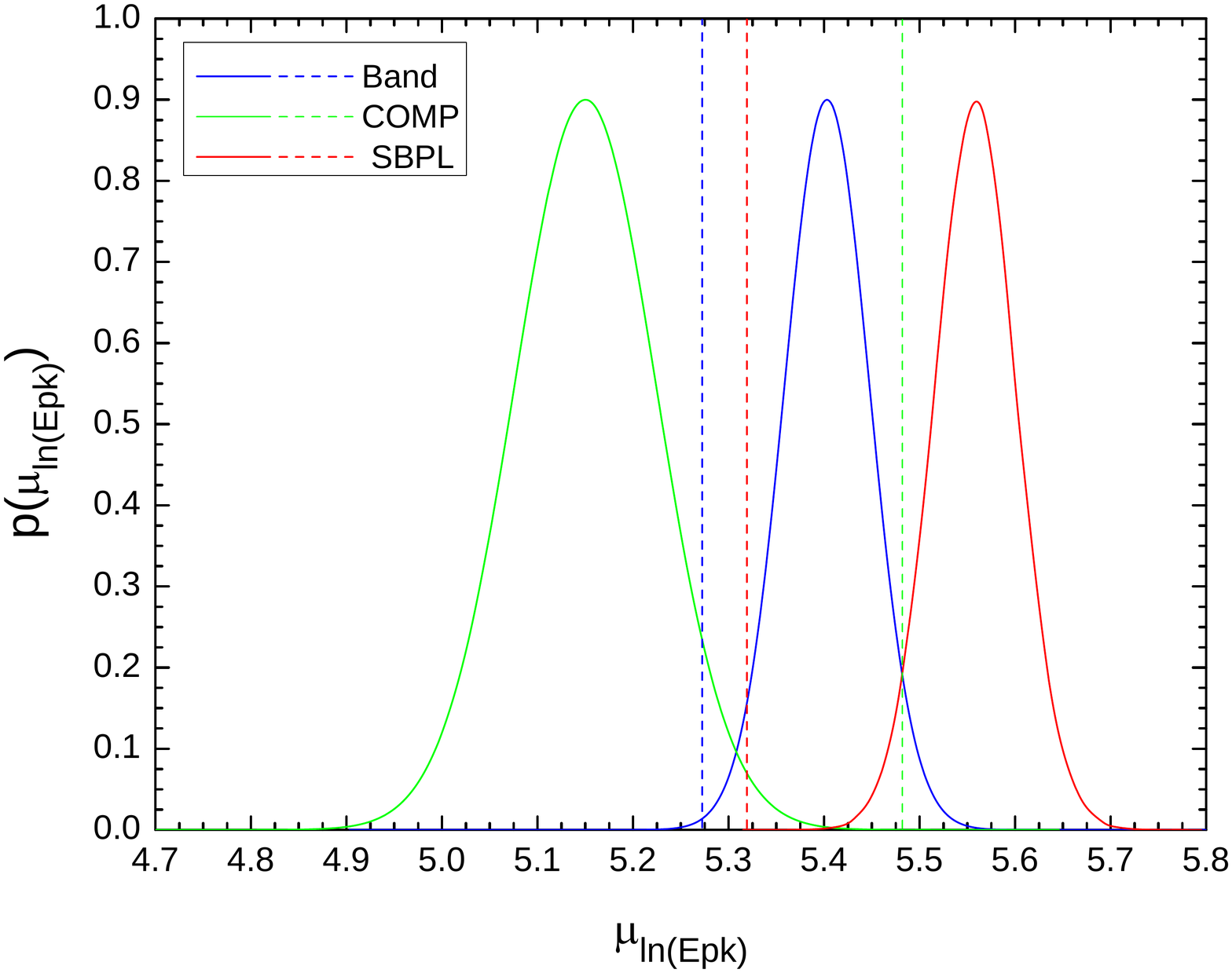}
\includegraphics[scale=0.20]{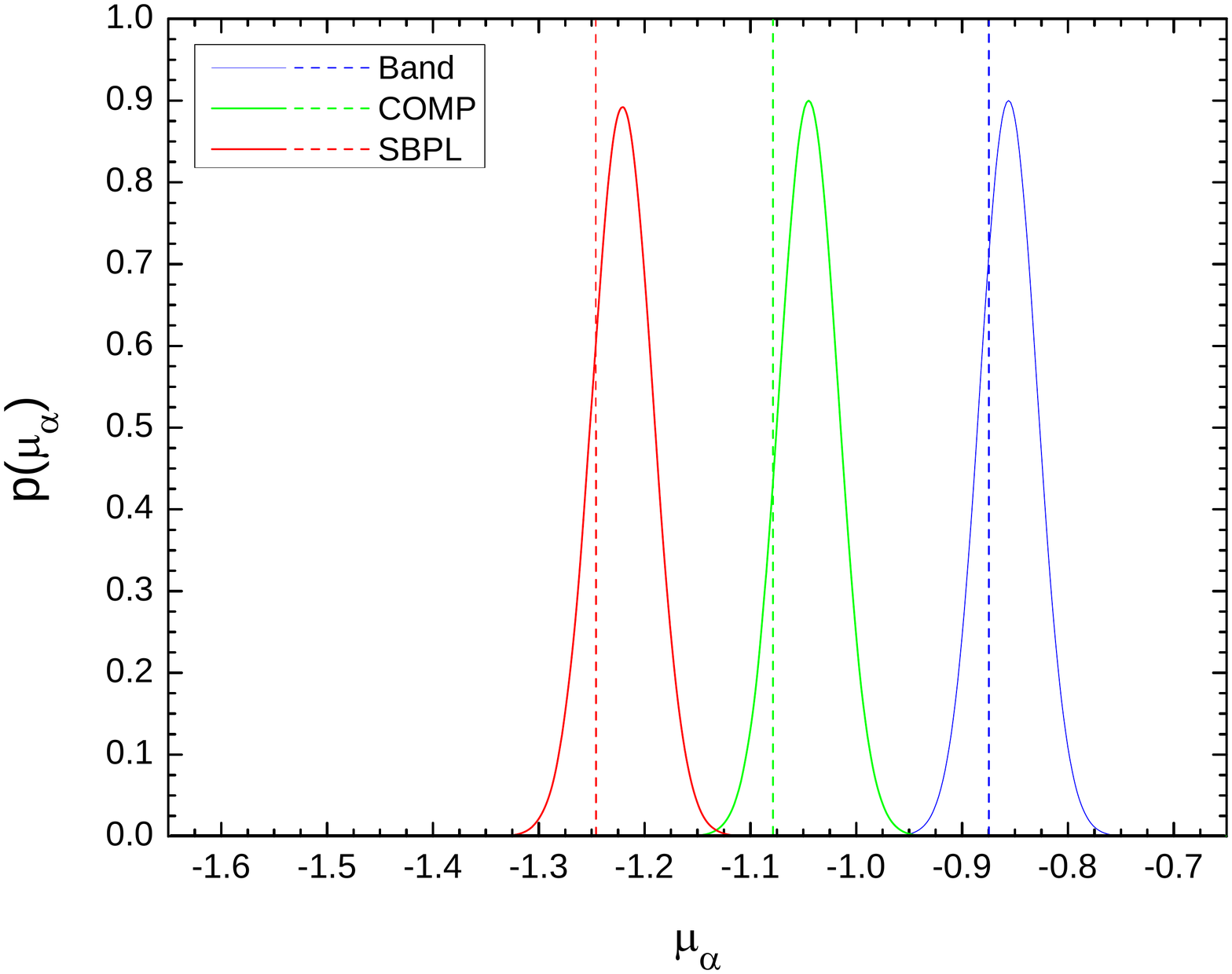}
\includegraphics[scale=0.20]{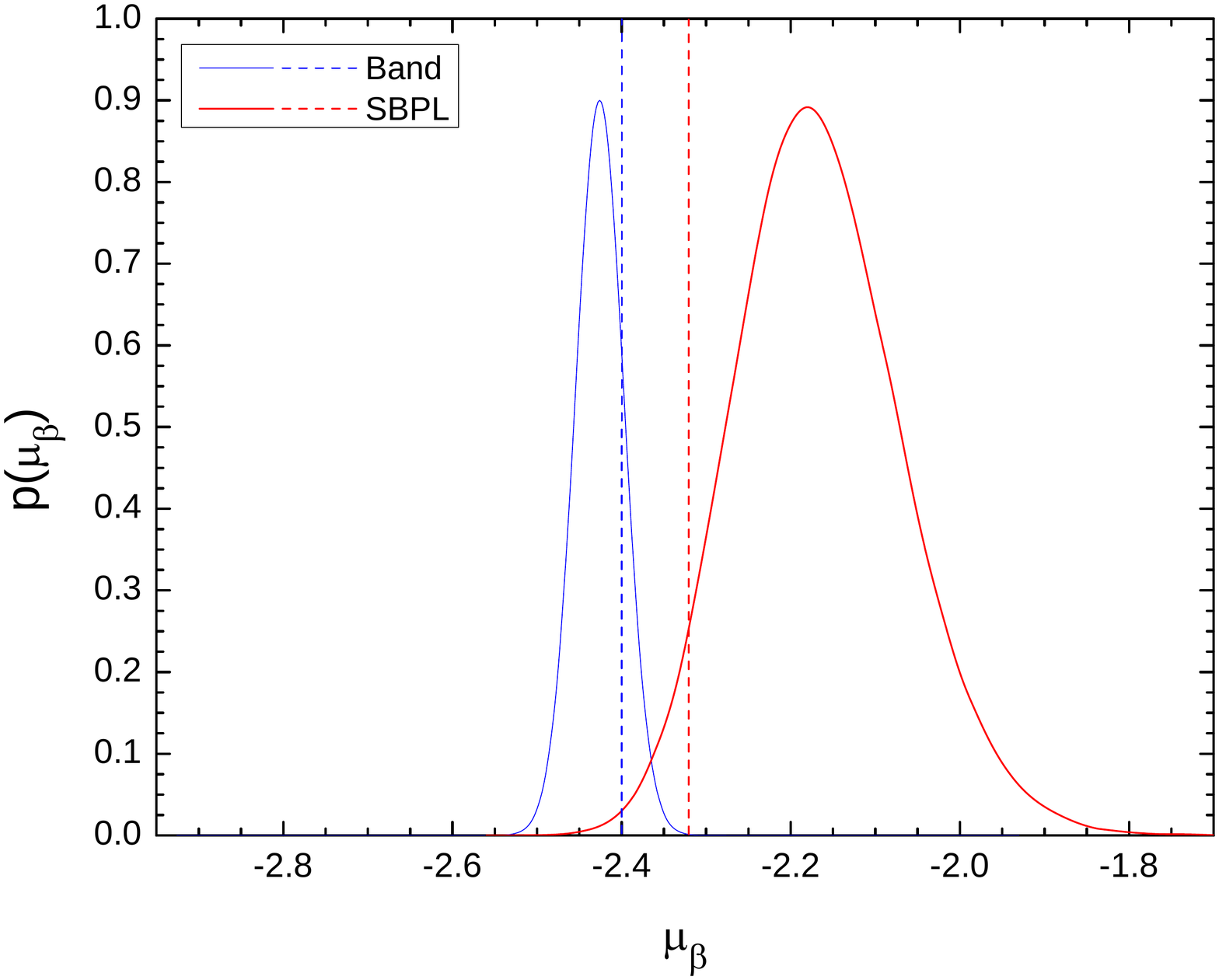}
\includegraphics[scale=0.20]{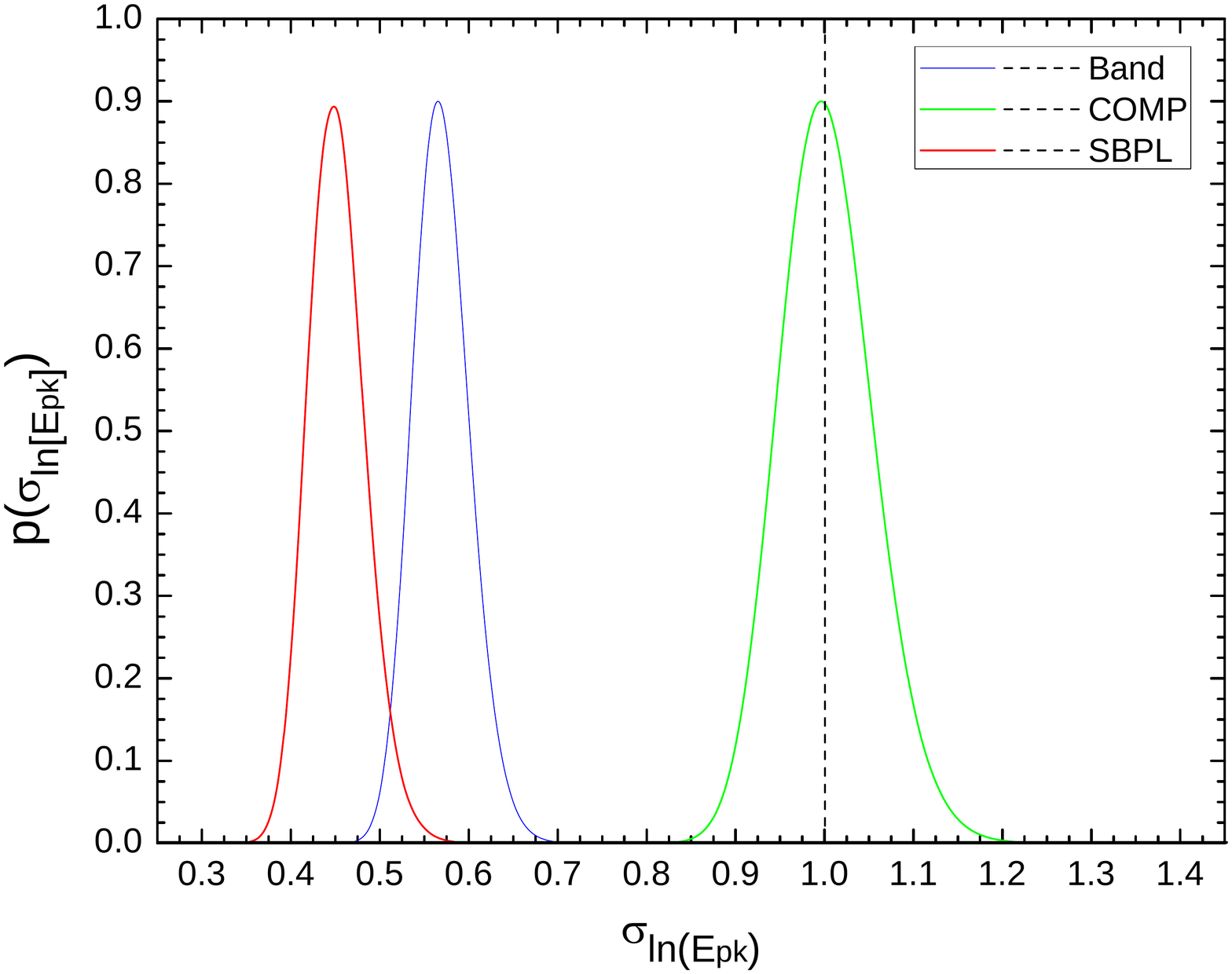}
\includegraphics[scale=0.20]{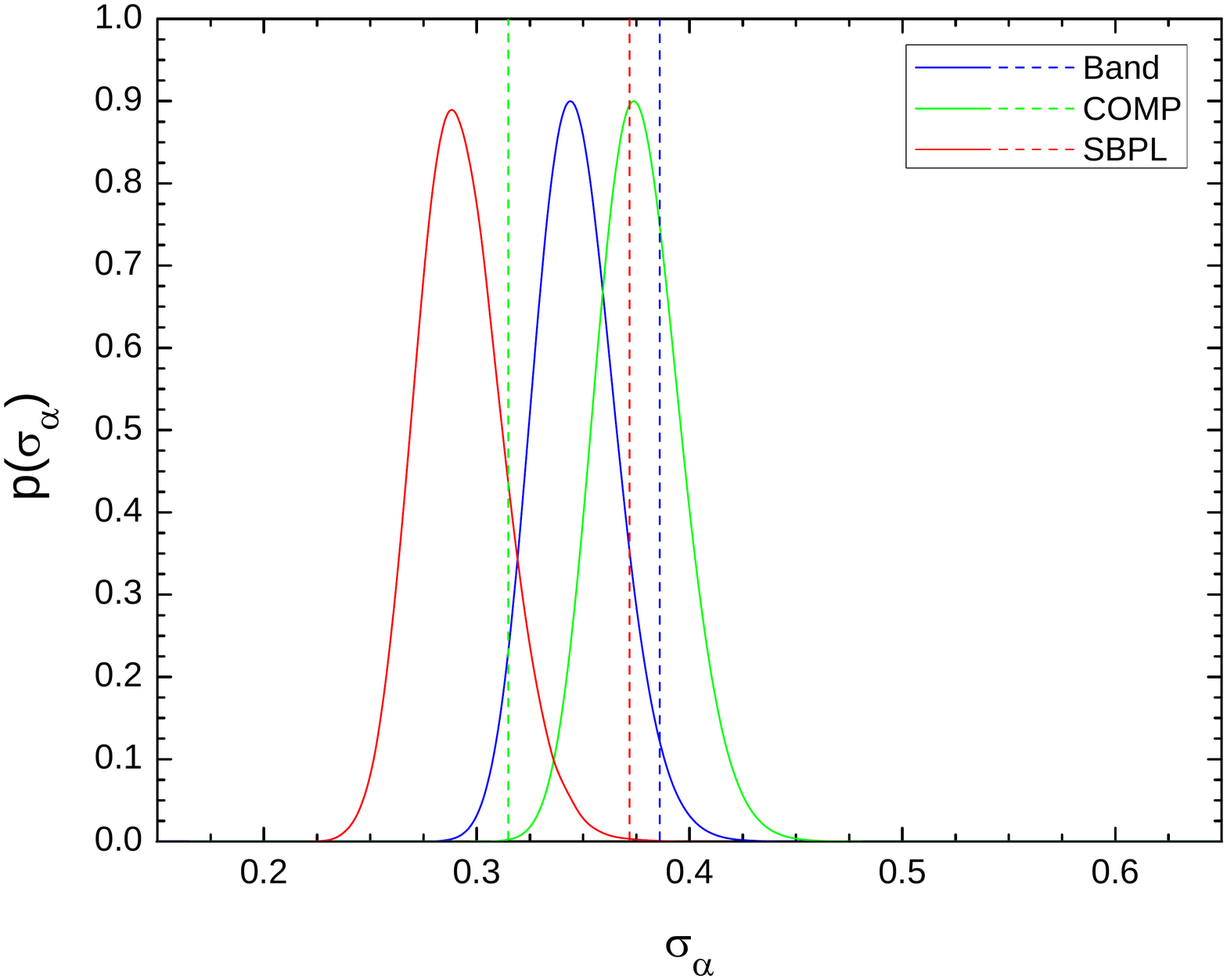}
\includegraphics[scale=0.20]{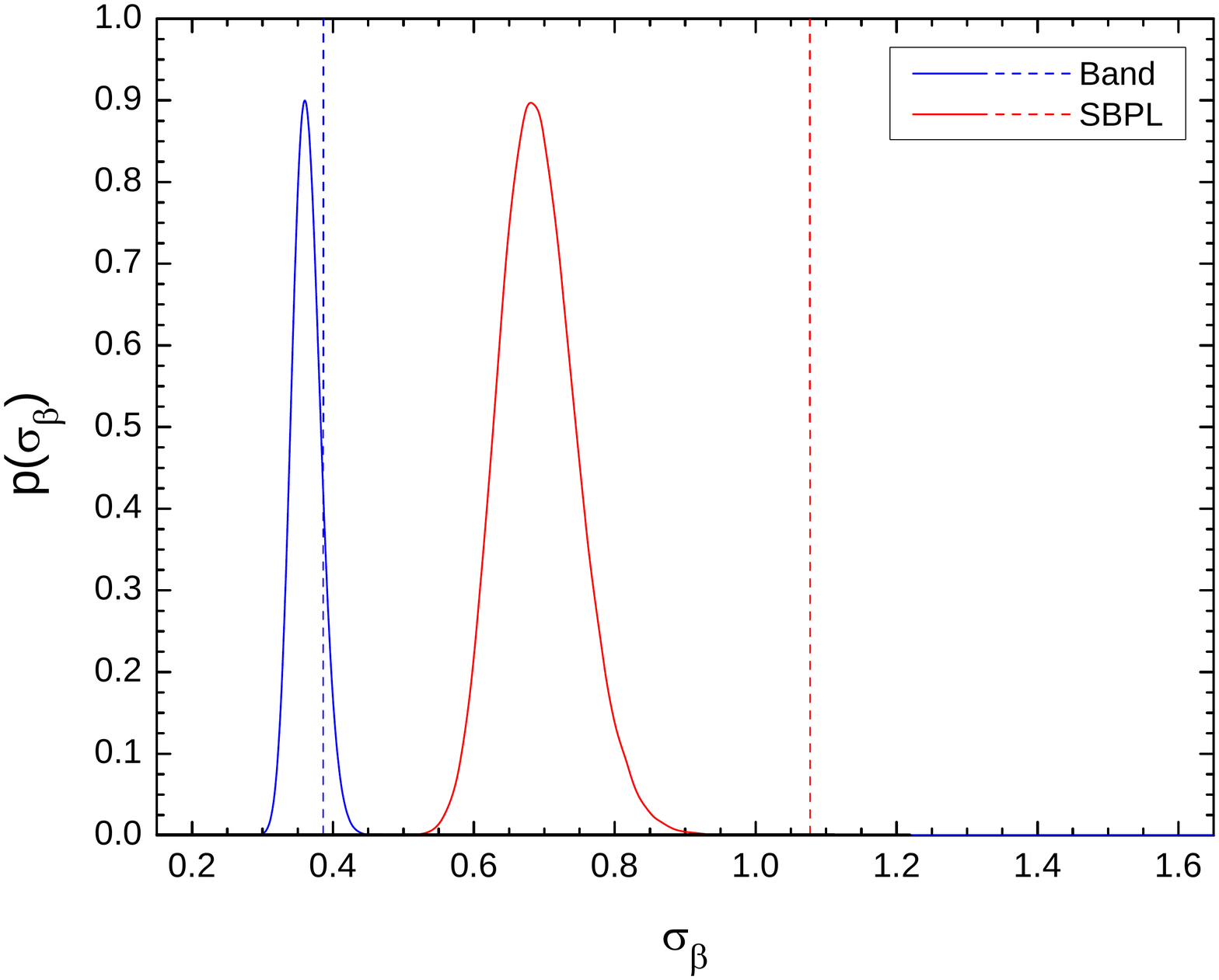}
\includegraphics[scale=0.20]{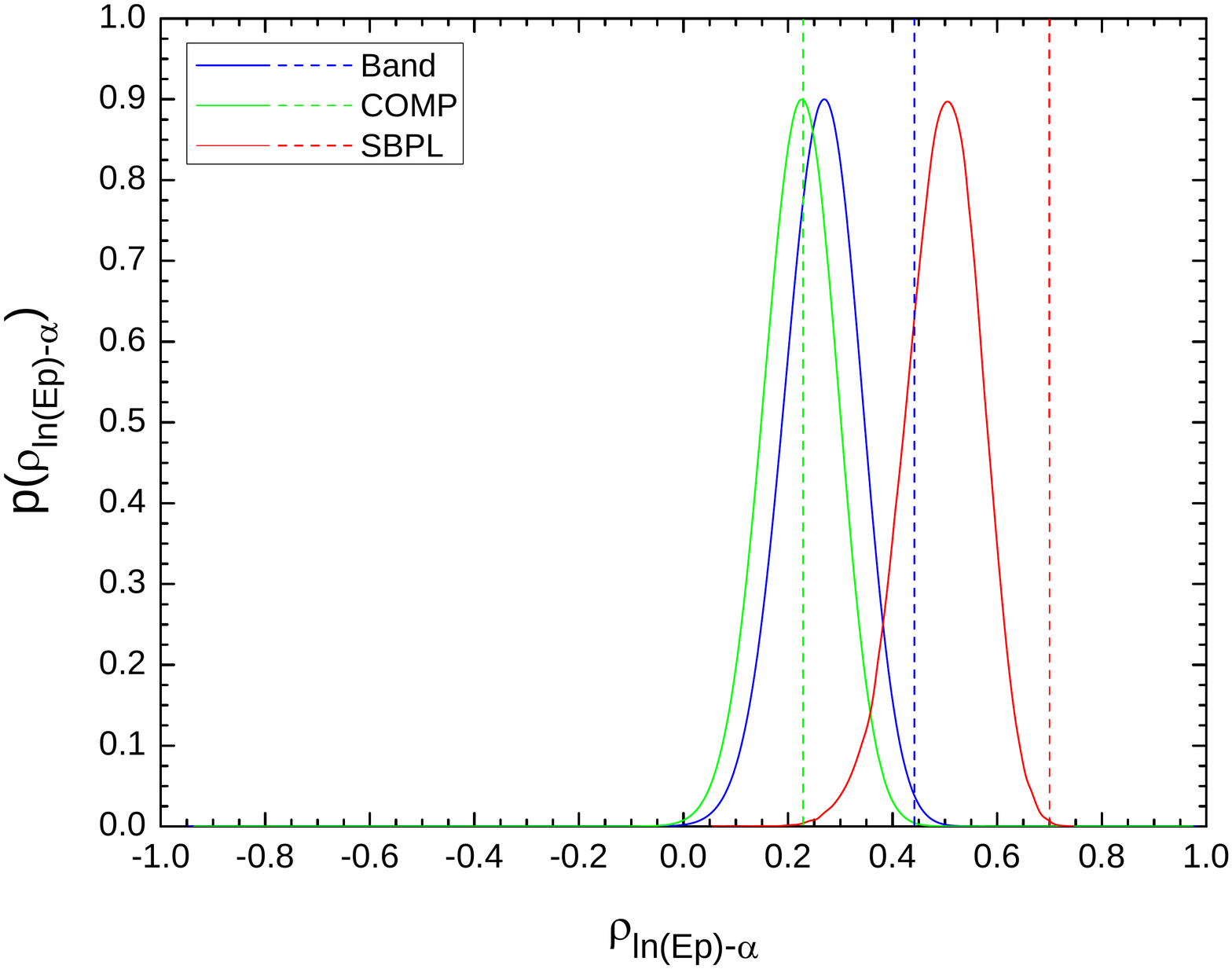}
\includegraphics[scale=0.20]{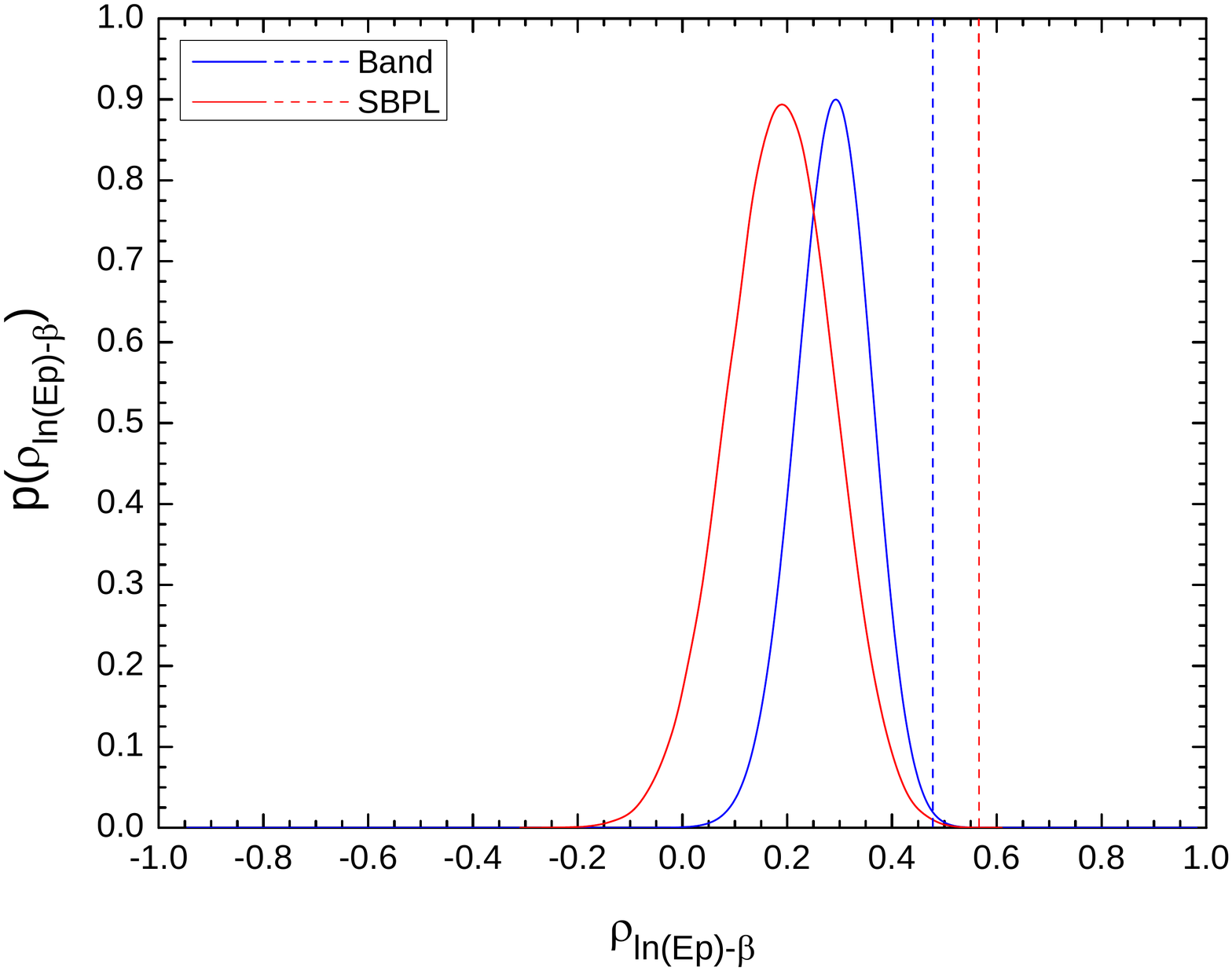}
\includegraphics[scale=0.20]{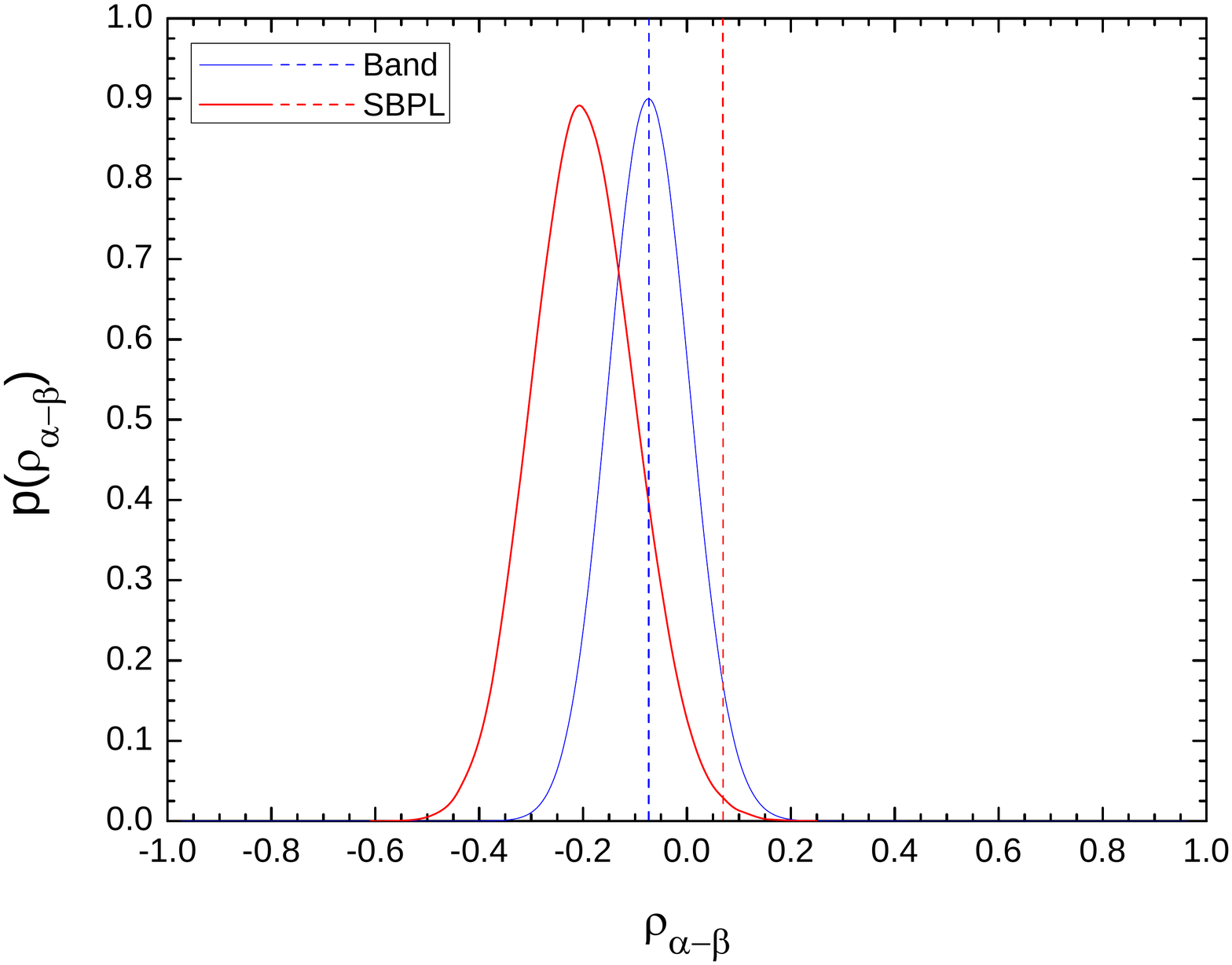}
\caption{Posterior distributions of the parameters of the truncated multivariate normal distributions considered for the spectral parameters of the 3 GRB models: Band, COMP (CPL) \& SBPL. The solid curves for each spectral model parameters are the marginal posterior pdfs derived from the Markov Chain Monte Carlo algorithm described in \S3, excluding the possible effects of data-truncation and sample incompleteness, mainly due to the flux-limits set by different authors on the observed-analyzed GRB samples (e.g. K06, G09, N08). The dashed-vertical lines in each graph represent the minimum $\chi ^2$ \& the minimum KS-distance estimates of the parameters obtained numerically via simulation including the effects of data-truncation and sample-incompleteness on the spectrally-analyzed GRB samples. \label{pdfs}}
\end{figure*}
\begin{figure*}
\includegraphics[scale=0.31]{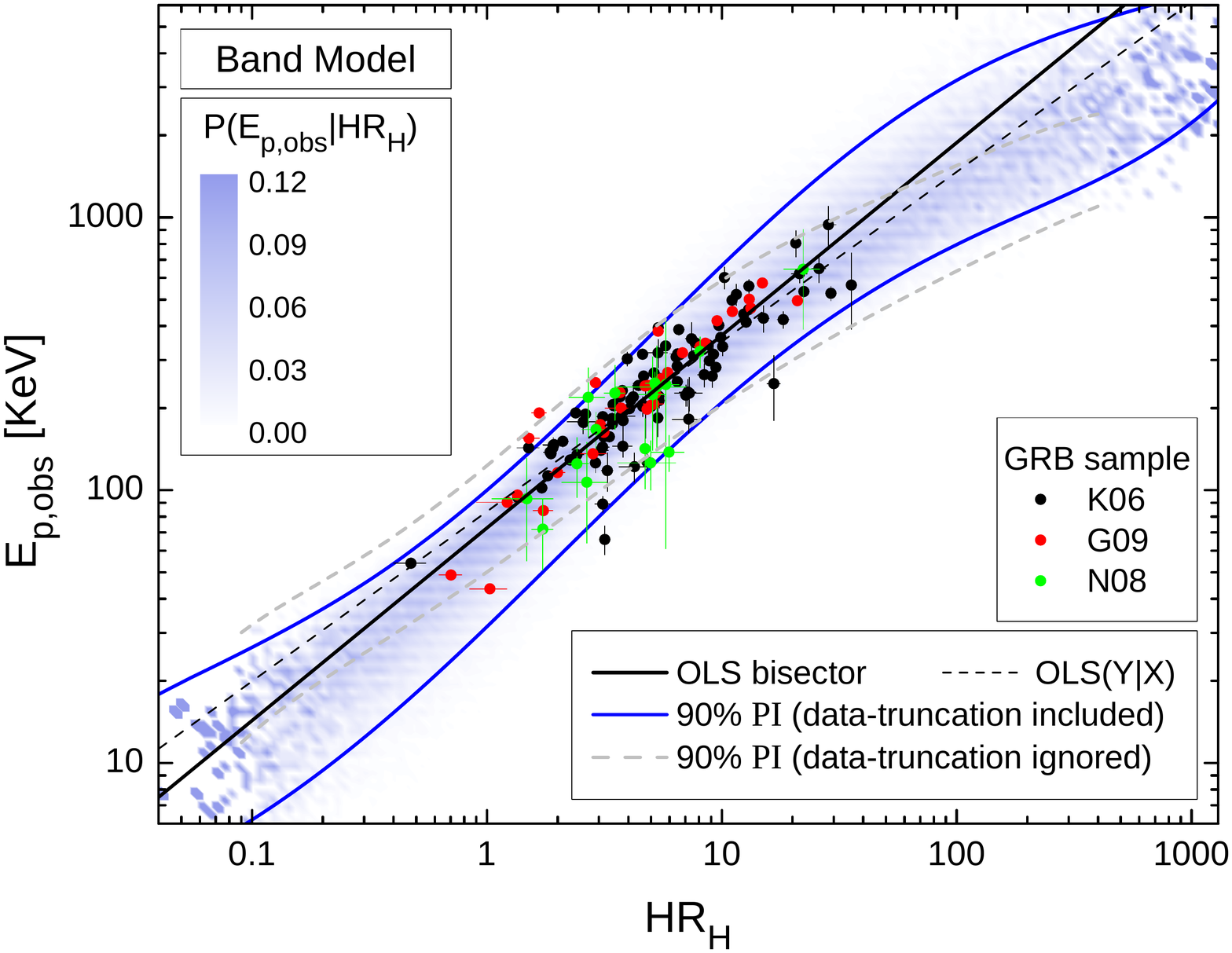}
\includegraphics[scale=0.31]{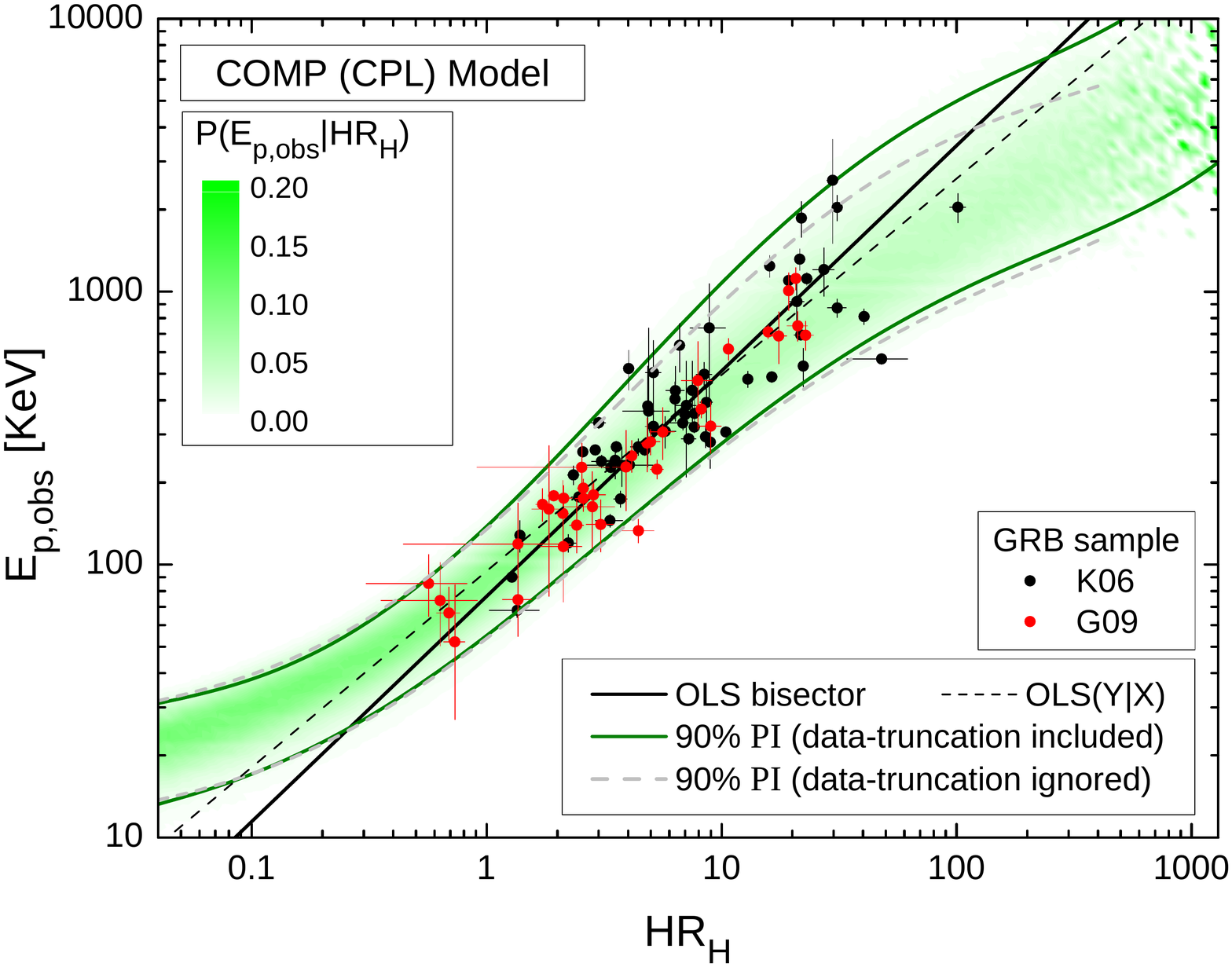} 
\includegraphics[scale=0.31]{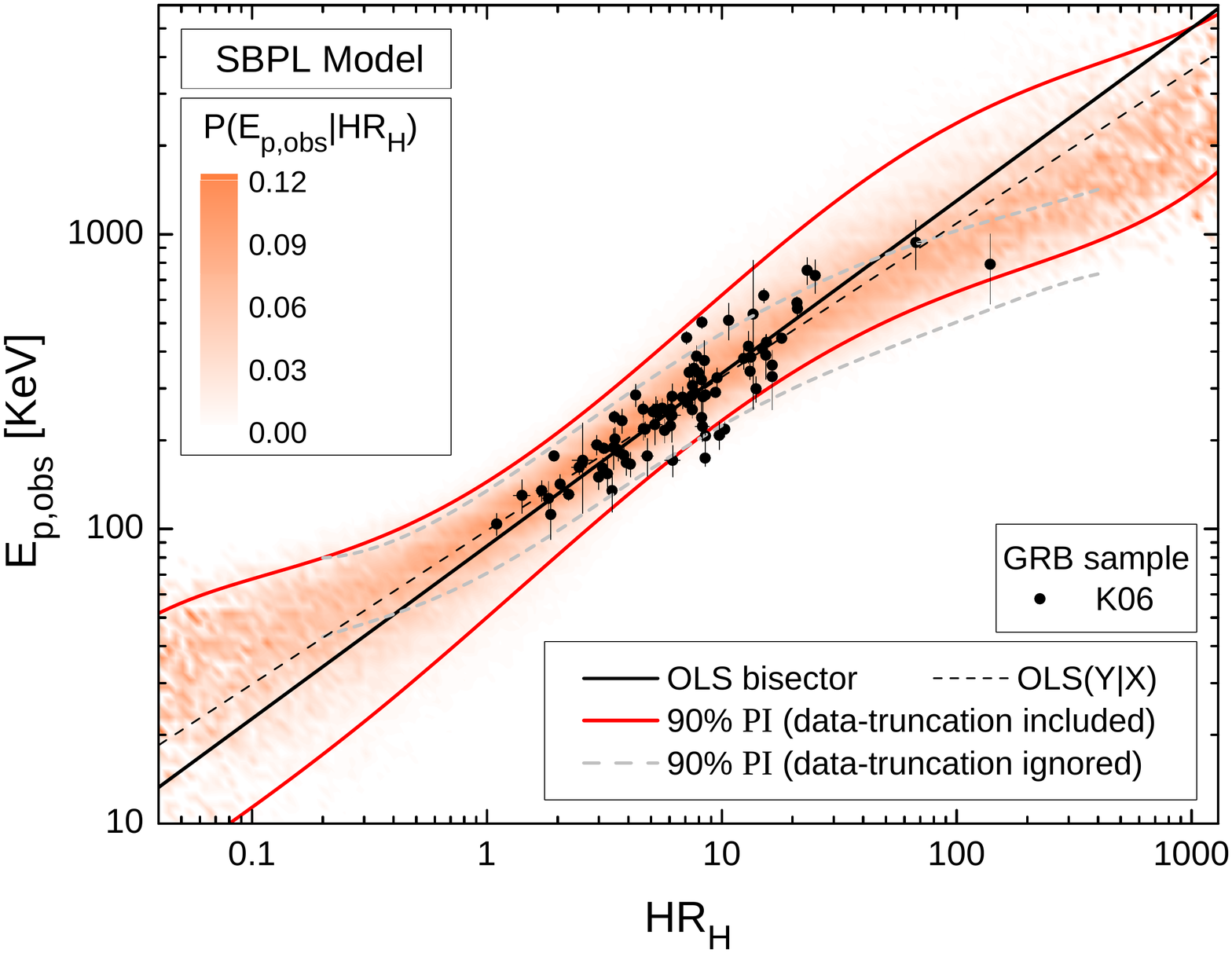} 
\includegraphics[scale=0.31]{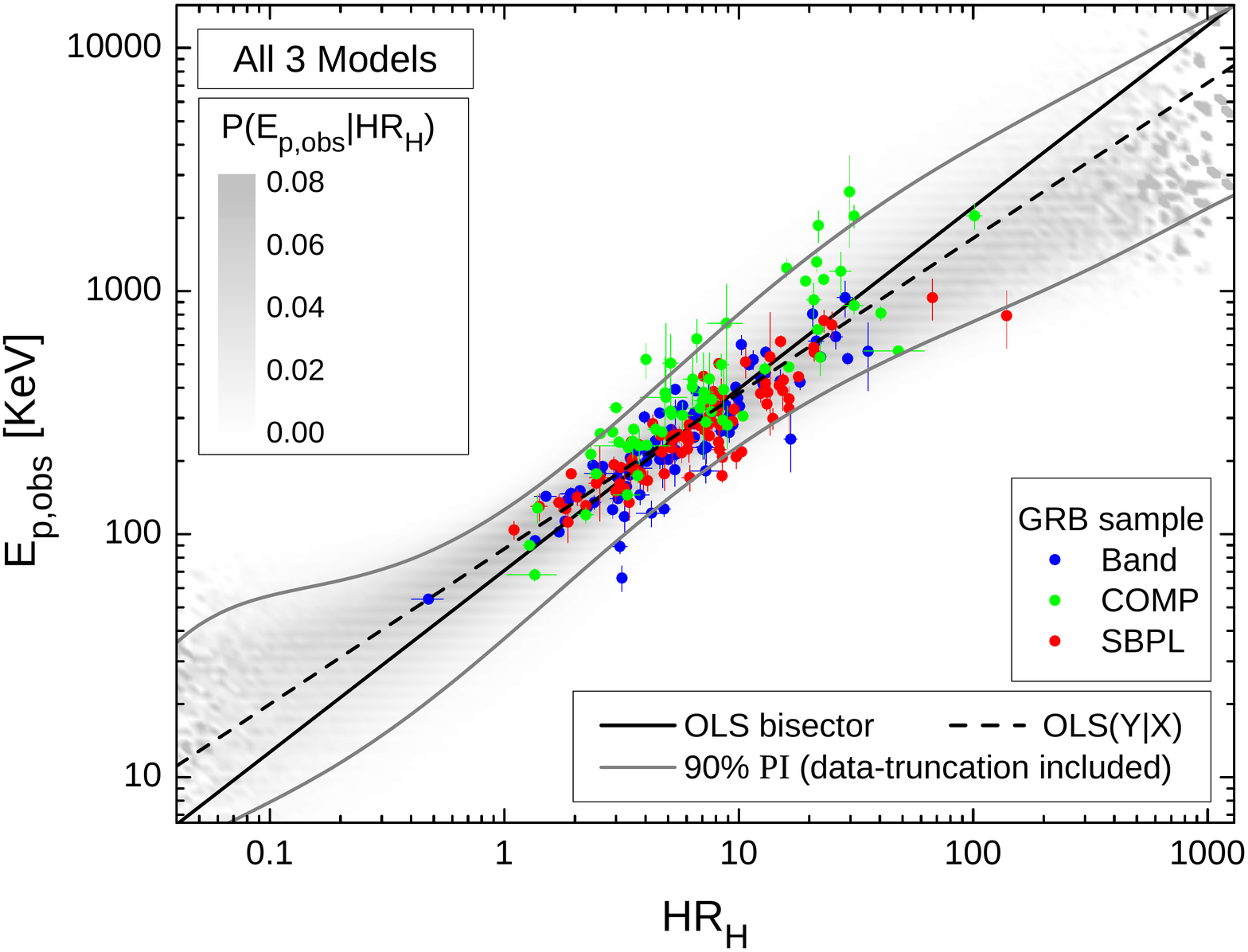} 
\caption{Plot of \epobs vs. \hrh depicting the $90\%$ Prediction Intervals (PI) for the three spectral models. The solid-curve PI in each graph, are based on the spectral parameters derived from the simulation including the effects of data truncation, mainly due to flux-limits set on the analyzed samples of GRBs by different authors (e.g. K06, N08, G09). In contrast, the gray dashed curves represent the $90\%$ PI for the case that no data-truncation was considered. The colorful dots in each graph represent the analyzed samples of BATSE GRBs by K06, N08 \& G09, used to constrain the spectral parameters distributions of the simulation. The background contour plots in the graphs indicate the conditional chances of observing a specific \epobs given $\hr$. The fourth plot ({\it bottom left}) depicts the $90\%$ prediction intervals derived for the conditional weighted average of $\ln(\epo)$ on $\ln(\hr)$ of the three GRB models as given in Eqn.~\eqref{eq:epkprob}. The prediction intervals are obtained by simulating 10,000,000 GRBs for each spectral model. \label{PIs}}
\end{figure*}
Similar to the Bayesian analysis of data in \S3.1.2, it is assumed throughout the simulation, that the spectral parameters of three GRB models (Eqns.~\eqref{eq:SP1},~\eqref{eq:SP2},~\eqref{eq:SP3}) are drawn from a truncated multivariate normal distribution. In addition to the truncations due to the physical constraints on the spectral parameters (\S3.1.2), we also consider the truncation due to sample-incompleteness and flux-limits imposed on the observational data (Figure~\ref{flux-limit}). Once the effects of sample-incompleteness are understood and modeled, the simulation algorithm is straightforward:
\begin{enumerate}
	\item Generate a random sample of the same size as the observational data from the truncated multivariate normal distribution with random mean vector and covariance matrix, subject to the truncation due to sample-incompleteness.\\
	
	\item Compare the similarity of the two simulated and observational samples by a proxy.\\
	
	\item Repeat the above steps to find the most probable set of parameters that would maximize the similarity of the simulated and the observational data. 
\end{enumerate}
	As seen in Figure~\ref{flux-limit}, the truncation on the observational data due to flux limits, primarily affects the distribution of $\ln(\epo)$ of BATSE GRBs. In particular, the variance of $\ln(\epo)$ distribution is likely significantly underestimated in K06 sample of GRBs. Although, our simulations indicate that the flux-limits might also directly affect the distributions of the GRB models' photon indices, we find it to be much weaker compared to the effects of truncation on $\ln(\epo)$ distribution. Thus, for simplicity we assume that sample-incompleteness only affects $\ln(\epo)$ directly, and through that, it might affect the probability distribution of any of the rest of spectral parameters that have nonzero covariance with $\ln(\epo)$.

To model the truncation due to flux-limits on the data, we first estimate the most probable mean and variance of $\ln(\epo)$ of BATSE GRBs for the 3 spectral models. This can be done, knowing that the hardness ratio distribution of the BATSE GRBs shows an approximately Gaussian behavior with a peak which is very close to the barycenter of the calibration sample that was used to derive the linear \hrep~ regression lines in \S2.2. Therefore, since \hrep ~relation is not a biased estimator of $\ln(\epo)$ close to the barycenter of the data, it can be used to convert the mode of $\ln(\hr)$ distribution to the mode of $\ln(\epo)$ normal distribution for each spectral model, which can then be used as an estimate of the mean of $\ln(\epo)$ of the entire BATSE GRBs for each spectral model {\it under the assumption of normality}.

To estimate the variance of $\ln(\epo)$ distribution, we rely on the fact that there is a partial linear correlation between $\ln(\epo)$ and the logarithm of bolometric $1-$second peak flux $\ln(P_{bol})$ of GRBs in the observer frame (e.g. N08; Lloyd \& Petrosian 2000). The hard-dim side of this correlation is likely affected by detector thresholds (e.g. Shahmoradi \& Nemiroff 2009a). However, the bright-soft region of the correlation is likely real, originated from the physics of prompt emission. Depending on the spectral model considered, the $3\sigma$ limit of this relation in the bright-soft region intersects BATSE's trigger threshold ($\sim0.3~ph\, s^{-1}$ for 1-second trigger in the energy range $50-300$ KeV) at some point which can be regarded as the $3\sigma$ lower limit for $\ln(\epo)$ of BATSE GRBs beyond which no burst could have been triggered by BATSE Large Area Detectors (LADs) (Shahmoradi \& Nemiroff 2009a). These estimates for different spectral models are depicted and compared with the estimates based on the observed samples in Figure~\ref{pdfs}. 

The effect of sample-incompleteness on the observational data can then be modeled by fixing the mean and variance of $\ln(\epo)$ distribution to the above estimates, and then creating simulated samples that have the same peak energies as in the observational sample. This means that the rest of spectral parameters of each model (i.e. photon indices) are drawn from their conditional distributions on $\ln(\epo)$. Once the simulated sample is ready, it can be compared to observational data to infer the similarity of the two samples.
\begin{figure*}
\includegraphics[scale=0.31]{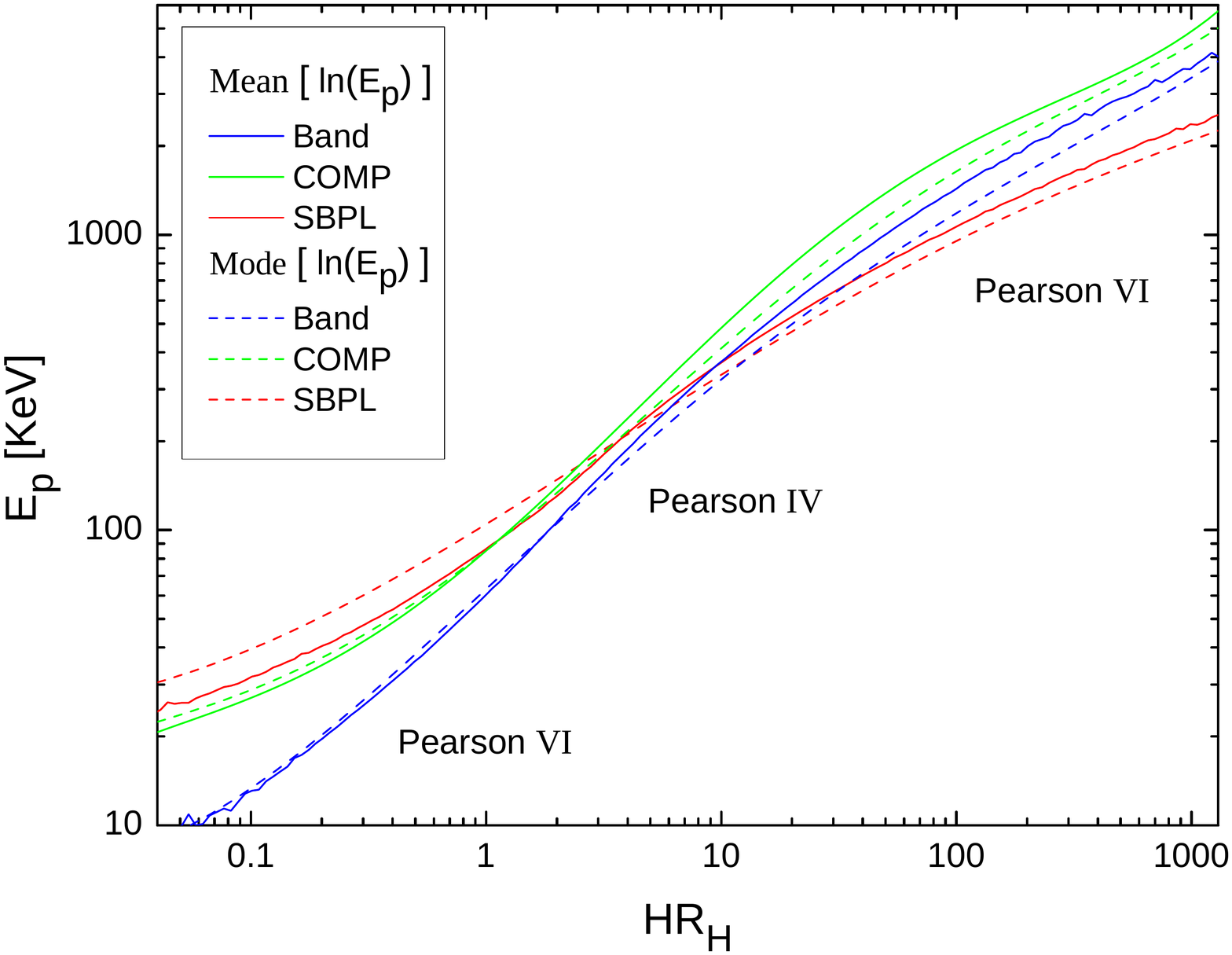}
\includegraphics[scale=0.31]{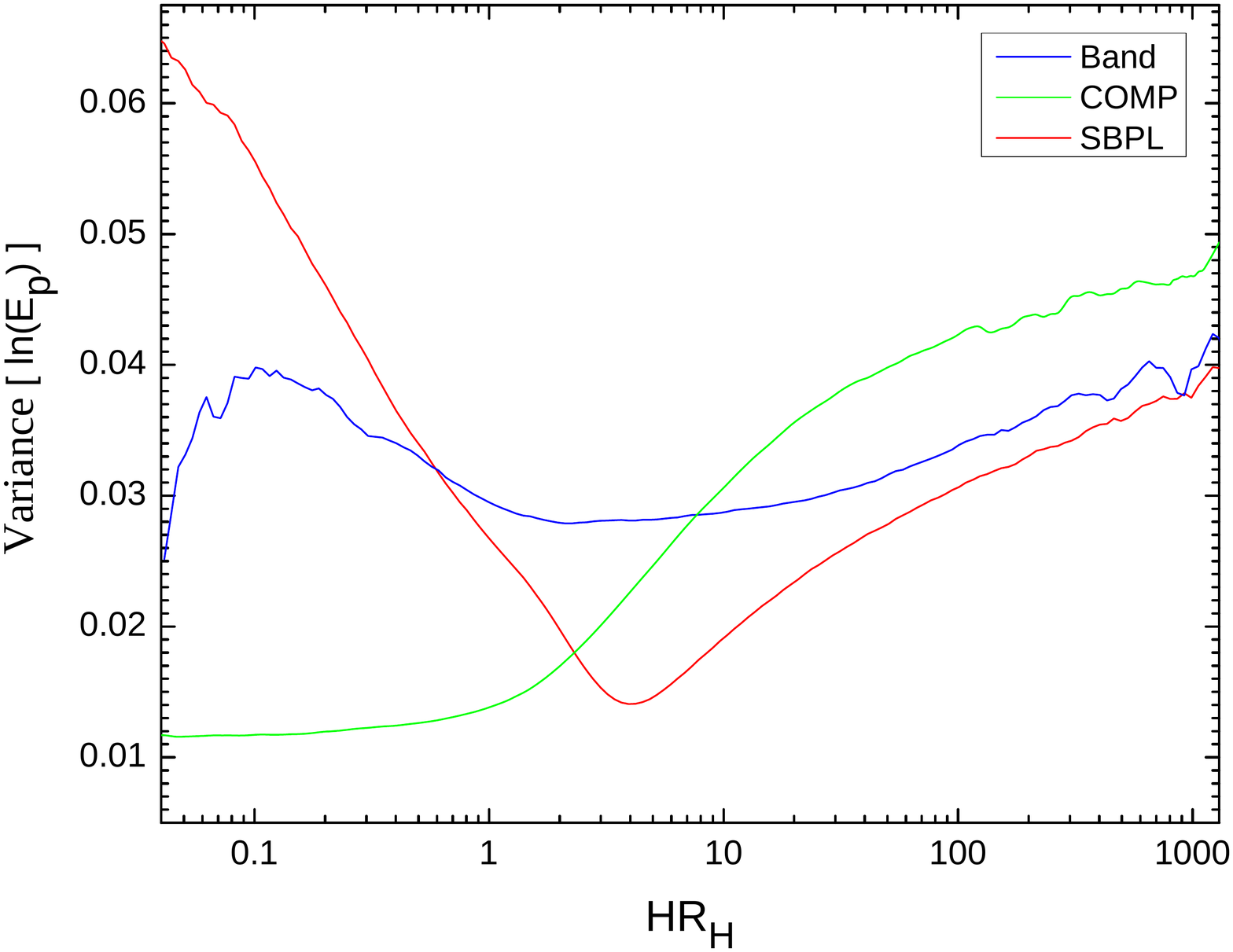} 
\includegraphics[scale=0.31]{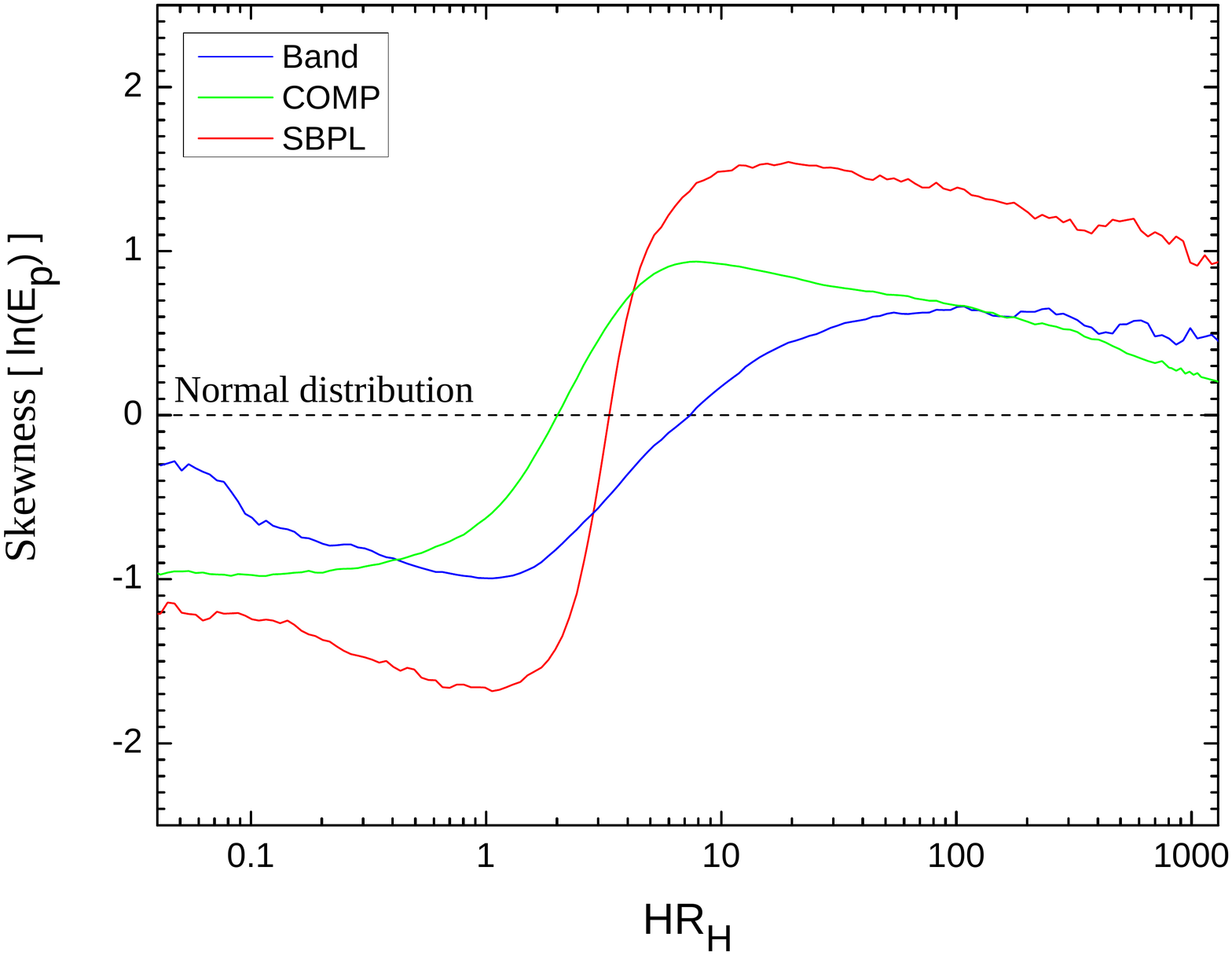} 
\includegraphics[scale=0.31]{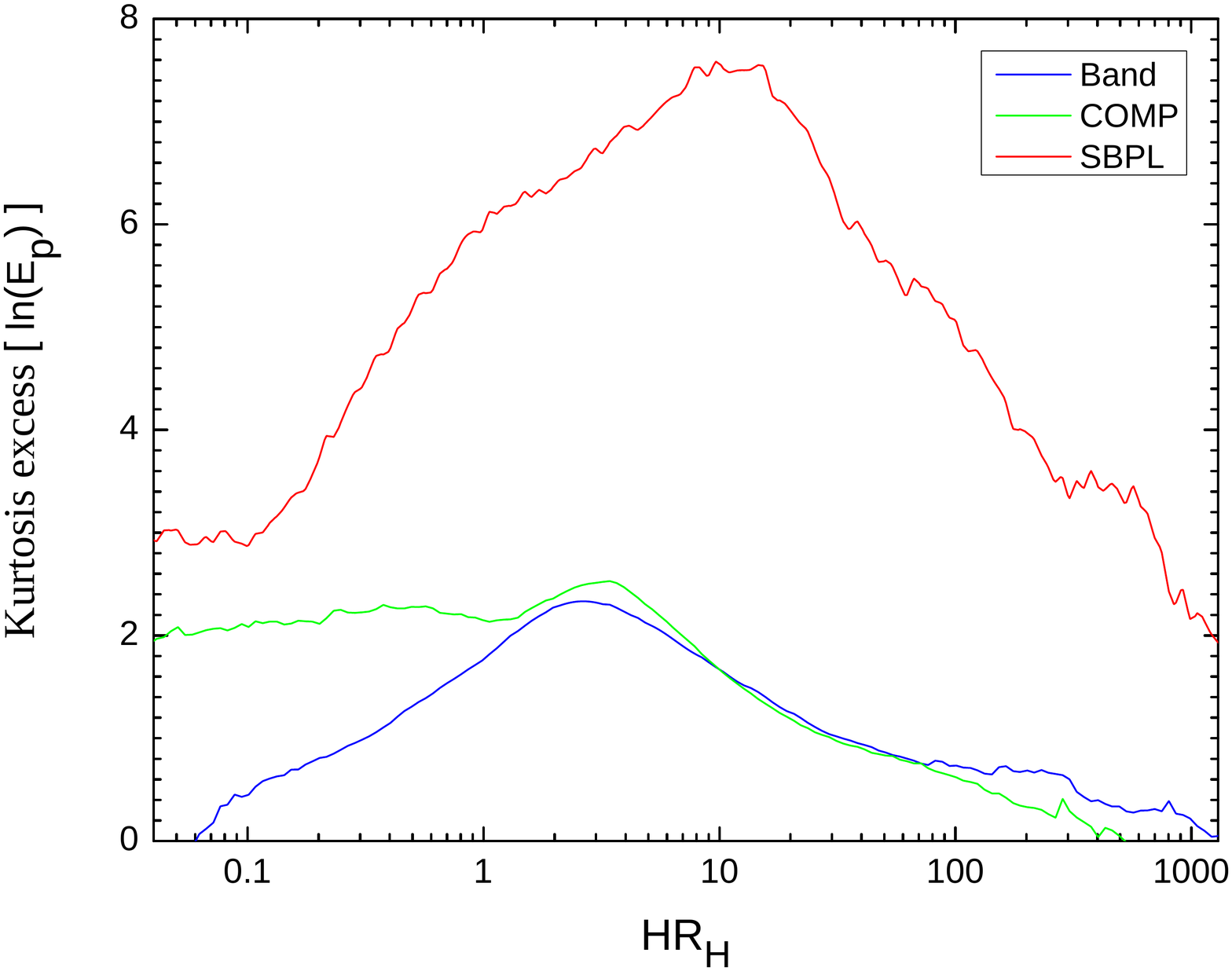} 
\caption{Plots of the first four moments of the conditional distribution $P\big(\ln(\epo)|\text{Model},\ln(\hr)\big)$ for the three GRB spectral models: Band, COMP(CPL) \& SBPL. {\bf Top Left:} Comparison of the modes (dashed curves) and means (solid curves) of \epobs given \hrh for the 3 spectral models. According to the Pearson's system of statistical distributions (Pearson 1901; Pearson 1895), $P\big(\ln(\epo)|\text{Model},\ln(\hr)\big)$ for the three GRB models is be best characterized by the Pearson's Type IV distribution (unlimited range, skewed) close to the center of BATSE $\ln(\hr)$ distribution and by the Pearson's Type VI distribution (limited range at one side, skewed) close to the tails of \hrh distribution. {\bf Top Right:}.  Variance of $\ln(\epo)$ given \hrh for the three GRB models. {\bf Bottom Left:} Variation of the third standardized moment (skewness) of $P\big(\ln(\epo)|\text{Model},\ln(\hr)\big)$ vs. \hrh for the three GRB models. The horizontal dashed line, represents the third standardized moment of normal distribution. {\bf Bottom Right:} Variation of the fourth standardized cumulant (Kurtosis excess) vs. \hrh. As it had been implied by the calibration sample before in \S2.2, $P\big(\ln(\epo)|\text{Model},\ln(\hr)\big)$ for all 3 GRB models exhibits a leptokurtic-type behavior by being more {\it peaked} and {\it heavy-tailed} compared to normal distribution which has zero Kurtosis excess. The fact that $P\big(\ln(\epo)|\text{Model},\ln(\hr)\big)$ has the highest Kurtosis excess close to the barycenter of simulated data for the Band \& SBPL models together with its low variance in the same region, indicates the relative accuracy of \hrep ~relation in estimating \epobs close to the barycenter compared to the extreme ends of the relation. By contrast, the uncertainty in \epobs estimation appears to increase continuously with \hrh for the COMP (CPL) model. \label{moments}}
\end{figure*}
Generally, the assessment of similarity, requires the use of nonparametric multivariate goodness-of-fit tests. Such tests, although exist, have been rarely discussed and treated in statistics due to difficulties in the interpretation of the test statistic (e.g. Justel et al. 1997; Peacock 1983). In general, one can always use the Pearson's $\chi^{2}$ goodness-of-fit test (Pearson 1900; Fisher 1924) for any multivariate distribution. However, for the special case considered here, one would need an observed sample consisting of $N\gg 1000$ GRBs to avoid serious instabilities that occur in $\chi^{2}$ tests due to small sample sizes (Cochran 1954). In addition, since it is already assumed that the data truncation mainly affects the distribution of $\ln(\epo)$, this parameter is not a random variable in the simulation and it should be regarded as an explanatory variable -- determined by the limitations of the analysis -- for the rest of the spectral parameters that are considered as the response variables to $\ln(\epo)$. In this case, the multivariate comparison of the spectral parameters can be reduced to a set of separate univariate goodness-of-fit tests for the marginal distributions of each of the spectral parameters. Then the likelihood of the data -- given the random mean vector and covariance matrix of the truncated multivariate normal distribution -- is assessed by the use of two separate nonparametric $\chi^{2}$ and KS test statistics (e.g. Weber, Leemis \& Kincaid 2006; Gy$\ddot{o}$rfi et al. 1996; Berkson 1980). To test for the significances of the similarities of the correlation coefficients between the spectral parameters, Fisher's $z$ statistic (Fisher 1921; Fisher 1915) is used. The simulation then runs for a large number of iterations until the likelihood function reaches its asymptotic form and the most probable set of parameters are found. The resulting marginal likelihoods are shown partly in Figure~\ref{Chisqsigs}.

It should be noted that although the most probable mean and variance of $\ln(\epo)$ distribution derived by the above methods might only be approximations to the true population mean and variance, we find that small deviations ($\lesssim 0.1$ dex) from these estimates result in negligible changes in the derived prediction intervals for \hrep ~relation. 

\subsection{Derivation of the Prediction Intervals for \hrep ~Relation}
Now, in order to construct the prediction intervals of \hrep ~relation, first the spectral parameters of a simulated GRB are randomly drawn from the truncated multivariate normal distribution with its mean vector and covariance matrix fixed to their most probable values derived in \S3.1.2 \& \S3.1.3. The hardness ratio $\hrs$ is then calculated for each simulated GRB based on its randomly given spectral parameters and mapped to \hrh via the linear relations \eqref{eq:olsband},\eqref{eq:olscomp} \& \eqref{eq:olssbpl} between $\hrs$ \& \hrh in \S2.2 (Figure~\ref{hrshrh}). The resulting $90\%$ Prediction Intervals (PI) for each spectral model (Figure~\ref{PIs}) are then fit by polynomials of the 5th order. The fits are summarized in table~\ref{PIfits}.

As seen in \S2.2, also Figure~\ref{PIs}, the conditional probability of having a particular $\ln(\epo)$ given a hardness ratio $\hr$, $P\big(\ln(\epo)|\ln(\hr)\big)$, at the tails of \hrep ~relation depends heavily on the spectral model that fits best the GRB spectrum. This means that in order to know the peak energy of a BATSE GRB given its hardness ratio \hrh from the BATSE catalog, we would also need to know the best fit spectral model for the GRB beforehand, that is,
\begin{eqnarray}
\label{eq:model1}
P\big(\epo|\hr\big) &\propto& \displaystyle\sum_{\text{Models}} P\big(\text{\textbf{Model}}\big) \\ 
&\times & P\big(\epo|\text{\textbf{Model}},\hr \big) \nonumber
\end{eqnarray}
where the probability of having a specific spectral model as the best fit, depends on the signal-to-noise ratio ($S/N$) of the GRB lightcurve and its hardness \hrh (K06; Band et al. 1993),
\begin{equation}
\label{eq:model2}
P\big(\text{\textbf{Model}}\big) = P\big(\text{\textbf{Model}}|S/N,\hr\big)
\end{equation}
For example, K06 found that the ability to fit GRBs spectra with the Band and SBPL models depends on $S/N$, also on whether the break energy of GRB in the two spectral models is inside the BASTE energy window. Generally, the most probable spectral model for a GRB can be found by simulation given its $S/N$ and \hrh. However, we leave this for the supplemental work to this paper (Shahmoradi \& Nemiroff 2009c), and assume here, that $P\big(\text{Model}\big)$ is independent of $S/N$ and \hrh of the burst. Therefore, the probability $P(\ln(\epo)|\ln(\hr))$ can be written as,
\begin{eqnarray}
\label{eq:epkprob}
&&P\big(\ln(\epo)|\ln(\hr)\big) \propto \\ \nonumber
&& P\big(\text{\textbf{Band}}\big)\times P\big(\ln(\epo)|\text{\textbf{Band}},\ln(\hr)\big) \\ \nonumber
&+& P\big(\text{\textbf{COMP}}\big)\times P\big(\ln(\epo)|\text{\textbf{COMP}},\ln(\hr)\big) \\ \nonumber
&+& P\big(\text{\textbf{SBPL}}\big)\times P\big(\ln(\epo)|\text{\textbf{SBPL}},\ln(\hr)\big)
\end{eqnarray}
where
\begin{eqnarray}
P\big(\text{\textbf{Band}}\big) &\sim & 118/325 \\
P\big(\text{\textbf{COMP}}\big) &\sim & 70/325 \\
P\big(\text{\textbf{SBPL}}\big) &\sim & 137/325 \label{eq:modelprob}
\end{eqnarray}
\begin{figure}
\includegraphics[scale=0.31]{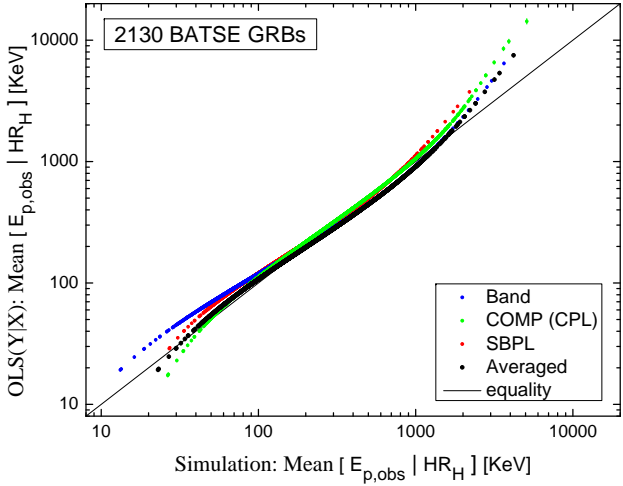}
\caption{Plot of Mean$[\epo]$ of 2130 GRBs in the BATSE catalog estimated by the OLS($\epo|\hr$) linear regression (\S2.2) versus Mean$[\epo]$ estimates of the same bursts derived from the simulation of \hrep ~relation (\S3.2), depicting the significance of the differences between the two estimates given different $\hr$. The blue/green/red circles correspond to the estimates of \epobs of the bursts assuming that all are best fit by the Band/COMP/SBPL models respectively. The black circles represent \epobs estimates from the OLS($\epo|\hr$) linear regression fit to the entire calibration sample versus weighted averages of \epobs estimates of the three GRB models derived form the simulation (\S3.2). The black solid line represents the equality. As seen, the two model-independent \epobs estimates (black circles) agree well with each other for almost the entire BATSE catalog GRBs. \label{SIMOLSEPK}}
\end{figure}
The above probabilities have been estimated from K06 time-integrated sample of GRBs assuming that the global fraction of GRBs best fit by each of the spectral models are the same as the fractions found in K06 sample of GRBs. This might not be far from the truth, since the brightness of GRBs in K06 sample, assures us that the spectral fits are not biased due to low $S/N$. The averaged $90\%$ Prediction Interval (PI) for \hrep ~relation is depicted in Figure~\ref{PIs} ({\it bottom right}). The $5th$ order polynomial fits to $90\%$ prediction intervals are also given in Table~\ref{PIfits}.

Based on the linear regression fits in \S2.2, and the simulated conditional probabilities $P\big(\ln(\epo)|\ln(\hr)\big)$ \& PIs for the three GRB models in \S3.2, the mean and the most probable $\epo$ of 2130 BATSE catalog GRBs are estimated and tabulated in Table~\ref{BATSEEPKs}.

It is useful to examine the behavior of the conditional probability P($\epo|\text{Model},\hr$) for the 3 GRB models for different $\hr$. Figure~\ref{moments} depicts the plots of the first four moments of P($\epo|\text{Model},\hr$) given $\hr$. As seen, the conditional densities appear to be significantly different from the normal distribution by being highly skewed and leptokurtic. The plots also clarify the patterns of heteroscedasticity in the linear regressions discussed in \S2.2. Although in such cases, the OLS linear regressions are not {\it BLUE}, comparison of {\it Mean}$[\epo]$ of 2130 BATSE GRBs derived from the simulation with their {\it Mean}$[\epo]$ estimates by the OLS($\epo|\hr$) linear regression in \S2.2, indicates the consistency of two methods over a wide range of BATSE $\hr$ (Figure~\ref{SIMOLSEPK})
\begin{figure*}
\includegraphics[scale=0.31]{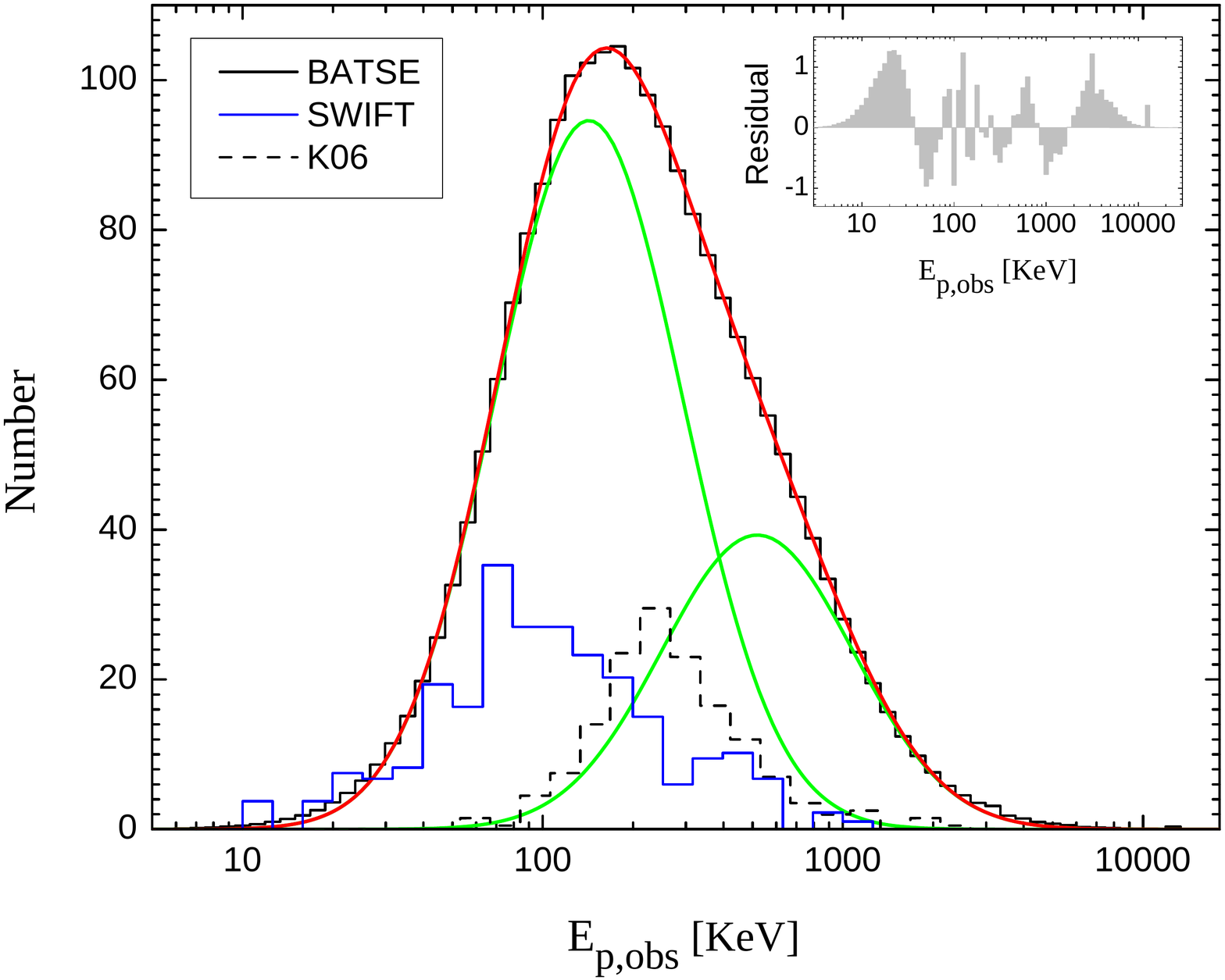}
\includegraphics[scale=0.31]{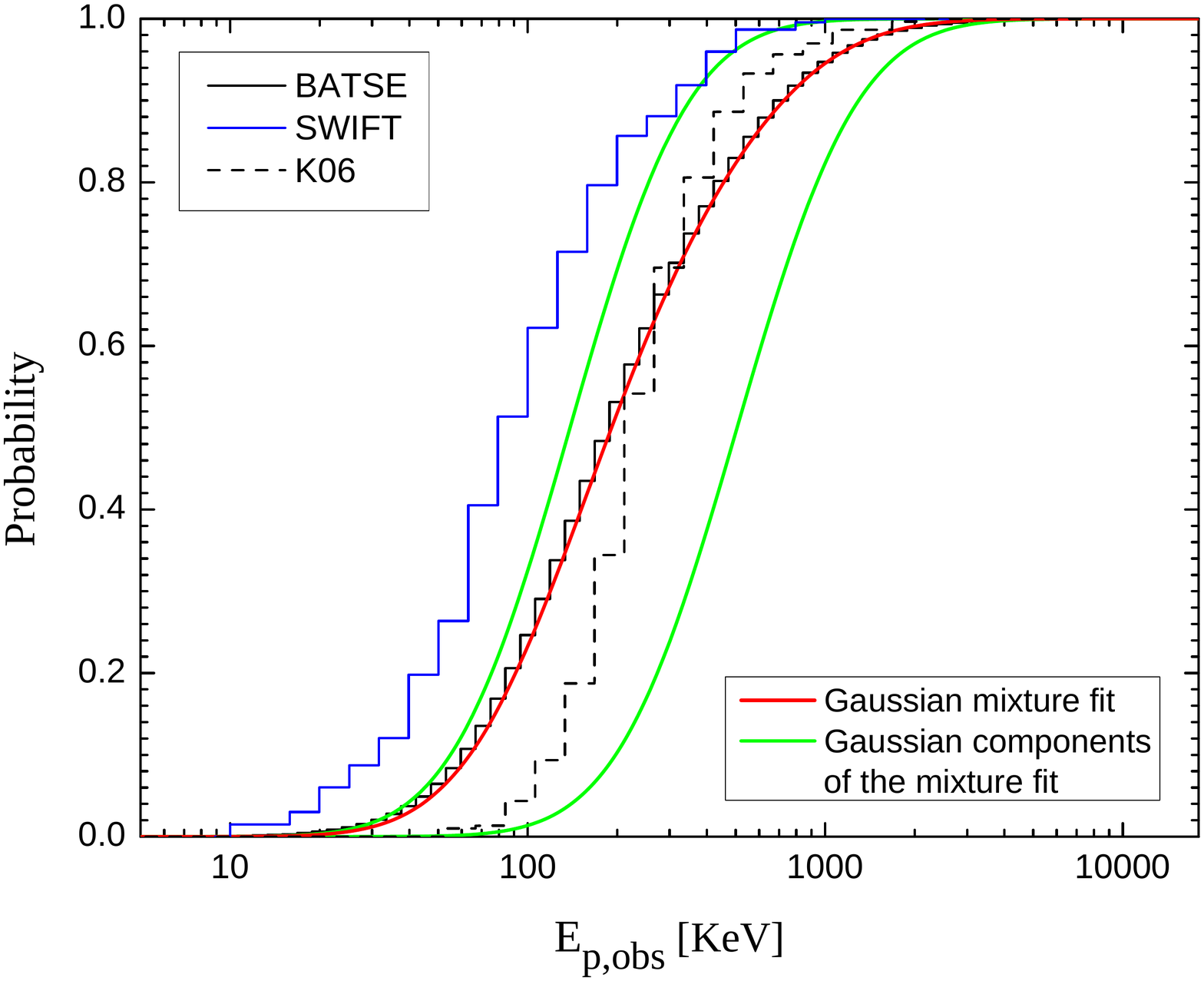}
\caption{{\bf Left:} The asymptotic $\epo$ distribution of 2130 BATSE GRBs derived from the integration of the conditional probability densities $P(\epo|\hr)$ over the entire BATSE catalog of GRB hardness ratios (Eqn.~\eqref{eq:epkbatse}). Mixture decomposition of the BATSE catalog $\epo$ distribution (black solid line) indicates that the distribution is well fit by the sum of two Gaussians (green solid lines) resulting in a reduced $\chi^2$ value of 0.26. The inset graph, represents the residuals of the fit. By contrast, a single Gaussian results in a poor fit to the BATSE catalog $\epo$ distribution at $>99.9\%$ confidence level with a resulting reduced $\chi^2$ value of $9.6$. The parameter estimates of the fits are given in Table~\ref{gaussianfits}. The two Gaussians might represent the $\epo$ distribution of BATSE LGRBs and SGRBs peaking at $\sim140$ KeV \& $\sim520$ KeV respectively. The dashed-line histogram represents the $\epo$ distribution of 299 bright BATSE GRBs in K06. Although K06 sample mainly consists of LGRBs, the $\epo$ distribution of the sample is biased towards harder bursts, as it had been predicted in Figure~\ref{flux-limit}. The solid blue histogram represents the $\epo$ distribution of SWIFT LGRBs taken from Butler et al. (2009b). As expected, the SWIFT GRBs appear to be much softer than BATSE GRBs, due to the relative energy sensitivity of SWIFT as compared to BATSE. {\bf Right:} Normalized cumulative $\epo$ distribution of 2130 BATSE catalog GRBs. The lines and colors have the same meaning as in the left plot. Assuming that the two Gaussians in the mixture fit to the BATSE $\epo$ distribution represent the $\epo$ density functions of BATSE LGRBs and SGRBs, then strong limits can be set on the fraction of BATSE LGRBs/SGRBs having an $\epo$ larger/smaller than a specified $\epo$. These limits might represent the global fractions of LGRBs/SGRBs having an $\epo$ larger/smaller than a specified $\epo$, provided that the effects of BATSE LAD's limited energy window ($20-2000$ KeV) on the global distribution of $\epo$ of GRBs are negligible. \label{BATSEEPKPDF}}
\end{figure*}

\section{Discussion and Summary}

Throughout this work, we presented a simple method of estimating the spectral peak energies of BATSE Gamma Ray Bursts via their hardness ratios $\hr$, defined as the sum of the fluences in channels 3 \& 4 of BATSE LADs (100-2000 KeV), divided by the sum of the fluences in channels 2 \& 1 (20-100 KeV), available in the BATSE GRB catalog.  Based on the strong correlation found between $\hr$ \& $\epo$ of 249 bright BATSE GRBs, the \hrep ~relation was constructed and shown to be linear over the wide energy range of BATSE (\S2.2). Although the relation is calibrated by a sample of 249 bright GRBs, since the hardness ratio is independent of GRB brightness, the relation can be used for nearly the entire BATSE sample: 2130 GRBs. 

\begin{table}
\caption{Coefficients of the $5th$ order polynomial fits to the means \& modes of P($\epo|\text{Model},\hr$) and the corresponding $90\%$ prediction intervals represented by Upper and Lower Confidence Levels (UCL \& LCL) for the three GRB spectral models: Band, COMP (CPL) \& SBPL. Also presented are the coefficients of the $5th$ order polynomial fits to the mean of the weighted average of P($\epo|\text{Model},\hr$) of the three GRB models and the corresponding $90\%$ UCL and LCL derived in \S3.2. \label{PIfits}}
\begin{tabular}{l l l l l l l l}
\hline
\hline
	& $a_0$ & $a_1$ & $a_2$ & $a_3$ & $a_4$ & $a_5$ \\
\hline
\multicolumn{7}{c}{Band} \\
\hline
UCL & 2.0 & 0.74 & 0.15 & -0.047 & -0.026 & 0.0067 \\
LCL & 1.5 & 0.83 & 0.065 & -0.071 & -0.008 & 0.0055 \\
Mean & 1.8 & 0.79 & 0.036 & -0.06 & 0.0078 & 0.00032 \\
Mode & 1.8 & 0.74 & 0.0094 & -0.048 & 0.0083 & 0.00021 \\
\hline
\multicolumn{7}{c}{COMP (CPL)} \\
\hline
UCL & 2.14 & 0.81 & 0.19 & -0.094 & -0.024 & 0.0098 \\
LCL & 1.74 & 0.66 & 0.11 & -0.06 & -0.012 & 0.0053 \\
Mean & 1.95 & 0.68 & 0.15 & -0.061 & -0.022 & 0.0074 \\
Mode & 1.92 & 0.62 & 0.12 & -0.046 & -0.013 & 0.0044 \\
\hline
\multicolumn{7}{c}{SBPL} \\
\hline
UCL & 2.16 & 0.5 & 0.19 & -0.027 & -0.038 & 0.0088 \\
LCL & 1.7 & 0.7 & 0.021 & -0.049 & -0.0094 & 0.0051 \\
Mean & 2.0 & 0.56 & 0.057 & -0.034 & -0.0011 & 0.00098 \\
Mode & 2.02 & 0.49 & 0.041 & -0.026 & 0.0003 & 0.00057 \\
\hline
\multicolumn{7}{c}{Weighted Average of the 3 GRB models} \\
\hline
UCL & 2.1 & 0.62 & 0.23 & -0.049 & -0.04 & 0.01 \\
LCL & 1.6 & 0.84 & 0.045 & -0.1 & 0.0054 & 0.0048 \\
Mean & 1.9 & 0.68 & 0.11 & -0.06 & -0.012 & 0.0055 \\
\hline
\hline
\end{tabular}
NOTE. -- The desired parameter can be estimated according to the equation:
\begin{equation}
y = a_0 + a_1x + a_2x^2 + a_3x^3 + a_4x^4 + a_5x^5, \nonumber
\end{equation}
where $y$ represents the logarithm of the parameter to be estimated and $x$ represents $\log(\hr)$. In general, the $90\%$ prediction intervals, represented by UCL \& LCL, should always be reported with the estimated mean/mode of $\epo$.
\end{table}

\begin{table}
\caption{Parameter estimates of the Gaussian mixture and single Gaussian fits to the derived asymptotic $\log(\epo)$ distribution of 2130 BATSE catalog GRBs based on Eqn.~\eqref{eq:epkbatse}. \label{gaussianfits}}
\begin{center}
\begin{tabular}{l l l l l l}
\hline
\hline
Model & $\hat A$ & $\hat \mu$ & $\hat{\sigma^2}$ & Area & $\chi^2/$d.o.f \\
 (1) & (2) & (3) & (4) & (5) & (6) \\
\hline
\hline
Mixture & --- & --- & --- & --- & 0.26 \\
Component 1 & 94.61 & 2.15 & 0.10 & 0.70 & --- \\
Component 2 & 39.28 & 2.71 & 0.10 & 0.30 & --- \\
\hline
Single Gaussian & 102.55 & 2.28 & 0.17 & --- & 9.6 \\
\hline
\hline
\end{tabular}
\end{center}
NOTE. -- column (2), (3), (4) \& (6) represent the maximum heights (normalization factors), means, variances and the reduced $\chi^2$ values of the Gaussian fits respectively. Column (5) represents the relative contributions of the two Gaussian components to the mixture fit. The standard errors on all estimated parameters are on the order of the third decimal places and therefore not reported.
\end{table}

Using Makov Chain Monte Carlo techniques, we also presented a careful multivariate analysis of GRB spectral parameters for the three main GRB spectral models: Band, COMP (CPL) \& SBPL, subject to the possible effects of data truncation and sample-incompleteness on the observational data (\S3.1). Similar to Shahmoradi \& Nemiroff (2009a), we find indications of significant -- and in some cases strong -- positive correlations among the high-$/$low-energy photon indices of the bright BATSE GRBs and their $\epo$. Investigation of the origins of such positive trends requires an accurate modeling of the possible role of the limited BATSE energy window in shaping the distribution of BATSE GRBs' spectral parameters. In general, the evolution of the photon indices with $\epo$, if intrinsic to GRBs, should be sought in the time-resolved spectral analyses of BATSE GRBs where the possible effects of the dispersion of $\epo$ distribution due to cosmological redshifts are eliminated (e.g. K06; Lloyd-Ronning \& Petrosian 2002; Crider et al. 1997).

Based on the results of the multivariate analysis of observational data we also derived the $90\%$ prediction intervals for \hrep ~relation (\S3.2 \& Table~\ref{PIfits}). These estimates were then used to create a complete catalog of $\epo$ estimates for 2130 BATSE GRBs with measured fluences available in the BATSE catalog (Table~\ref{BATSEEPKs}). These estimates might be very useful in population studies of Long- \& Short-duration GRBs, also in the studies of GRB hardness-brightness correlations (e.g. Shahmoradi \& Nemiroff 2009a) and the possible use of GRBs as cosmological tools.

As an immediate outcome of the analyses presented here, we have derived a probabilistic $\epo$ distribution of the entire BATSE GRBs for the first time by integrating the simulated conditional distributions $P(\epo|\hr)$ (Eqn.~\eqref{eq:epkprob}) in \S3.2 over the entire BATSE catalog GRB hardness ratios (Figure~\ref{BATSEEPKPDF}),
\begin{equation}
\label{eq:epkbatse}
P(\epo) \propto \sum_{i=1}^{2130}P(\epo|H\!R_{H,i})
\end{equation}

As it was expected (e.g. Kouveliotou et al. 1993), the resulting distribution is significantly different from a single Gaussian distribution. By contrast, the sum of two Gaussians provides a better fit than the single Gaussian at $0.001$ significance level. The two fits are summarized in Table~\ref{gaussianfits}. Assuming a log-normal distribution for the $\epo$ of BATSE LGRBs and SGRBs, the two Gaussian components of the mixture likely represent the $\epo$ probability density functions of Long- \& Short-duration BATSE GRBs with peaks at {\bf $\sim$ 140 KeV \& $\sim$ 520 KeV} respectively.

It should be noted that although we focused our attention only on a specific definition of hardness ratio, other definitions might also show a strong correlation with $\epo$. In particular, the hardness ratios defined by the raw photon counts of GRBs might even appear as better and more straightforward estimators of $\epo$, since their calculations does not require folding of GRBs spectra with the Detector Response Matrices.

The authors would like to thank Robert Preece at NASA MSFC and David Band (UMCB) for several useful communications and comments on BATSE Large Area Detectors and the BATSE catalog of Gamma Ray Bursts. We also thank Lara Nava (SISSA) and Giancarlo Ghirlanda (INAF) for their comments and timely feedback on the manuscript.

\onecolumn


\clearpage

\begin{center}
\topcaption{Summary of the spectral properties of 249 BATSE GRBs used to calibrate the linear \hrep ~relation in \S2.2.$^\star$ \label{Calsample}}
\tablefirsthead{%
	\hline
	\hline
	Trigger $^a$ & Model $^b$ & \hrh $^c$ & $A$ $^d$ & $\epo$ $^e$ & $\alpha$ $^f$ & $\beta$ $^g$ & $E_{break}$ $^h$ & $\Lambda$ $^i$ \\
	(1) & (2) & (3) & (4) & (5) & (6) & (7) & (8) & (9) \\
	\hline}
\tablehead{%
	\hline
	\hline
	Trigger $^a$ & Model $^b$ & \hrh $^c$ & $A$ $^d$ & $\epo$ $^e$ & $\alpha$ $^f$ & $\beta$ $^g$ & $E_{break}$ $^h$ & $\Lambda$ $^i$ \\
	\hline}
\tabletail{%
	\hline \multicolumn{9}{r}{\emph{Continued on next page}}\\}
\tablelasttail{%
	\hline \hline \\}
\begin{supertabular}{lllllllll}
105	&	SBPL	&	$	1.41	\pm	0.12	$	&	$	3.28	\pm	0.03	$	&	$	130	\pm	17	$	&	$	-0.97	\pm	0.08	$	&	$	-2.96	\pm	0.13	$	&	$	126	\pm	14	$	&	$	0.4	$	\\
109	&	COMP	&	$	4.02	\pm	0.08	$	&	$	0.81	\pm	0.04	$	&	$	523	\pm	87	$	&	$	-1.24	\pm	0.06	$	&		---				&		---				&		---		\\
130	&	BAND	&	$	3.80	\pm	0.10	$	&	$	1.1	\pm	0.15	$	&	$	180	\pm	23	$	&	$	-1.23	\pm	0.13	$	&	$	-2.33	\pm	0.22	$	&	$	141	\pm	31	$	&		---		\\
143	&	SBPL	&	$	20.95	\pm	0.13	$	&	$	4.82	\pm	0.01	$	&	$	586	\pm	28	$	&	$	-1.06	\pm	0.01	$	&	$	-2.22	\pm	0.03	$	&	$	420	\pm	13	$	&	$	0.2	$	\\
219	&	SBPL	&	$	3.49	\pm	0.11	$	&	$	1.98	\pm	0.01	$	&	$	240	\pm	12	$	&	$	-1.23	\pm	0.02	$	&	$	-2.28	\pm	0.03	$	&	$	191	\pm	8	$	&	$	0.2	$	\\
226	&	COMP	&	$	6.34	\pm	0.25	$	&	$	0.37	\pm	0.03	$	&	$	404	\pm	63	$	&	$	-1.05	\pm	0.10	$	&		---				&		---				&		---		\\
249	&	SBPL	&	$	7.09	\pm	0.04	$	&	$	3.55	\pm	0.01	$	&	$	446	\pm	22	$	&	$	-1.06	\pm	0.01	$	&	$	-3.30	\pm	0.07	$	&	$	537	\pm	21	$	&	$	0.5	$	\\
298	&	SBPL	&	$	15.45	\pm	0.47	$	&	$	1.31	\pm	0.03	$	&	$	389	\pm	67	$	&	$	-1.24	\pm	0.04	$	&	$	-2.40	\pm	0.28	$	&	$	387	\pm	66	$	&	$	0.01	$	\\
\end{supertabular}
\end{center}
NOTE. -- \\
\noindent $^\star$ The hardness ratios are calculated via the BATSE catalog data, while the rest of spectral parameters are taken from K06. \\
\noindent $^a$ Burst's trigger number as reported in the BATSE catalog. \\
\noindent $^b$ \hrh represents Hardness Ratio as defined in \S2.2. No attempt was made to keep the significant digits. Values are rounded off at the $2nd$ decimal places. \\
\noindent $^c$ $A$ represents the normalization factor of the assumed spectral model for each GRB in units of $0.01\text{{\it ph}}\, s^{-1}cm^{-2}$. \\
\noindent $^d$ All spectral peaks \epobs and the corresponding $1\sigma$ uncertainties are in units of KeV. \\
\noindent $^e$ $\alpha$ represents the low-energy photon index of the Band Model for GRBs best described by the Band model, and the low-energy photon index of SBPL Model for GRBs best described by SBPL, also the photon index of COMP model for GRBs best described by COMP. \\
\noindent $^f$ $\beta$ represents the high-energy photon index of the Band Model for GRBs best described by the Band model, and the low-energy photon index of SBPL Model for GRBs best described by SBPL. \\
\noindent $^g$ $E_{break}$ represents the break-energy of the Band Model for GRBs best described by the Band model, and the break-energy of SBPL Model for GRBs best described by SBPL. All reported break-energies and the corresponding $1\sigma$ uncertainties are in units of KeV. \\
\noindent $^h$ $\Lambda$ represents the break scale of SBPL Model for GRBs best described by SBPL.


\clearpage

\begin{center}
\topcaption{\epobs estimates for 2130 GRBs in the BATSE catalog.$^{\star}$ \label{BATSEEPKs}} 
\tablefirsthead{%
	\hline
	\hline
	Trigger$^a$ & \hrh $^b$ & Mean[$\epo$]$^c$ & Mean[$\epo$]$^d$ & Mode[$\epo$]$^e$ & Mode[$\epo$]$^f$ & Mode[$\epo$]$^g$ & Mean[$\epo$]$^h$ \\
		& Hardness & OLS(Y\textbar X) & OLS bisector & Band & COMP(CPL) & SBPL & Expected \\
	(1) & (2) & (3) & (4) & (5) & (6) & (7) & (8) \\
	\hline}
\tablehead{%
	\hline
	\hline
	Trigger$^a$ & \hrh $^b$ & Mean[$\epo$]$^c$ & Mean[$\epo$]$^d$ & Mode[$\epo$]$^e$ & Mode[$\epo$]$^f$ & Mode[$\epo$]$^g$ & Mean[$\epo$]$^h$ \\
		& Hardness & OLS(Y\textbar X) & OLS bisector & Band & COMP(CPL) & SBPL & Expected \\
	\hline}
\tabletail{%
	\hline \multicolumn{8}{r}{\emph{Continued on next page}}\\}
\tablelasttail{%
	\hline \hline \\}
\begin{supertabular}{llllllll}
$ 105^*$ & $   1.41 \pm   0.12$ & --- & --- & $  81^{+  49}_{-  39}$ & $ 104^{+  80}_{-  34}$ & $ 124^{+  49}_{-  60}$ & $ 101^{+  58}_{-  48}$ \\
$ 107$ & $   1.22^{\dagger}$ & $   98.78^{+ 0.10}_{- 0.10}$ & $   83.01^{+  0.10}_{-  0.10}$ & $  73^{+  43}_{-  36}$ & $  94^{+  68}_{-  32}$ & $ 116^{+  45}_{-  58}$ & $  91^{+  53}_{-  44}$ \\
$ 108$ & $   64.20 \pm  28.70 $ & $  1242.15^{+ 3.86}_{- 3.85} $ & $  1595.74^{+  5.92}_{-  5.90} $ & $  949^{+1573}_{- 298} $ & $ 1280^{+2725}_{- 463} $ & $  795^{+1152}_{- 242} $ & $ 1297^{+1730}_{- 633} $ \\
$ 109^*$ & $   4.02 \pm   0.08$ & --- & --- & $ 174^{+ 134}_{-  72}$ & $ 212^{+ 262}_{-  66}$ & $ 212^{+ 120}_{-  80}$ & $ 218^{+ 152}_{-  92}$ \\
$ 110$ & $    2.37 \pm   0.80 $ & $   150.98^{+ 0.05}_{- 0.05} $ & $   136.22^{+  0.06}_{-  0.06} $ & $  119^{+  78}_{-  54} $ & $  147^{+ 145}_{-  47} $ & $  161^{+  74}_{-  70} $ & $  147^{+  90}_{-  65} $ \\
$ 111$ & $    0.68 \pm   0.15 $ & $    67.68^{+ 0.12}_{- 0.12} $ & $    53.38^{+  0.12}_{-  0.12} $ & $   47^{+  28}_{-  24} $ & $   66^{+  36}_{-  23} $ & $   87^{+  34}_{-  48} $ & $   61^{+  39}_{-  32} $ \\
$ 114$ & $   0.53^{\dagger}$ & $   57.99^{+ 0.13}_{- 0.13}$ & $   44.57^{+  0.13}_{-  0.12}$ & $  40^{+  25}_{-  21}$ & $  57^{+  28}_{-  20}$ & $  77^{+  32}_{-  45}$ & $  53^{+  36}_{-  29}$ \\
$ 121$ & $    2.35 \pm   0.22 $ & $   150.16^{+ 0.06}_{- 0.06} $ & $   135.36^{+  0.06}_{-  0.06} $ & $  118^{+  78}_{-  53} $ & $  146^{+ 144}_{-  47} $ & $  161^{+  73}_{-  70} $ & $  146^{+  90}_{-  65} $ \\
\end{supertabular}
\end{center}
NOTE. -- \\
\noindent $^{\star}$ All \epobs estimates are in units of KeV. Overall, we recommend the use of either \epobs estimates from OLS($\epo|\hr$) (column 3) or the expected \epobs estimates from the simulation (column 8) together with the $90\%$ upper and lower prediction intervals given in column (8). \epobs by the OLS-bisector (column 4) might be useful in cases where both \hrh and \epobs need to be treated impartially (e.g. Shahmoradi \& Nemiroff 2009a; Isobe et al. 1990). The model-dependent simulation-based estimates of \epobs (columns 5 \& 6 \& 7) might be used only in cases where the best fit spectral model of the GRB is known independently. In general, the $90\%$ lower and upper prediction intervals on the estimated \epobs {\it should always} be reported and considered in analyses. \\
\noindent $^a$ Burst's trigger number as reported in the BATSE catalog. Triggers that are labeled by $*$ in column (1), represent GRBs used to derive the regression lines in \S2.2. Therefore, their \epobs estimates from the linear regressions are not given. \\
\noindent $^b$ \hrh represents hardness Ratio as defined in \S2.2. No attempt was made to keep the significant digits. Values are rounded off at the $2nd$ decimal places. For 262 GRBs marked by $\dagger$ in column (2), the uncertainties in the fluences are greater than their reported fluences in BATSE catalog. Therefore, for these GRBs, the error propagation also results in \hrh uncertainties that are larger than the value of \hrh. In these cases, the uncertainties on \hrh are not reported. \\
\noindent $^c$ Mean response of OLS($\epo|\hr$) (Eqn.~\eqref{eq:olsyx}) with the corresponding $1\sigma$ uncertainties on the mean response. No attempt was made to keep the significant digits. Values are rounded off at the $2nd$ decimal places. \\
\noindent $^d$ Mean response of OLS-bisector (Eqn.~\eqref{eq:olsbisector}) with the corresponding $1\sigma$ uncertainties on the mean response. No attempt was made to keep the significant digits. Values are rounded off at the $2nd$ decimal places. \\
\noindent $^e$ Most probable \epobs with $90\%$ Prediction Interval (PI) derived from simulation in case of the Band model as the best spectral fit. \\
\noindent $^f$ Most probable \epobs with $90\%$ PI derived from simulation in case of the COMP (CPL) model as the best spectral fit. \\
\noindent $^g$ Most probable \epobs with $90\%$ PI derived from simulation in case of the SBPL model as the best spectral fit. \\
\noindent $^h$ The weighted average of \epobs of the three GRB models with $90\%$ PI derived from simulation according to Eqns.~\eqref{eq:epkprob}-\eqref{eq:modelprob}. 

\label{lastpage}

\end{document}